\documentclass[a4paper,fleqn,useAMS,usenatbib]{mnras}

\pdfoutput=1

\usepackage{natbib}
\usepackage{mathptmx}

\usepackage[T1]{fontenc}
\usepackage{ae,aecompl}

\usepackage[dvips]{graphicx}
\usepackage{amsmath}
\usepackage{amsfonts}
\usepackage{amssymb}
\usepackage{comment}
\usepackage{bm} 
\usepackage[dvipsnames]{xcolor}
\usepackage[]{hyperref}
\hypersetup{
	colorlinks=true,	
	urlcolor=MidnightBlue,		
	hypertexnames=true,
	plainpages=false,
	naturalnames=true,
	pdftitle={Diving orbits galactic center},    
	pdfauthor={Zephyr Penoyre},     
	linkcolor=WildStrawberry,          
	citecolor=Periwinkle,        
}

\usepackage[percent]{overpic}
\usepackage{subfig}
\usepackage{csvsimple}
\usepackage{lscape}
\usepackage[shortlabels]{enumitem}

\DeclareMathAlphabet{\mathcal}{OMS}{ntxsy}{m}{n}   
\SetMathAlphabet{\mathcal}{bold}{OMS}{ntxsy}{b}{n} 

\makeatletter
\def\env@matrix{\hskip -\arraycolsep 
  \let\@ifnextchar\new@ifnextchar
  \array{*{\c@MaxMatrixCols}c}}
\makeatother

\newcommand{\cz}{\mathrm{c}}
\newcommand{\sz}{\mathrm{s}}

\title[Disruptions in axisymmetric systems]{Disruptions of stars and binary systems on chaotic orbits in an axisymmetric Milky Way center}
\author[Z. Penoyre et al.]{Zephyr Penoyre$^{1}$\thanks{E-mail:
\href{mailto:zephyrpenoyre@gmail.com}{zephyrpenoyre@gmail.com}}, Elena Maria Rossi$^{1}$, Nicholas C. Stone$^{2,3}$\\
$^{1}$Leiden Observatory, Leiden University, P.O. Box 9513, 2300 RA Leiden, the Netherlands\\
$^{2}$Department of Astronomy, University of Wisconsin, Madison, WI, 53706, USA\\
$^{3}$Racah Institute of Physics, The Hebrew University, 91904, Jerusalem, Israel
}

\date{Submitted 17 April 2025;}

\pubyear{2024}

\begin{document}
\label{firstpage}
\pagerange{\pageref{firstpage}--\pageref{lastpage}}
\maketitle

\begin{abstract}
Non-spherical potentials allow a wide range of  trajectories, both regular and chaotic, whose periapse distances can vary orbit to orbit. In particular chaotic trajectories can bring a system arbitrarily close to the central massive black hole leading to a disruption. In this paper, we work with an observationally benchmarked model of the innermost 200 pc of the Milky Way and show that low z-angular momentum trajectories are commonly chaotic. We compute the timescales and properties of close pericenter passages, and compare the implied \textit{collisionless} disruption rate to the well-studied \textit{collisional} rate from 2-body scatterings. We find that the relative \textit{collisionless} rate can dominate by orders of magnitude. Our calculations are relevant for a wide range of disruption phenomena, including the production of hypervelocity stars (HVSs) and tidal disruption events (TDEs).
Most of these disruptions involve stars come from the Nuclear Stellar Cluster, with a pericenter distribution that strongly favours shallow encounters, and a preference for high inclination interactions. The latter implies that unbound disrupted material --whether ejected stars or stellar debris -- would be preferentially directed towards the galactic poles. Many of our conclusions apply generally to any galaxy with a non-spherical galactic centre potential and central massive black hole.
\end{abstract}


\section{Introduction}

When a bound system, be it an individual star, a binary, higher-multiple or cluster, passes close to a Massive Black Hole (MBH) it can be disrupted by the extreme tidal force. This liberates some of the mass and much of the energy (internal and orbital), leaving a tightly bound  remnant and expelling energetic ejecta. Depending on the system of interest, one or both of these may be observable.

Disrupted stars are observed as Tidal Disruption Events (TDEs), for which the bound debris forms a tightly bound accretion disk around the MBH \citep{Hills75, Rees88}. The formation and accretion of this disk causes a bright transient flare, observable in distant galaxies \citep{vanVelzen20, Gezari21}, which fades over the course of months to years.  Early efforts \citep{Frank76} to understand TDE rates led to the formulation of the stellar loss-cone theory, which quantifies \citep{Lightman77, Cohn78} the \textit{collisional} rate of disruption of stars which are are scattered (by two-body encounters) onto those radial orbits vulnerable to disruption.


If a binary system (in most cases we imagine two stars, but one or both could be a stellar mass compact object) is disrupted, one of the pair tends to be ejected at very high speeds, producing a Hyper-Velocity Star (HVS, \citealt{Hills88,Sari10}). The other becomes tightly bound to the MBH, on the kind of tens or hundreds of AU orbit observed in the S-stars around Sagittarius A$^{*}$ \citep{Ghez05, Gillessen09}. Thus, when such a disruption occurs in the Milky Way Galactic Center (GC) both remnants can be observable \citep{Verberne25} - one on a very tight orbit around the MBH, and the other moving through the galaxy at anomalously high speeds (see e.g. S5-HVS1 \citealt{Koposov20}, which is believed to have originated by this mechanism, alongside many other, harder to confirm candidates \citealt{Brown05,Brown18,Marchetti22}).  As with TDEs, HVS rates have often been computed in the framework of a \textit{collisional} loss cone \citep{Perets07}, although other mechanisms to produce HVSs have also been proposed \citep{Yu03,Fragione19}.


Other types of systems can also be vulnerable to tidal disruption if they experience near-radial orbits, for example star clusters and other extended objects such as Giant Molecular Clouds (GMCs). 
The outcome of this process is evidenced in the dearth of clusters and other extended systems in the inner few parsecs of the Milky Way center \citep{Neumayer20}. 

Close pericenter passages can even be relevant for a variety of non-tidal astrophysical phenomena: for example, if the massive central object is not a single MBH but a binary, stars scattered onto radial orbits will be ejected by the binary's time-dependent potential, carrying away the orbital energy of the binary MBH and slowing it \citep{Milosavljevic03,Yu03,Vasiliev15}.

Likewise, the orbital energy of a stellar mass compact object can be radiated (in gravitational waves) via a close passage and eventual merger with a MBH. These Extreme Mass Ratio Inspirals (EMRIs; \citealt{Sigurdsson97, Pau07}) could spend a long time (years, and thousands of orbits) in the observable frequency range of upcoming long-baseline interferometers such as {\it LISA} \citep{AmaroSeoane23} - which would allow precise measurements of the mass and spin of the host MBH \citep{Ryan95, Barack04}.

All of these disparate phenomena require a close encounter with the MBH. The rate and mechanism of such interactions in the Milky Way is the focus of this paper. Specifically we show that, due to the flattened potential of the GC, orbits which spend most of their time at large characteristic radii can occasionally dive to very small distances and disrupt. This can occur without interactions with other bodies, purely as a consequence of the non-conserved angular momentum and the chaotic motion possible in non-spherical potentials. 

We can compare this \textit{collisionless} channel for disruptions, to the much more widely studied \textit{collisional} mechanism of scattering systems via a random walk of 2-body encounters \citep{Frank76,Lightman77,Merritt13}. The two channels produce not just different rates, but also different properties of the disruption and observational consequences.  Collisionless loss-cone refilling has been previously studied in \citet{Magorrian99,Merritt04,Khan13,Vasiliev13,Vasiliev15} and \citet{Kaur24}, and is sometime visualized as a \textit{loss wedge} - an expanded version of the usual \textit{loss cone} that 
has grown to encompass a much larger part of angular momentum space.  We expand on the method and scope of these previous works, highlighting some specific changes in methodology that better capture the true physics and subsequent rates. We do this, for the first time, in the context of a detailed and up to date model of the Milky Way galactic center drawn from \citet{Sormani22}.

In section \ref{sec:coord} we map out the basics of orbits in axisymmetric potentials and in the near vicinity of an MBH. Section \ref{sec:model} introduces our model of the Milky Way galactic center. In section \ref{sec:orbits} we integrate orbits directly in this potential, highlighting which have the capacity to dive to arbitrarily small radii. Section \ref{sec:prob} derives the properties and timescales of encounters with the MBH. Finally in section \ref{sec:rates} we find the steady state rate of \textit{collisionless} disruptions and compare this to a simple calculation of the rate arising from the \textit{collisional} channel.

All of our calculations depend directly on the critical angular momentum of an orbit for which a system will tidally disrupt, $l_{\rm t}$, which is an indirect proxy for the mass and physical size of the system. This makes all the calculations presented generic for any type of system, single object or multiple, depending only on $l_{\rm t}$. However, as we do not specify the type or other properties (namely mass) of the systems we cannot account for the true population or lifetime. We will extend this formalism, addressing these limitations, in an upcoming work (Penoyre et al. 2025, in prep.) to provide realistic rates of disruptions for Milky Way
like galaxies.








\section{Motion in axisymmetric potentials}
\label{sec:coord}

We will consider orbits in an axisymmetric potential, $\Phi=\Phi(R,z) \leq 0$ where $R$ is the cylindrical radius, $z$ is the height above or below the equatorial plane, and the azimuthal angle $\phi$ does not enter into the potential. 

It will also be useful to consider the spherical coordinate system $r,\ \theta,\ \phi$ where $R=r\sz_\theta$ and $z=r \cz_\theta$ ($\sz_x$ and $\cz_x$ are shorthands for $\sin(x)$ and $\cos(x)$ that we'll use throughout). In this coordinate system the polar angle, $\theta$, ranges from 0 to $\pi$, and the azimuthal angle, $\phi$, from 0 to $2\pi$. The unit vectors $\bm{\hat{r}}$, $\bm{\hat{\theta}}$, $\bm{\hat{\phi}}$ are in the direction in which each quantity is increasing locally and obey a right handed convention in this order (i.e. $\bm{\hat{r}}\wedge \bm{\hat{\theta}}=\bm{\hat{\phi}}$).  It is also useful to note the time derivatives of each of these vectors, $\bm{\dot{\hat{r}}}=\dot{\theta} \bm{\hat{\theta}}+ \sz_\theta \dot{\phi}\bm{\hat{\phi}}$, $\bm{\dot{\hat{\theta}}}=-\dot{\theta} \bm{\hat{r}}+ \cz_\theta \dot{\phi}\bm{\hat{\phi}}$ and $\bm{\dot{\hat{\phi}}}=-\sz_\theta \dot{\phi}\bm{\hat{r}} - \cz_\theta \dot{\phi}\bm{\hat{\theta}}$.

In terms of these coordinates, a particle has position, velocity, and specific angular momentum 
\begin{align}
& \mathbf{r}=r \mathbf{\hat{r}} \\
& \mathbf{v}=\mathbf{\dot{r}}=\dot{r} \bm{\hat{r}} + r \sz_\theta \dot{\phi} \bm{\hat{\phi}} + r \dot{\theta} \bm{\hat{\theta}} \\
& \mathbf{l}=\mathbf{r}\wedge \mathbf{v}= r^2 \left( \dot{\theta} \bm{\hat{\phi}} - \sz_\theta \dot{\phi} \bm{\hat{\theta}} \right) = l_\theta \bm{\hat{\phi}} - \frac{l_z}{\sz_\theta}\bm{\hat{\theta}}
\label{eq:Lvect}
\end{align}
such that the magnitude of the three relevant angular momenta\footnote{What we write here as $l_\theta$ is sometimes denoted as $p_\theta$. For a given Lagrangian $\mathcal{L}$ that depends on variable $x$, the generalized momentum is $p_x =\partial_{\dot{x}} \mathcal{L}$. In these terms the angular momentum would be written as $l^2 = p_\theta^2 +\frac{p_\phi^2}{\sz_\theta^2}$.} are
\begin{align}
& \label{eq:angmom} l^2=l_\theta^2 +\frac{l_z ^2}{\sz_\theta^2} \\
& l_z = r^2 \sz_\theta^2 \dot{\phi} \\ 
& l_\theta = r^2 \dot{\theta}.
\end{align}
The (conserved) specific energy obeys
\begin{align}
& \label{eq:epsilon} \epsilon = \frac{v^2}{2} + \Phi = \frac{\dot{r}^2}{2} + \Phi_{\rm eff} \\
& \Phi_{\rm eff} = \Phi +\frac{l^2}{2 r^2}
\end{align}
which can be rearranged and differentiated to give
\begin{align}
& \dot{r}^2 = 2(\epsilon - \Phi_{\rm eff}) \\
& \ddot{r}=-\partial_r \Phi_{\rm eff}.
\end{align}

\subsection{Variation in angular momentum}

The variation of the potential as the orbit moves in the polar direction causes a torque, changing the overall angular momentum via
\begin{equation}
\bm{\dot{l}}=\bm{r} \wedge \bm{\dot{v}} = - \bm{r} \wedge \nabla \Phi = - \partial_\theta \Phi \bm{\hat{\phi}} + \frac{1}{\sz_\theta} \partial_\phi \Phi \bm{\hat{\theta}}
\label{eq:l_torque}
\end{equation}
(where for now we have made no assumption about the symmetry of the potential).
We can similarly work from equation \ref{eq:Lvect} to find
\begin{equation}
\bm{\dot{l}}= \dot{l}_\theta \bm{\hat{\phi}} + l_\theta \bm{\dot{\hat{\phi}}} + \left(\frac{\cz_\theta l_z}{\sz_\theta^2} \dot{\theta}  - \frac{\dot{l}_z}{\sz_\theta}\right)\bm{\hat{\theta}} - \frac{l_z}{\sz_\theta}\bm{\dot{\hat{\theta}}}=\left( \dot{l}_\theta - \frac{\cz_\theta}{\sz_\theta} \dot{\phi} l_z \right) \bm{\hat{\phi}} - \frac{\dot{l}_z}{\sz_\theta} \bm{\hat{\theta}}
\end{equation}
(all other terms in $\bm{\hat{r}}$ and $\bm{\hat{\theta}}$ directions cancel directly) and thus
\begin{equation}
\dot{l}_\theta= \frac{\cz_\phi}{\sz_\phi}\dot{\phi}l_z - \partial_\theta \Phi
\end{equation}
and
\begin{equation}
\dot{l}_z=-\partial_\phi \Phi.
\end{equation}

Invoking axisymmetry ($\partial_\phi \Phi = 0$) we recover $\dot{l_z}=0$. The conservation of the z component of the angular momentum can be seen as almost axiomatic of axisymmetric potentials, and can be derived in a number of (more direct) ways -- including via Noether's theorum. However this formulation is a useful framing here, as it shows that said conservation is a consequence of the absence of torque in the (local) $\bm{\hat{\theta}}$ direction.

Note that if $|l_z|>0$ then $l_\theta$ can vary even for a spherically symmetric potential ($\partial_\theta \Phi =0$) but in such a way that $\dot{l}=0$.

\subsection{Region explorable by an orbit}

The maximum instantaneous angular momentum for a given $\epsilon$ and $\theta$ can be found by rearranging the energy equation into the form
\begin{equation}
l^2 = \omega(r,\theta) - r^2 \dot{r}^2
\end{equation}
where
\begin{equation}
\omega(r,\theta)=2 r^2 \left( \epsilon - \Phi(r \sz_\theta, r \cz_\theta)\right)
\end{equation}
Across all possible orbits this is maximised, at a given position, by those with $\dot{r}=0$ (with $\dot{\phi}$ and $\dot{\theta}$ set by $r$, $l_z$ and $\epsilon$).
Thus for a given $\theta$ we can therefore find a latitudinal maximum angular momentum, 
\begin{equation}
\label{eq:l_max_theta}
l_{\rm max}(\theta)^2 = \omega(r_{\rm mid},\theta),
\end{equation}
where $r_{\rm mid}$ is the radii which maximises $\omega$.

$\Phi$ is negative and it's magnitude decreases with increasing $r$. For a bound orbit, $\epsilon<0$, and at large radii $\omega<0$ (corresponding to an imaginary $\dot{r}$ and thus a region the orbit cannot traverse). At small $r$ the potential term dominates and $\omega$ is positive. As we approach the origin ($r\rightarrow 0$) the potential cannot be steeper than $r^{-1}$ (a point mass) and the whole expression is proportional to $r^{n}$ for $n\geq 1$ such that if $r\rightarrow 0$ then $\omega\rightarrow 0$. Thus the radius which maximises the expression is intermediate (hence the choice of subscript) and can be solved numerically for $r_{\rm mid}$ by maximising $\omega$, or by solving $\partial_{r} \omega|_{r_{\rm mid},\theta}=0$.

For a fixed $l_z$ bound orbits can only exist, at a given $\theta$, for $l_{\theta}$ less than or equal to
\begin{equation}
\label{eq:l_theta_max_theta}
l_{\theta, \rm max}^2(\theta) = l_{\rm max}^2(\theta) - \frac{l_z^2}{\sz_\theta^2}
\end{equation}
and thus for any $l_z>0$ there is a limiting minimal $\sz_\theta$ that the orbit can reach.  In other words, any orbit with $l_z \neq 0$ cannot pass through the poles.

In an oblate potential (as we might assume for any reasonable disky galaxy-like model) at a fixed spherical radius, the magnitude of $\Phi$ (<0) is maximised in the disk plane ($z=0, \ \theta=\frac{\pi}{2}$) and thus the largest possible angular momentum an orbit can achieve is when passing through the equatorial plane.

For any given energy, there exists a circular orbit in this equatorial plane\footnote{Outside of the equatorial plane, a circular path would vary in $z$ and thus the potential could not in general be constant, giving time-varying non-radial acceleration inconsistent with circular motion.} at some $r=R=R_{\rm c}(\epsilon)$ such that $\dot{r}=\ddot{r}=0$ with a corresponding circular angular momentum
\begin{equation}
l_{\rm c}^2(\epsilon) = 2 R_{\rm c}^2 \left(\epsilon - \Phi(R_{\rm c},0) \right) = R_{\rm c}^3 \partial_r \Phi |_{R_{\rm c},0}.
\end{equation}
Thus $l_{\rm c} = l_{\rm max}(\frac{\pi}{2})$ and $R_{\rm c}$ is the corresponding maximising radius.
$R_{\rm c}(\epsilon)$ and $l_{\rm c}(\epsilon)$ are thus useful proxies for the physical size of the orbit and the magnitude of angular momentum -- although one should bear in mind that the latter is a true maximum magnitude, whilst some orbits can explore radii out to a few $R_{\rm c}$. The period associated with this circular orbit, $T_{\rm c}$, is also useful as a characteristic timescale.

As $|l_z/\sz_\theta| \leq l \leq l_{\rm max}(\theta)$ the orbit must stay within a region centred on the equatorial plane ($\theta=\frac{\pi}{2}$) such that $\sz_\theta \geq \sz_{\theta_{\rm min}}$ where $l_{\rm max}(\theta_{\rm min})\sz_{\theta_{\rm min}} = |l_z|$.

The minimum radius for a given polar angle, $r_{\rm min}(\theta)$, satisfies
\begin{equation}
\left(\frac{l_z}{\sz_\theta} \right)^2 = \omega(r_{\rm min},\theta)
\end{equation}
with the solution in the range $0<r \leq r_{\rm mid}(\theta)$. The equivalent maximum radius, $r_{\rm max}(\theta)$, satisfies the same expression but with the solution in the range $r_{\rm mid}(\theta)\leq r < \infty$. At $\theta=\theta_{\rm min}$ (or equivalently below the equatorial plane at $\theta=\pi-\theta_{\rm min}$) both limiting radii converge at $r_{\rm mid}$.


\subsection{Close encounters with a central massive object}
\label{sec:closeenc}

In a general potential there is no analytic expression for the apoapse or periapse radii, though they can be solved numerically by finding $r$ that satisfies equation \ref{eq:epsilon} when $\dot{r}=0$. However, if the innermost part of the potential is dominated by a point-like MBH of mass $M$, then the local potential is approximately Keplerian.\footnote{At very small distances, a factor of a few Schwarzchild radii, relativistic effects can become important -- and similarly for speeds approaching a significant fraction of the speed of light. Such conditions are not modeled here.}  This small region is often denoted the \textit{sphere of influence}, with $r \ll r_{\rm inf}$, the radius of influence. 

This radius is defined in a variety of different ways in the literature. For example, it might be natural to define it as the radius at which the potential of the MBH, $\Phi_{\rm MBH}$ 
is equal to that of the rest galaxy excluding the MBH, $\Phi_{\rm gal}$. The latter is normally non-singular, and thus tends to some constant value $\Phi_0=\Phi_{\rm gal}(0,0)$. Thus a reasonable point of transition would be where $\Phi_{\rm MBH}(r_{\rm inf}) =\Phi_0$.

It is more common to use a simpler definition, namely that the radius of influence describes the (approximate in an axisymmetric system) radius at which the stellar mass contained is equal to the mass of the black hole -- i.e. $M(r \le r_{\rm inf})=2M$.  In most cases both these methods yield similar radii.

For $r \ll r_{\rm inf}$, $\Phi(R,z) \approx -\frac{G M}{r}$ and thus equation \ref{eq:epsilon}, with $\dot{r}=0$, can be solved to give the pericenter radius
\begin{equation}
\label{eq:rperi}
r_{\rm p} \approx \frac{G M}{2 \epsilon_0} \left( \sqrt{1+\frac{2 \epsilon_0 l^2}{G^2 M^2}}-1\right)
\end{equation}
where 
\begin{equation}
\label{eq:eps0}
\epsilon_{0}=\frac{v^2}{2} + \Phi_{\rm MBH} = \epsilon-\Phi_0
\end{equation}
(i.e. the energy associated with just the local orbit around the MBH).

We can examine equation \ref{eq:rperi} in two regimes:
\begin{itemize}
\item $\epsilon_0 l^2 \gg G^2 M^2$ which gives $r_{\rm p} \rightarrow \sqrt{\frac{l^2}{2\epsilon_0}}$
\item $\epsilon_0 l^2 \ll G^2 M^2$ which gives $r_{\rm p} \rightarrow \frac{l^2}{2 G M}$.
\end{itemize}
In both cases, $r_{\rm p}$ decreases with decreasing $l$, thus if we are interested in the smallest possible periapse, we are necessarily interested in small $l$, which means that a close passage of interest is much more likely to occur in the second regime.

If there exists some critically small periapse distance, which we will here call $r_{\rm t}$ (the tidal radius) then there is some corresponding total angular momentum (at periapse) below which the orbit enters this critical regime,
\begin{equation}
\label{eq:lperi}
l_{\rm t} \approx \sqrt{2 G M r_{\rm t}}.
\end{equation}
We might also be interested in the (locally valid) eccentricity of such a passage close to the MBH. This obeys
\begin{equation}
e^2 = 1 + \frac{2 \epsilon_0 l^2}{G^2 M^2}
\end{equation}
and again assuming we are in the periapse-minimizing $\epsilon_0 l^2 \ll G^2 M^2$ regime we can write this as
\begin{equation}
e - 1 \approx \frac{\epsilon_0 l^2}{G^2 M^2}
\end{equation}
where $|e-1|  \ll 1$. Orbits with $\epsilon_0>0$ ($R_{\rm c} \gtrsim r_{\rm inf}$) will be mildly hyperbolic, and those with $\epsilon_0<0$ mildly eccentric.

\section{Model of the Galactic center}
\label{sec:model}

\begin{figure*}
\includegraphics[width=\textwidth]{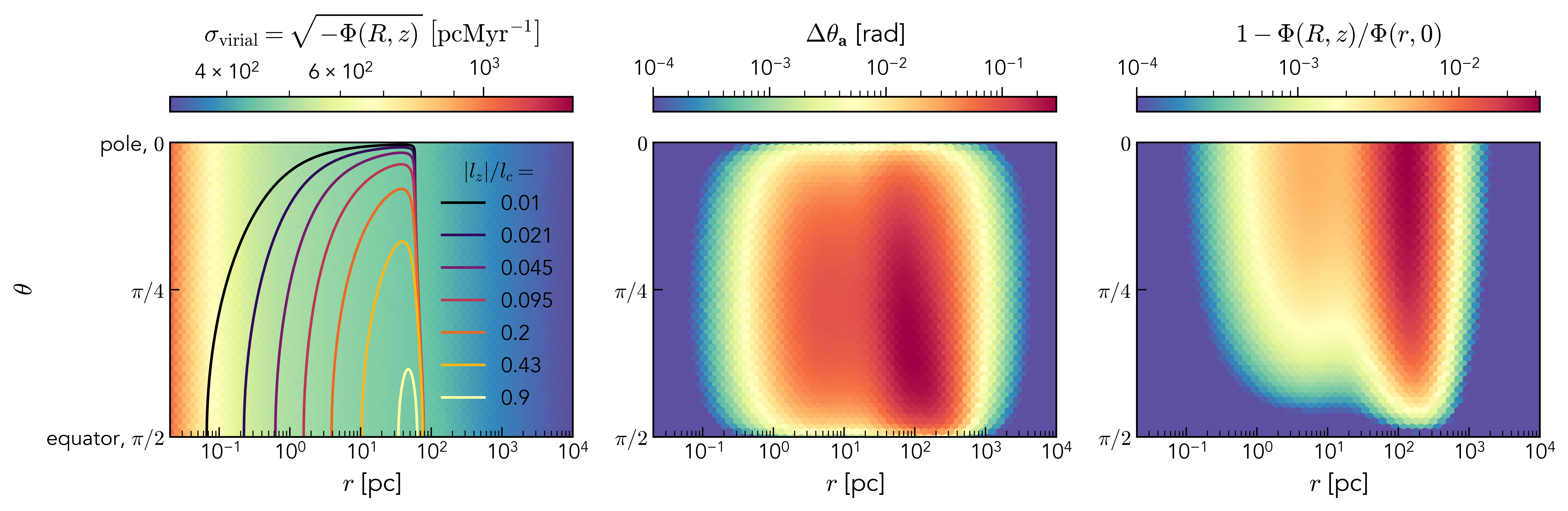}
    \caption{The potential, $\Phi(R,z)$, of the galactic center model used to integrate orbits throughout the rest of this paper. The \textit{left panel} expresses the potential through the virialized velocity dispersion (i.e. if all motion is random, obeying the virial theorem). We work in units of pc, Myr and $M_\odot$ throughout this work, but note that 1.0227 pc Myr$^{-1}$ is equal to 1 km s$^{-1}$ allowing easy conversion. We also show the limiting radii that a trajectory with fixed energy (corresponding to a circular radius of 50 pc) and z-angular momentum (see legend), can reach. 
    The two right-hand panels show the deviation of the potential from spherical. The \textit{middle panel} shows the difference in angle between the local acceleration and the radial vector (the direction of acceleration in a spherical potential) -- any deviation is always towards the disk-plane (steeper angles). The \textit{right panel} shows the difference between the local potential at some point, and the potential at the same spherical radius on the disk plane, which is always the same or larger. The right-hand panels show the largest deviations from sphericity around 100 pc, corresponding to the highly flattened Nuclear Stellar Disk (NSD). Interior to this ($\sim$ 1 to 10 pc), the potential is dominated by the slightly flattened Nuclear Stellar Cluster (NSC). At smallest radii the point-mass Massive Black Hole (MBH) dominates and within 0.1 pc the potential is essentially spherical. Note that while we only show the upper hemisphere ($\theta \leq \frac{\pi}{2}$), the behaviour is symmetric above and below the equatorial plane.}
    \label{fig:GCmodel}
\end{figure*}

\begin{figure}
\includegraphics[width=0.95\columnwidth]{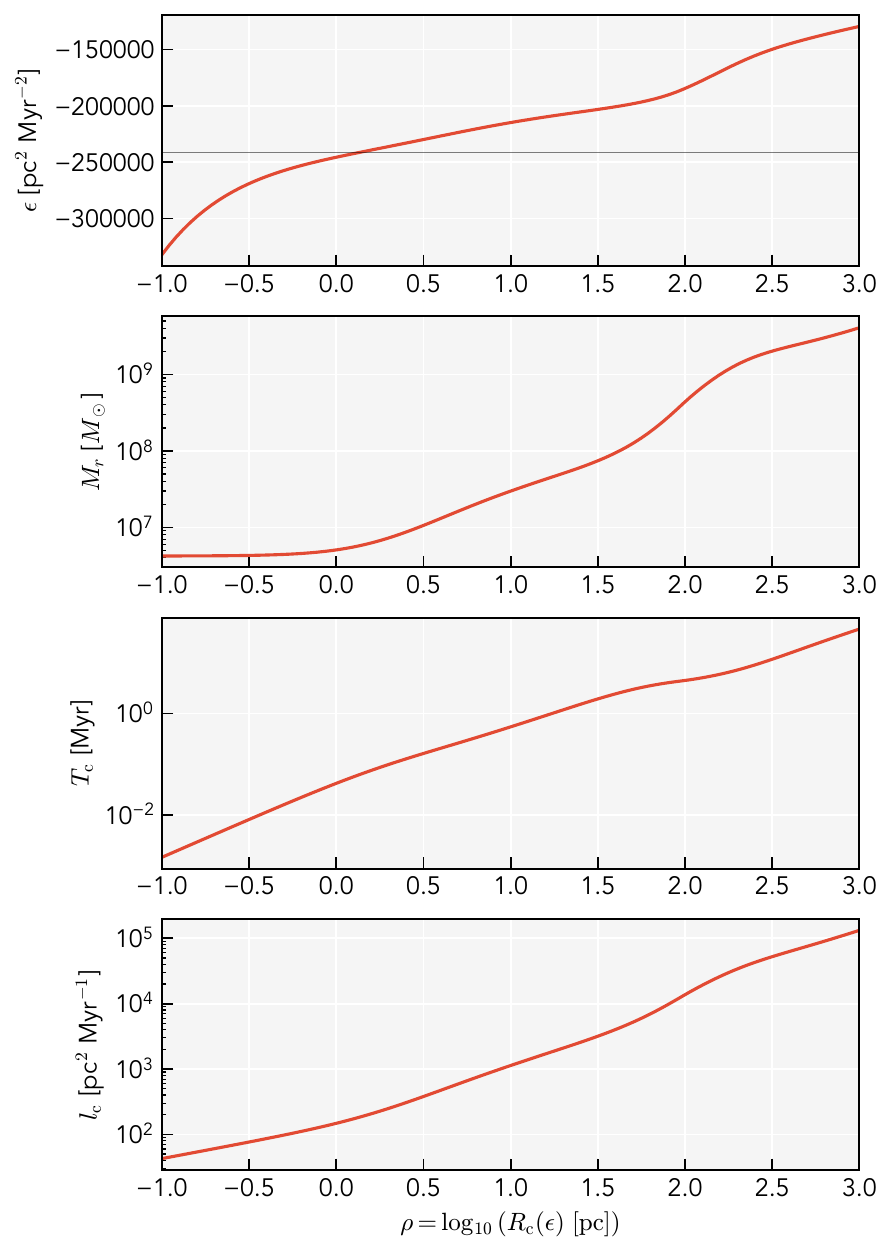}
    \caption{Relevant energy, mass, time and angular momentum scales of our GC model shown as a function of (log) circular radius. From top to bottom we show: the energy, $\epsilon$, the total mass enclosed within spherical radii $r=R_{\rm c}$, the period of a circular orbit, $T_{\rm c}$, and the angular momentum of a circular orbit, $l_{\rm c}$. The horizontal line in the $\epsilon$ plot shows $\Phi_0$ (see equation \ref{eq:eps0}), the central potential of our model excluding the MBH.  We work in units of $M_\odot$, pc and Myr but note that 1 pc Myr$^{-1}$ is almost exactly 1 km s$^{-1}$ allowing easy conversion.}
    \label{fig:galproperties}
\end{figure}

For the purpose of practical computation, we need a model of the Milky Way's central regions. Our model must specify the density, potential, and velocity structure (in the form of a distribution function, or DF) of the region in which our systems of interest reside.  The Galactic Center (GC) of the Milky Way is dominated by 3 components on scales of hundreds of parsecs or smaller:
\begin{itemize}
\item A significantly axisymmetric Nuclear Stellar Disk (NSD), which dominates by mass from hundreds of parsecs down to tens of parsecs;
\item A slighty axisymmetric Nuclear Stellar Cluster (NSC), which dominates from tens of pc down to around 1 pc;
\item A central MBH which dominates the inner few parsecs.
\end{itemize}
Other sources of mass, such as dark matter and interstellar gas, are assumed to be subdominant, and as our model is based on dynamical observations any contribution is implictly inlcuded in the stellar components.

We use a static density-potential pair, such that any orbit in our potential conserves energy. For much of this work, we employ the fiducial model of \citet[][S22 hereafter]{Sormani22}, which uses a NSC model from \citet{Chatzopoulos15} and a NSD model fitted to the actions of stars observed in the GC. Following S22 we use the \texttt{AGAMA} code \citep{Vasiliev18}, which has many capabilities that will be crucial for this work.\footnote{\texttt{AGAMA} includes a publicly accessible reproduction of the computation of the self-consistent NSD potential and DF performed in S22 - \url{https://github.com/GalacticDynamics-Oxford/Agama/blob/master/py/example_mw_nsd.py}} Firstly, \texttt{AGAMA} allows us to iteratively converge a DF and potential-density pair to reach self-consistency.

In S22, and in this work, a GC model is constructed by defining a fixed density structure for all components except the NSD. The NSD is definied by its DF, with parameters set by observable stellar actions, for which we can find the consistent density profile  iteratively (see S22 for details). 

The S22 model does not include central MBH, which is irrelevant for their NSD fit, as few NSD stars pass through the inner few parsecs where the MBH potential is significant. However, the MBH is important for our examination of orbits that traverse very small radii. We introduce the MBH after performing the self-consistency iterations (as we find that a point-like contribution to the potential can lead to instability in the iterative procedure).  We model the MBH as a Plummer sphere with $M_{\rm BH}=4\cdot 10^6 M_\odot$ and a scale radius $R_{\rm BH}=10^{-3} ~{\rm pc}$ (i.e. not an actual point mass, for numerical stability). As we add an extra mass not included in the iterative procedure, we correspondingly subtract $M_{\rm BH}$ from the NSC, such that the total mass internal to the NSD is constant.

The orbits considered in this work will generally be confined to the innermost few hundred parsecs, but on larger scales (which they sometimes probe), the galactic bulge dominates the potential. Given the focus of this work, we model this very simply as a Plummer sphere with mass $M_{\rm Bulge}=2 \cdot 10^{10} M_\odot$ and scale radius $R_{\rm Bulge}=0.5$ kpc \citep{Miyamoto75}. Similarly to the central MBH, we subtract the masses of the internal components ($M_{\rm NSD} = 10^9 M_\odot$ and $M_{\rm NSC}=6\cdot 10^7 M_\odot$) from this bulge mass to avoid double counting of any component.

We also define a fixed NSC DF from an updated model for the MW NSC (Vasiliev et al. in prep) which follows a similar method to S22 fitting NSC stars in action space. As discussed in Appendix \ref{ap:n} this produces a distribution function that is centrally tangentially biased and which we will use to sample the orbital properties of systems in the NSC later. We do not perform the same iterative updating of the NSC potential to reach a self-consistent solution. Therefore, this NSC DF and the NSC potential from S22 that is used in our GC model are likely to be mildly inconsistent. However, as this DF is fitted to NSC data, we use it as a best available approximation to the real kinematics in the NSC.


The resulting combined potential for the GC model is shown in Figure \ref{fig:GCmodel}. 
At large radii the potential is dominated by the bulge, and thus shows no signs of asphericity. Moving within a few hundred parsecs, the NSD comes to dominate, from which we see a significant dampening of the potential towards the poles. Away from the poles and disk plane we see significant non-radial acceleration. Below a few tens of parsecs the NSC dominates, which is still slightly flattened but much less so than the NSD. Finally, within parsec scales the MBH dominates, restoring spherical symmetry to the potential and aligning the acceleration vector closer and closer to the radial direction.

Even in the NSD-dominanted regions, the degree of asphericity is small -- the maximum deviation of the acceleration is of order 0.1 radians: i.e. acceleration is still close to radial, and only slightly directed down towards the disk plane. Similarly, the potential away from the plane is at most a few percent less than in-plane. This is not itself surprising, as even a razor thin-disk density model has a roughly (though not exactly) spherical potential.As we will see, however, even these modest potential asphericities produce a broad range of orbital behaviour, that ultimately stems from non-conservation of $\mathbf{l}$.

In Figure \ref{fig:GCmodel}, we also show the limiting region that an orbit of given $\epsilon$ and $l_z$ can traverse, using the example of $R_{\rm c}(\epsilon)=$50 pc and a range of different $l_z$. These lines correspond to the apo- and periapse for an orbit with $l_\theta = 0$; for $l_\theta>0$, the accessible region is further reduced. As we will see, particular trajectories only explore a certain range of $l_\theta$ and thus some orbits may not reach $l_\theta=0$ and explore the full region. 

For $|l_z| \rightarrow l_{\rm c}$, the orbit is confined to the disk plane and a narrow range of radii, whilst for smaller $|l_z|$ it can explore closer to the poles and potentially much smaller radii (the maximum radius, when $l_z=0$, asymptotes to a constant $r_{\rm max}$ such that $\Phi(r_{\rm max},0)=\epsilon$). Only orbits with $l_z=0$ can pass through the poles. 

We also show in Figure \ref{fig:galproperties} the physical scales of our GC model, and how they vary with $R_{\rm c}$. The first panel shows the conversion between energy, $\epsilon$, and $R_{\rm c}$. We also show $\Phi_0$, the central potential excluding the MBH. Where the two lines meet roughly defines the influence radius and where the MBH potential comes to dominate on sub-parsec scales. The mass profile similarly shows the MBH (with mass $4\cdot 10^6 M_\odot$) dominates the inner parsec, but is quickly superseded by the NSC (a few $10^7 M_\odot$) up to about 10 parsec, the NSD ($10^9 M_\odot$) out to hundreds of parsecs, beyond which the bulge dominates. The characteristic timescale of orbits, $T_{\rm c}$, is the period of a circular orbit. $T_{\rm c}$ is a little under 100 kyr at the radius of influence and increases rapidly moving outwards ($\underset{\sim}{\propto} R_{\rm c}$). The charecteristic (and maximal) angular momentum of a circular orbit, $l_{\rm c}$, also increases sharply with radius (also $\underset{\sim}{\propto} R_{\rm c}$).

\section{Trajectories}
\label{sec:orbits}

\begin{figure}
\includegraphics[width=0.9\columnwidth]{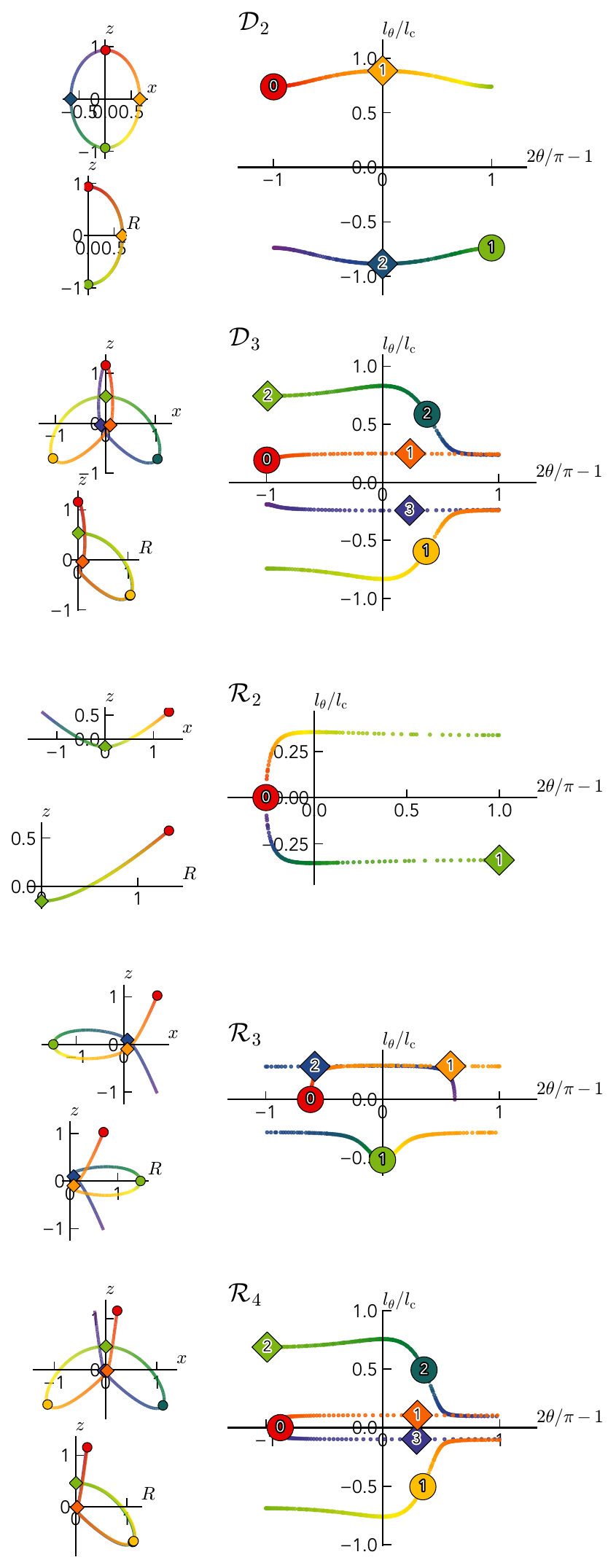}
    \caption{Examples of 5 closed trajectories in the GC potential, all with $R_{\rm c}(\epsilon)=50$ pc and $l_z=0$. We label them by their classification, $\mathcal{D}_{\rm n}$ or $\mathcal{R}_{\rm n}$ (see section \ref{sec:sos}). For each we show the trajectory, in units of $R_{\rm c}$, in the $x,\ z$ plane (upper left) and $R,\ z$ plane (lower left). The larger right-hand panel shows the evolution in $\theta$ (normalized to run from 1 at the north pole, at the 0 equator and -1 at the south pole) and $l_\theta$ (normalized by the maximum angular momenta for a given energy $l_{\rm c}(\epsilon)$).  Apoapsides are denoted by circles, and periapsides by diamonds, and numbered sequentially with 0 being the initial apoapse. For reversible ($\mathcal{R}_{\rm n}$) trajectories we only show the first half of a full closed trajectory, as after this they double back on themselves in physical space and they mirror (through $l_\theta=0$) their path in $(\theta,l_{\theta})$ space.}
\label{fig:traj_examples_grid}
\end{figure}


In the following, we will refer to \textit{orbit} as a single radial period of the full \textit{trajectory}, that conversely is the long-term behaviour over many individual \textit{orbits}.



We are particularly interested in finding those trajectories that reach very small pericenter distances and thus could be disrupted by the central MBH. This can be characterised by a critical \textit{tidal angular momentum} at periapse, $l_{\rm t}$, below which a disruption is expected to occur (see equation \ref{eq:lperi}).

As $l \ge l_z$ always, any trajectory that can experience disruption must have $l_z<l_{\rm t}$. Because tidal radii are small compared to galactic scales (for example $r_{\rm t}$ is typically $\lesssim$ 100 AU for a hypervelocity star progenitor, or 1 AU for a TDE), $l_{\rm t}$ is tiny compared to typical angular momenta of most trajectories. However, while the average angular momentum of a trajectory may be large, for diving orbits $l_\theta$ can vary over orders of magnitude and may approach values $<l_{\rm t}$ for some orbits. 

In order to examine which kind and fraction of trajectories manage to reach this disruption angular momentum parameter space , we use the \texttt{AGAMA} package \citep{Vasiliev18} which can integrate trajectories in the potential defined in section \ref{sec:model} using a modified version of the 8th order adaptive-timestep Runge-Kutta scheme DOP853 \citep{Hairer93}.

We have two conserved quantities, $\epsilon$ and $l_z$, and a coordinate which contains no information, $\phi$, and thus the 6D phase space reduces to 3D. Many authors take these dimensions to be $R,\ z$ and $v_{R}$ (the velocity in the $R$ direction).\footnote{Using conservation of energy, we could substitute any of these out for $v_z$, the vertical velocity.}  Here we instead employ a different set of coordinates, inspired in part by \citet{Magorrian99}, examining the polar angle $\theta$, and its corresponding angular momentum $l_\theta$ at apo- or peri-apse (where $\dot{r}=0$ and $\ddot{r}$ is negative or positive respectively). Either set of coordinates is sufficient to fully define the state (except for an arbitrary offset in $\phi$).


\texttt{AGAMA} returns both the timesteps it used for computation and an interpolator for finding the state of the system between those timesteps. This allows us to resample the time series, for example to find which original timesteps, $t_i$ and $t_{i+1}$, correspond to a change in sign of $v_r = \mathbf{\hat{r}}\cdot \mathbf{v}$ from positive to negative, and then to find accurately the moment of apoapse corresponding to $t_i < t_{\rm a} < t_{i+1}$ where $v_r(t_a)=0$. Across an integrated trajectory we can thus find $N$ apoapsides or periapsides 
at times $t_{\mathrm{a},n}$ (or $t_{\mathrm{p},n}$ for periapse) and corresponding properties of the orbit at those times, e.g. $z_{\rm a}$ or $l_{\theta, \rm p}$. Thus the initial accuracy of the orbit need only be sufficient to resolve the trajectory, and then the state at key times can be found more exactly from the interpolator.

Ideally we would like the number of apoapsides, $N$, to be high enough to reliably diagnose the orbit but not so high as to require excessive computation. For a general potential, the 
radial period $T_{r,n} = t_{\mathrm{a},n+1} - t_{\mathrm{a},n}$ is neither constant nor simply related to the  azimuthal period. However we can use the fact that in general $\frac{1}{2} \lesssim \frac{T_{r,n}}{T_{\rm c}} \lesssim 1$. We use an initial guess as an ideal runtime, with $N_{\rm ideal}$ apoapsides, of $\frac{2}{3} N_{\rm ideal} T_{\rm c}$. 

\subsection{Surfaces of Section (SoS)}
\label{sec:sos}

In order to visualise and analyse the aforementioned 3D space for our trajectories, we will use the \textit{Surface of Section} (SoS) as a tool. A SoS projects a higher dimensional trajectory onto a 2D plane by examining the behaviour only at some fixed point on the orbit.  For example, SoS are commonly made for axisymmetric trajectories by examining $R$ and $v_R$ as the orbit crosses the $z=0$ equatorial plane. Our focus on the angular momentum of orbits at periapse instead motivates us to make our SoS in $\theta$ and $l_{\theta}$ when $\dot{r}=0$, thus specifiying the state at apoapse ($\ddot{r}<0$) and periapse ($\ddot{r}>0$).

In Figure \ref{fig:traj_examples_grid}, we show 5 example trajectories in our GC potential, all with $R_{\rm c}(\epsilon)=50$ pc.\footnote{If we compare families of trajectories with the same specific energy $\epsilon$ then $R_{\rm c}(\epsilon)$ and thus $l_{\rm c}(\epsilon)$ are fixed for all trajectories in this family.  We generally use $R_{\rm c}$ as a more intuitively accessible proxy for $\epsilon$ in our plots and discussion.} We show only examples with $l_z=0$, which prevents the trajectory from rotating around the z-axis. Hence we can set $y=0$ and $\dot{y}=0$ at all times, allowing easier visualization. For $l_z \neq 0$ trajectories, the $(x, z)$ plane becomes a less useful visualization, but the $(R, z)$ plane retains its clarity.

The trajectories in $(x,\ z)$ space are the most intuitive, with orbits oscillating above and below the disk plane ($z=0$), often with periapsides coming close to the origin. We can see that $l_\theta$ changes little during these rapid close passages, and that most of the torquing occurs during the longer times (and with the longer orbital lever arms) spent near apoapse. The apparent exception to this trend is the sign flip of $l_\theta$ when the orbit passes either pole ($\theta =0$ or $\pi$) but this is an effect of the coordinate system rather than the underlying dynamics. Conversely, the fast periapse passages are associated with rapid changes in $\theta$, partly due to the high Cartesian velocity, but also to the amplification of $\dot{\theta}=\frac{v_\theta}{r}$ that would occur at low $r$ even at constant velocity.

The trajectories shown in Figure \ref{fig:traj_examples_grid} are all \textit{regular} and \textit{closed} paths repeating the same shape indefinitely, and thus returning to the same points in the SoS each time. Slightly displaced from these initial conditions we would see a broader family of \textit{regular} paths, which are not closed and precess within a small volume of phase space around the closed trajectory (and similar occupy finite regions of the SoS adjacent to the repeating points of the closed orbit). Within the family of regular orbits associated with a given closed can also be found \textit{harmonic} closed orbits, which are closed trajectories that modify their parent trajectory with a secondary and subdominant distinct closed trajectory (and thus also move between a finite, but larger, set of points on the SoS).

Regular trajectories 
posses three integrals of motion (confining the region that they can explore). However, axisymmetric potentials 
only guarantee two integrals of motion ($\epsilon$ and $l_z$) and there also exist \textit{chaotic} trajectories that 
lack a third integral and thus explore a much wider region of state space.

Another distinct 
way to separate trajectories is by whether they always have a \textit{definite} or \textit{reversible} sense of polar rotation near apoapse. Some trajectories are capable of passing through $l_{\theta}=0$ at or near apoapse, corresponding a change of sign of $\dot{\theta}$ and a doubling back of the trajectory, which we would say have \textit{reversible} polar rotation. Other trajectories maintain non-zero $l_\theta$ and $\dot{\theta}$ near apoapse, and thus cannot double back on themselves and continue with the same \textit{definite} sense of polar rotation.\footnote{There is a sub-class of \textit{definite} trajectories, that are  called \textit{centrophillic} because they pass directly through the origin. They have many interesting properties, including mimicking the orbital frequency harmonics of the \textit{reversible} trajectories. However as our potential is singular at the origin they are not stable and do not appear in our analyses.}

Regular closed trajectories (excluding higher harmonics) can be described by their sense of rotation and
the number of apoapsides. Definite (reversible) trajectories with n apoapsides are
denoted as $\mathcal{D}_n$ ($\mathcal{R}_n$). In Figure \ref{fig:traj_examples_grid} we show, in order, the first two \textit{closed definite} trajectories and the first three \textit{closed reversible} trajectories: $\mathcal{D}_2$, (also commonly termed a box or loop trajectory), $\mathcal{D}_3$, $\mathcal{R}_2$ (or banana trajectory), $\mathcal{R}_3$ (or fish trajectory), and $\mathcal{R}_4$ (or pretzel trajectory). Definite trajectories require $n$ orbits to close, whilst reversible trajectories need $2(n-1)$ orbits.

This is a distinct but compatible schema for classifying orbits to that found in works such as \citet{Laskar90,Valluri98,Laskar99} and \citet{Nieuwmunster24}, which focus on harmonics of orbital frequencies. We favour our scheme here as it is topological in nature, rather than based on the (non-trivial) task of extracting a representative frequency from a simulated orbit.

Not all closed trajectories (and associated islands of regular trajectories) exist for a given potential and integrals of motion. In particular, for a potential with a central singularity, such as ours, trajectories that would come near the origin are likely to diverge and become chaotic.\footnote{The exception to this is orbits that remain deep within the sphere of influence (i.e. $R_{\rm c} \ll r_{\rm infl}$) which experience an essentially Keplerian potential and are thus regular.}  

These divergences are often associated with periapsides close to $l_\theta = 0$, where $r_{p}$ can be very small and the orbit can experience large variation in the acceleration that is specific to particular, small regions of phase space. This is a hallmark of a transition to chaos, where a strong influence from a small region of state space can lead to families of initially similar trajectories dispersing.

In general divergence of trajectories near the origin suppresses orbits of higher $n$, reducing the size of the associated island of regular trajectories, and potentially eradicating them. As we shall see, it is the chaotic trajectories that have the greatest importance for low-$l$ disruptions, and we therefore focus our attention in this section on diagnosing and quantifying populations of chaotic trajectories. In the SoS, regular trajectories are the `negative space' to the chaotic trajectories, and thus building intuition about the presence of one informs us about the absence of the other.

\begin{figure}
\includegraphics[width=0.95\columnwidth]{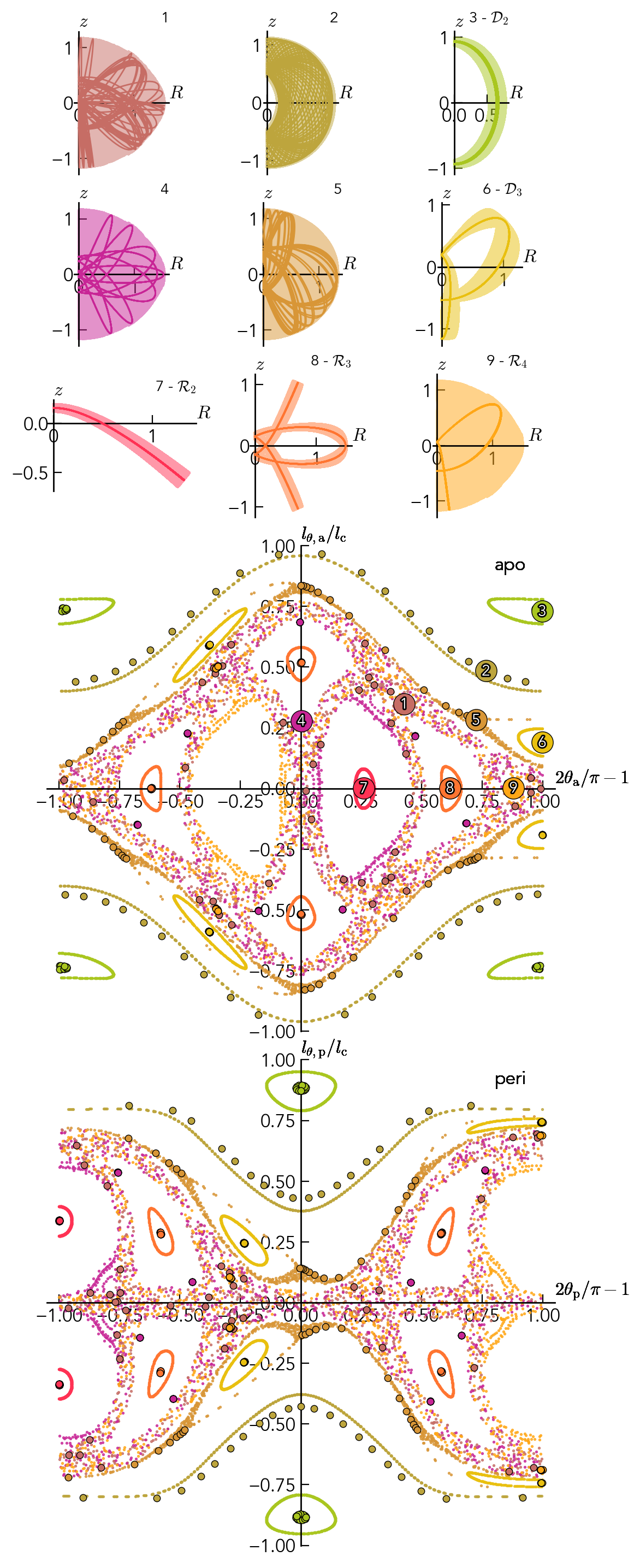}
    \caption{Examples of 9 trajectories integrated in our GC potential with $R_{c}= 50$ pc and $l_z=0$. We integrate each over $\sim$100 orbits (solid line, large dots), then perturb the initial conditions slightly and integrate over $\sim$1000 orbits (translucent line, small dots). In the 9 small panels at the top we show each trajectory in $R$ and $z$ space (in units of $R_{\rm c}$). 
    We also show the $(l_\theta, \theta)$ SoS at apoapse (middle panel) and periapse (bottom panel), for all integrated trajectories. The large numbered points in the apoapse SoS show the initial state of each unperturbed trajectory. 
    }
    \label{fig:traj_examples_9}
\end{figure}

In Figure \ref{fig:traj_examples_9}, we show examples of a handful of trajectories and their SoS, to demonstrate how the SoS can allow for easier identification of chaos. In the top panels of this Figure, we show nine trajectories in the $(R, z)$ plane (with trajectories 3, 6, 7, 8 and 9 corresponding to those shown in Figure \ref{fig:traj_examples_grid}) along with a shadowed region produced by long-term integration of the same trajectories with slightly perturbed initial conditions. Trajectories 2 and 4 are examples of higher harmonics (associated with $\mathcal{D}_2$ and $\mathcal{R}_4$ repsectively). We can see now how perturbing some of the closed trajectories (3, 6, 7 and 8) leads to regular trajectories filling a broader but still confined region of phase space. Some trajectories are initially chaotic (1 and 5) and remain so under perturbations, but others (4 and 9) become chaotic only after the perturbation. This suggests that there must only be a small island of regularity around those closed trajectories.

The space-filling nature of chaotic trajectories is somewhat apparent in the $(R,z)$ plane, but more clearly so in SoS shown in the bottom panels of Figure \ref{fig:traj_examples_9}. Closed trajectories can be seen as distinct points, regular trajectories as occupying well-constrained surrounding regions, and chaotic trajectories fill a large and complex ``chaotic sea''.

We also see that the chaotic perturbed trajectories 1, 4 and 9 overlap (in fact any two chaotic orbits that occupy the same region of the SoS are the same orbit separated only by time), but trajectory 5 is distinct. This exemplifies the dichotomy between definite and reversible trajectories; they are topologically distinct and thus cannot overlap.

\subsection{Classifying chaotic orbits}
\label{sec:chaotic_orbits}

We are interested in trajectories that come arbitrarily close to the 
minimum permissible pericenter (MPP) set by $l_z$ conservation.  For the dynamical processes motivating this paper (TDEs and hypervelocity stars), the relevant pericenters for disruption lie well inside the MBH sphere of influence, so the MPP is $r_{\rm MPP} \approx l_z^2 / (2GM)$.  
As $l \geq l_\theta$ at all times, we can define a necessary criterion for disruption to be that $l_{\theta, \rm p}$ can attain arbitrarily small values.  Note that this is not a sufficient criterion for disruption; this would also require that $r_{\rm MPP}$ is beneath the relevant tidal disruption radius.

We find that regular trajectories never satisfy these conditions. Chaotic trajectories, however, can. In other words, trajectories which traverse arbitrarily close to the center of the potential are necessarily chaotic. One way to explain this is to notice that an orbit approaching the center of galaxy from large distances follows a course almost independent of the potential (or gradient thereof) near the origin -- but there is a bifurcation in the rest of the orbit depending on whether it passes just above, or just below, the MBH. Thus 
there is a large change in the orbit's final state stemming from a small change in initial conditions. 


There can also be chaotic behaviour in orbits which do not pass arbitrarily close to the galactic center, so we start by classifying all orbits showing chaotic behaviour, and then defining a subset that pass near the MBH, which we will term \textit{diving orbits}.

\begin{figure*}
\includegraphics[width=\textwidth]{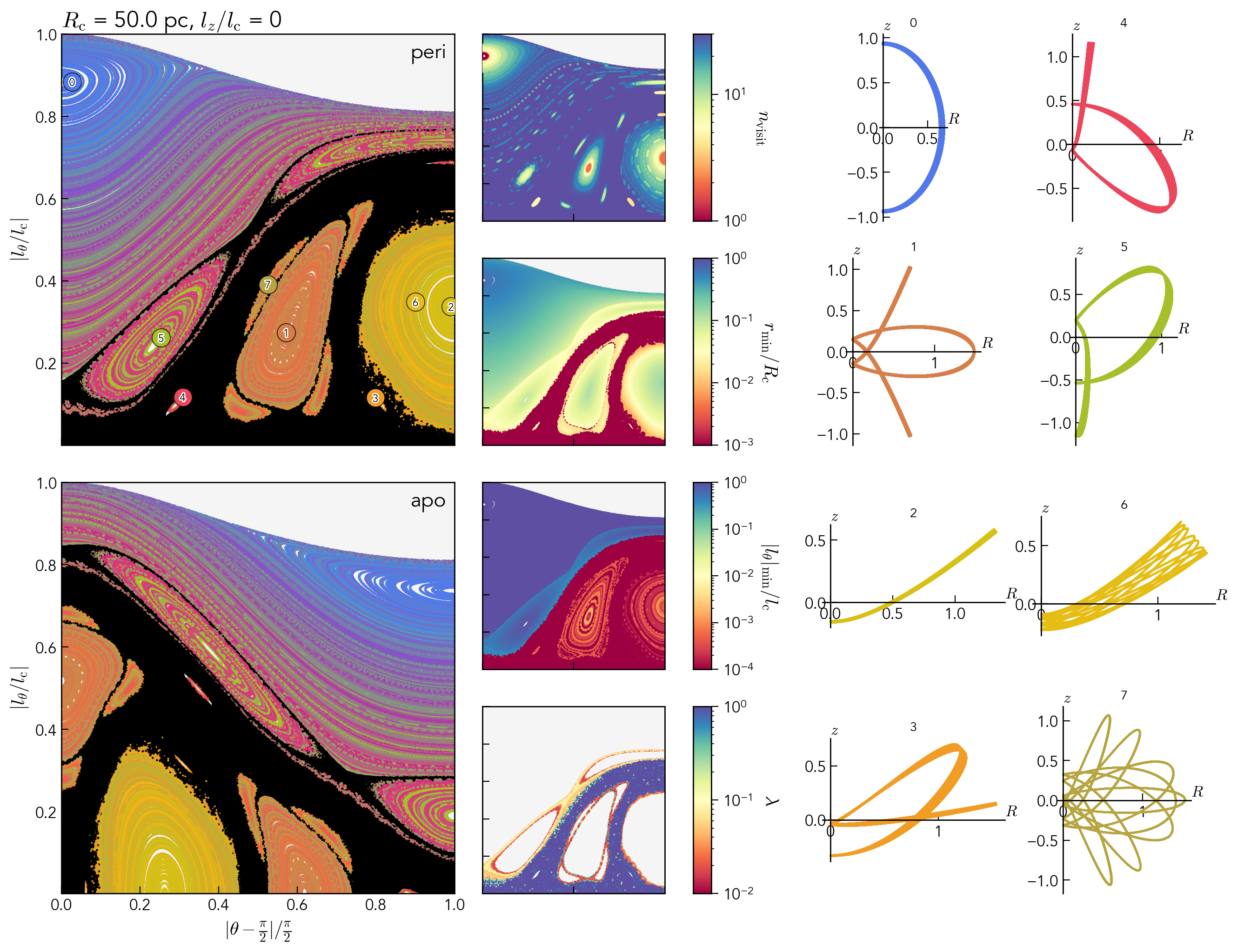}
    \caption{The full range of possible orbits with $R_{\rm c}$ of 50 pc and $l_z=0$, shown in the reduced space of $|l_\theta/l_{\rm c}|$ and $|\theta-\frac{\pi}{2}|/\frac{\pi}{2}$ (which is 0 at the equator and 1 at the poles). We show 2,000 random trajectories integrated over approximately 300 orbits. The \textit{leftmost column} shows the state at apo- and peri- apse, coloured by the initial $l_{\theta, \rm p}$ and $\theta_{\rm p}$, with chaotic orbits ($\lambda>0.01$) in black. The \textit{second column} shows the periapse space coloured by various properties of the orbit. $n_{\rm visit}$ is the number of cells, in a 30 by 30 grid over the space shown, visited by the trajectory - small values correspond to regular orbits that repeat their motion. $r_{\rm min}$ is the smallest radius visited by the orbit, here in units of $R_{\rm c}$. We also show the smallest (magnitude)  $|l_\theta|_{\rm min}$ that the trajectory attains, and the value of the lyapunov exponent, $\lambda$, (in units of $T_{\rm c}^{-1}$) for chaotic orbits. In the \textit{righthand columns} we show, in order, the 8 trajectories in $R$ and $z$ space (in units of $R_{\rm c}$) which visit the smallest number of cells - showing the regular orbits (and in some cases their harmonics) which are correspondingly numbered in the top left panel. Of these 5 are closed trajectories (0 - $\mathcal{D}_2$, 1 - $\mathcal{R}_3$, 2 - $\mathcal{R}_2$, 4 - $\mathcal{R}_4$ and 5 - $\mathcal{D}_3$) whilst the others are higher harmonics (3 \& 6 - of $\mathcal{R}_2$ and 7 - of $\mathcal{R}_3$).}
    \label{fig:bigmap}
\end{figure*}

\subsubsection{Lyapunov exponents}

We use two tools for identifying chaotic behaviour in this work -- one qualitative, the other quantitative. 

Qualitatively, we can examine the SoS to identify regions where regular orbits exist and conversely where the orbits are chaotic, as discussed regarding Figure \ref{fig:traj_examples_9}. 

Unfortunately, more detailed and quantitative analysis based on the SoS is difficult, as this would require mapping out the (often complex) topology of the chaotic region, excluding any (potentially infinite) number of regular orbit islands nested within and around it.

To progress further, we instead turn to a quantitative method for identifying chaotic orbits: the Lyapunov exponent, a measure of how quickly two initially similar orbits diverge \citep{Skokos10}. It is defined as the exponent, $\lambda$, for which the difference between the  state, $\mathbf{s}$, of two trajectories separated by an initially infinitesimal displacement, $\mathbf{\delta s}_0$ at some time $t_0$, will diverge with time $t$, i.e.
\begin{equation}
|\mathbf{\delta s}|(t) = |\mathbf{\delta s}_0|e^{\lambda (t-t_0)}.
\end{equation}
In general all trajectories can be expected, for small enough displacements, to have an initial phase where the pertubation grows linearly. If the trajectory exhibits chaotic behaviour the exponential chaotic mode will eventually come to dominate and the above equation applies, with $t_0$ corresponding to this moment of transition. Regardless, as $|\mathbf{\delta s}| \rightarrow |\mathbf{s}|$, i.e. approaches the position and velocity scale of the orbit, the displacement saturates as the two trajectories are now completely detached and represent two distinct orbits, and $|\mathbf{\delta s}|$ remains roughly constant.

$\lambda$ should only be calculated for the section of a trajectory for which the displacement is growing exponentially. It is common to use the convention of $\lambda=0$ for trajectories which show no such exponential growth and thus are assumed to be regular.  Following the scheme outlined in \citet{Vasiliev13b}, we can efficiently estimate $\lambda$  in units of the  circular period $T_{\rm c}$, such that its value is normally between 0 and a maximum close to 1 (corresponding to an trajectory that diverges over a single orbit). 

A completely general and robust quantification of $\lambda$ is a difficult task and we note two edge cases that infrequently occur.\footnote{These issues have been fixed in the most recent update to \texttt{AGAMA}, however our results are still based on analysis using an older version of the code where they are still present.} The first comes from demanding too high an integration accuracy from the integrator, which can result in 'ghost' timesteps (with ${\rm d}t_i = t_{i+1}-t_i =0$) causing $\lambda$ to be undefined. The second comes from lower integration accuracy, leading to larger timesteps which are still well optimized for resolving the orbital path but which may now be insufficiently spaced for calculating $\lambda$, and thus give erroneous $\lambda>0$ for regular orbits (as identified, for example from the SoS).

\texttt{AGAMA}'s integration accuracy is set by requiring the variation in the orbital energy to be conserved up to some fractional tolerance - i.e. $t_{i+1}$ must obey $|E_{i+1}-E_{i}|< \eta E_{i}$ where $\eta$ is a small number.  If this condition is not satisfied by some initial proposed ${\rm d}t'_i$, then the trial timestep is reduced and the state is recalculated. We find setting $\eta = 10^{-12}$ to be a reasonable compromise between these two edge cases, though both may still occur occasionally. We also slightly soften the requirement for a regular orbit (formally $\lambda=0$) to $\lambda<0.01$ which salvages some regular orbits classified as showing chaotic behaviour.  


In figure \ref{fig:bigmap} we now show a large (2000) number of trajectories to examine the behaviour of the full SoS. Here we colour chaotic trajectories in black, showing clearly the presence of constrained islands of regularity in the chaotic sea. It is also now clear to see that the SoS at periapse and apoapse are homotopic transforms of each other, i.e. they have the same features and one could be reconstructed by shifting the other.

A qualitative inspection of Figure \ref{fig:bigmap} reveals one trend which is general for small $l_z$, and of key importance for our calculations: in the pericentric SoS, {\it regions of low $l_\theta$ are uniformly chaotic}. At higher $l_z$ some regular orbits can survive at low $l_{\theta}$, likely because they avoid a close periapse with the MBH, and eventually chaotic orbits dissapear entirely as $l_z \rightarrow l_{\rm c}$.

We focus now on the behaviour at periapse (the moment at which disruption is possible) and the smaller SoSs show features of the trajectories projected onto that space. To find the closed orbits we employ a novel method: we divide the SoS at periapse into a grid, and for each trajectory ask how many grid cells it visits ($n_{\rm visit}$). Closed orbits generally minimise this number, and we use this to pick the 8 example trajectories that we show on the right-hand side. These include the five closed trajectories we have already discussed, and three higher-harmonics. We can see that the higher harmonics occupy neighbouring yet distinct regions to their base closed trajectory.

We show the minimum radius an orbit reaches ($r_{\rm min}$), which is essentially just a measure of the lowest $l_{\theta, \rm p}$ that it's region of the SoS spans. $l_{\theta, \rm min}$ allows us to separate definite and reversible rotating orbits (with the former constrained to only moderate values or higher - the top $\sim$half of the SoS). Finally we show the calculated lyapunov exponent, $\lambda$, which we see is lowest at the boundaries of regular islands - equivalent to saying that trajectories can remain pseudo-regular for many orbits before dispersing and exploring the entire chaotic sea.

We see here the importance of separating chaotic orbits by whether they have definite or reversible polar rotation - as only the latter can have \textit{diving orbits}, i.e. access the arbitrarily low $l_{\theta}$ values needed to have a close pericenter passage.

\subsection{Uniform filling of the chaotic region}
\label{sec:filling}

\begin{figure*}
\includegraphics[width=\textwidth]{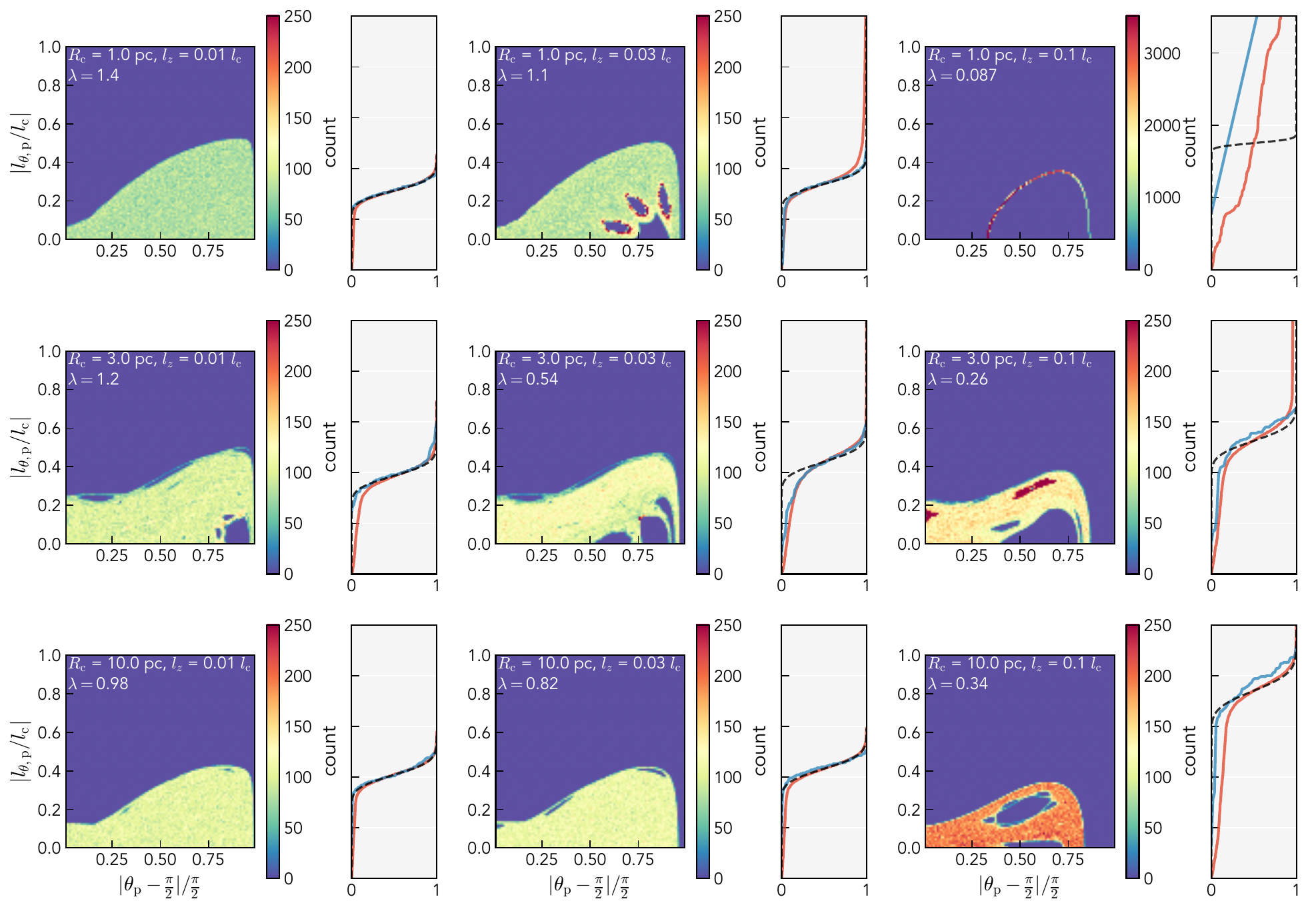}
    \caption{For a given $R_{\rm c}$ and $l_z$, we simulate one trajectory (with initial $l_{\theta, \rm p}=0.01 \ l_{\rm c}$ and $\theta_{\rm p}=\frac{\pi}{3}$) over approximately 300,000 orbits. Each coloured SoS panel shows the count of how many periapsides access a given region of $(\theta_{\rm p},l_{\theta,\rm p})$ space, digitized over a 100x100 grid. We also note the Lyapunov exponent, $\lambda$, in units of $T_{\rm c}$, which can help distinguish between chaotic trajectories and regular ones (we use $\lambda>0.01$ as a rough criterion for chaos, though the third panel in the top row shows an example of a regular orbit that still exceeds this). We also show the CDF of the counts (excluding 0's) aligned with each colourbar. The lines in the CDF show counts over the whole grid (red) and over just the bottom row (blue). This CDF is plotted alongside a Poisson distribution with the same median  (black, dashed line), which we would expect to characterize numerically sampled counts when occupancy is physically uniform.  
    The generally good agreement between count CDFs and the Poisson distribution at low $l_z$ justifies our later assumption of uniform occupancy in this regime.}
    \label{fig:chaosdensity}
\end{figure*}

A useful behaviour of chaotic orbits is that they are ergodic, meaning that they fill their accessible region of phase space uniformly. Because $\theta$ and $l_\theta$ are a canonical coordinate pair this means that for a given $\dot{r}$ (e.g. at periapse where $\dot{r}=0$ and $\ddot{r}>0$) the accessible $(l_{\theta, \rm p}, \theta_{\rm p})$ is sampled $\sim$uniformly and randomly. This behaviour is shown in Figure \ref{fig:chaosdensity} where we integrate a range of trajectories over 100,000 radial periods. From this we count the number of times any given periapse has a certain $\theta_{\rm p}$ and $l_{\theta,\rm p}$. 

From left to right, the panels show trajectories of increasing $l_{z}$, in the range $0.01 \le l_{z}/i_{c} \le 0.1$. These low-$l_z$ trajectories  occupy a complex region, that is bounded from above in $l_{\theta, \rm p}$ and also avoids islands of regular orbits. Some of these excluded regions may be fractal in nature (corresponding to harmonics of the dominant local regular orbit) and thus smaller than what is resolved by this plot, while other parts of the chaotic region may be narrow enough to not span an entire cell - all of which can lead to a decreased number density.
Nevertheless, by visually examining the whole parameter space of low-$l_z$ trajectories, we see that the occupation fraction within the chaotic sea is close to uniform. To make this analysis quantitative, we take the counts from the plots and show their Cumulative Distribution Function (CDF).  We compare each trajectory's CDF to that of a Poisson distribution, which we would expect if all cells contained the same number of counts $N_{\rm c}$ and varied only by shot noise $\sqrt{N_{\rm c}}$. We see that the Poisson distribution agrees very well with the observed counts at smaller $l_z$. Generally speaking, we find excellent agreement with the Poisson distribution (and thus uniform occupancy of the chaotic region) for $l_z/l_{\rm c} \lesssim 0.1$. Cells spanning an edge of the chaotic region may have a slightly diminished count, leading to an excess of low counts - as we see in some panels, where the red line exceeding the Poisson distribution at low counts.

Other boundaries display a `sticky' behaviour (for example in the top middle panel) with trajectories oversampling the edge of a regular region. This may be due to numerical integration error allowing the trajectory to drift into a regular region for a while, completing a number of orbits before drifting back into the chaotic region.

If we restrict ourselves to the lowest values of $l_{\theta, \rm p}$ (those most important for close passages; we show these as blue lines in the CDF panels) the agreement is even better, partly because there is much less influence from edges of regular regions. 

For $l_z \gtrsim 0.3$, few or no chaotic trajectories exist; we find only regular orbits. An example can be seen in the top right panel of figure \ref{fig:chaosdensity}, with a 1-dimensional loci in the SoS and their highly non-Poissonian CDFs.


\subsection{Diving orbits}
\label{sec:diving}

\begin{figure}
\includegraphics[width=0.95\columnwidth]{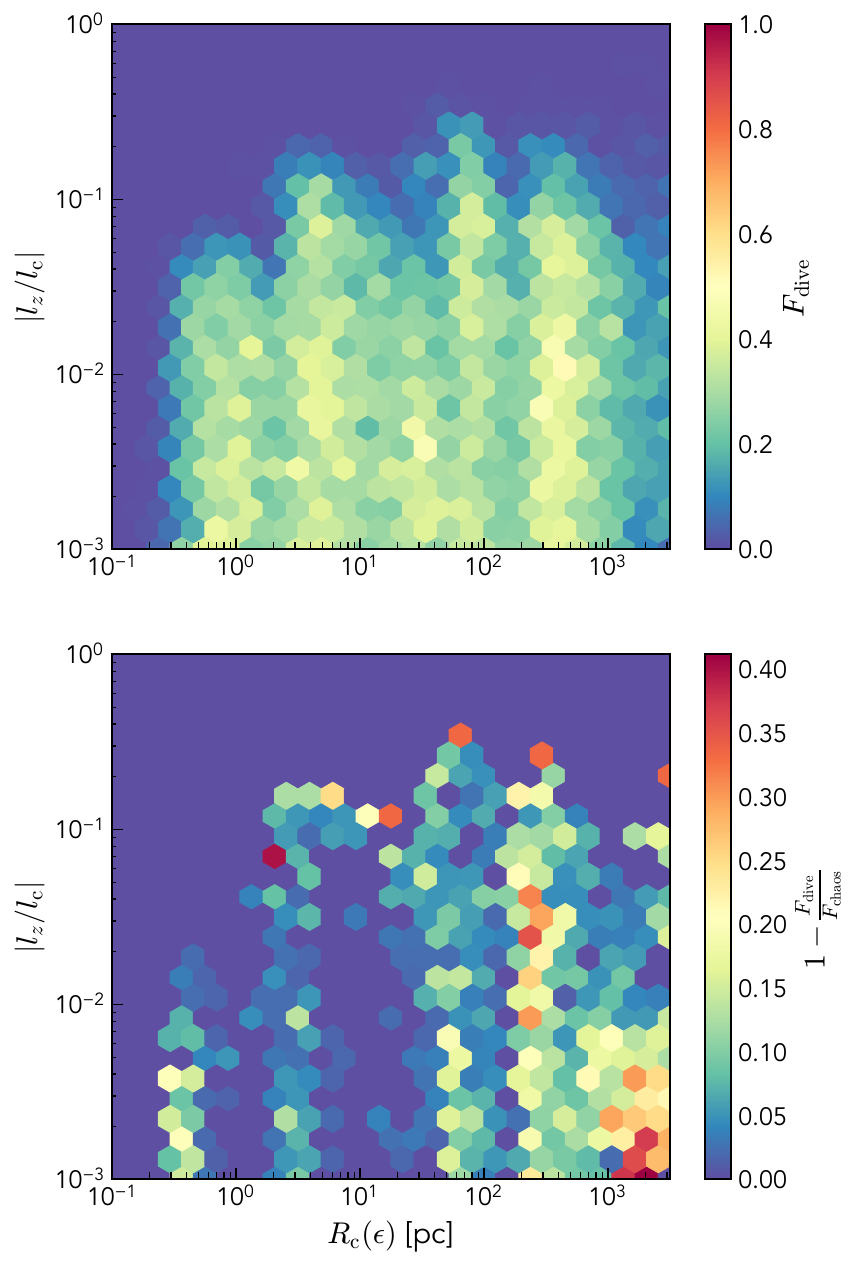}
    \caption{Top panel: the fraction ($F_{\rm dive}$) of trajectories which dive to arbitrarily small $l_{\theta, p}$ and thus reach the minimum pericenter permitted by $l_z$ conservation. Middle panel: the fraction of chaotic trajectories that do not dive ($1-\frac{F_{\rm dive}}{F_{\rm chaos}}$). Bottom panel: the normalising fraction of bound trajectories ($F_{\rm phys}$). The top two panels are found by simulating 100,000 trajectories at random $R_{\rm c}$ and $l_z/l_{\rm c}$ values (uniformly sampled in the space shown), initialized with uniform random $l_{\theta, \rm p}/l_{\rm c}$ and $\theta_{\rm p}$, and integrated for approximately 400 orbits each. We see that diving orbits are almost always a majority of chaotic orbits.}
    \label{fig:divingfractiontimes}
\end{figure}

\begin{figure}
\includegraphics[width=0.98\columnwidth]{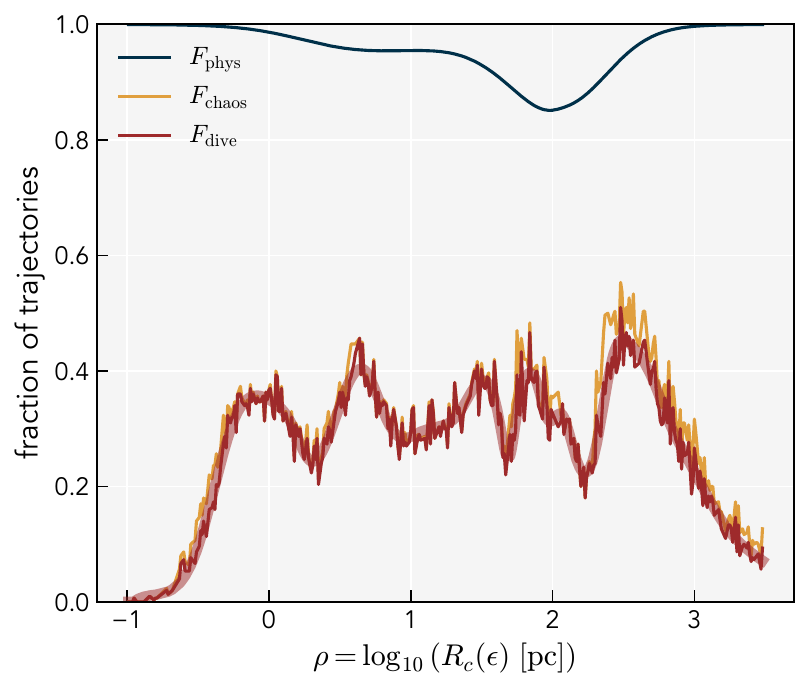}
    \caption{Fraction of $l_z=0$ trajectories that are physical, chaotic, and exhibit diving behaviour, shown as a function of $R_{\rm c}(\epsilon)$. For each $R_{\rm c}$ simulated, we sample 300 trajectories with random $\theta_0$ and $l_{\theta,0}/l_{\rm c}$ and follow them for approximately 500 radial periods. For  $F_{\rm dive}$ we also create a smoothed spline version of the curve (thick translucent red line) which will be useful for sampling these at arbitrary radii in later sections.}
    \label{fig:fractions}
\end{figure}

Not every chaotic trajectory comes as close as possible to the MPP, as some chaotic trajectories occupy the definitely rotating region of the SoS. We use a simple criterion to separate these less relevant chaotic trajectories from the diving trajectories of interest, namely that $\min(|l_{\theta, \rm p}/l_{\rm c}|) < N_{\rm p}^{-\frac{1}{2}}$ where $N_{\rm p}$ is the number of periapse passages. This criterion is motivated by the assumption of ergodicity (uniform occupation of the chaotic region) meaning that if the region extends to $l_{\theta,\rm p}=0$ the minimum observed value will be small, but limited by Poisson noise (hence the comparison to $N_{\rm p}^{-\frac{1}{2}}$). This ignores the detailed shape and size of the full chaotic region, but we have found that this simple criterion cleanly separates diving orbits from confined chaotic orbits. One caveat is that a reliable measure requires $N_{\rm p} \gg 1$
; we will mostly use 400 or more periapse passages for our analysis.

We are now able to calculate the fraction of orbits which are chaotic, $F_{\rm chaos}$, and, by using the above criterion, the subset $F_{\rm dive}$ that show diving behaviour. We do this by sampling a large number of orbits, with initial conditions at apoapse chosen uniformly from all possible values of $l_{\theta,a}$ and $\theta_a$ for a given $\epsilon$ and $l_z$. 

Near the poles of a flattened potential, some values of $|l_{\theta}| < l_{\rm c}$ may result in unbound trajectories,\footnote{We remind the reader that $l_{\rm c}$ is the maximum angular momentum for the whole system but that it corresponds to orbits in the equatorial plane, where the potential is itself maximzed.} i.e. those with initial $l_\theta > l_{\theta,\rm max}(\theta)$ (see equation \ref{eq:l_theta_max_theta}). We can discard these unphysical orbits in two different ways. 
 The efficient but more complex way is to choose random $\theta$ values, weighted by $l_{\theta,\rm max}(\theta)/l_{\rm c}$, and then to choose uniform $|l_\theta|<l_{\theta, \rm max}(\theta)$. The simpler method is to choose uniform random $\theta$ and $|l_\theta|<l_{\rm c}$ and then reject initial conditions with $|l_\theta|>l_{\theta,\rm max}(\theta)$. We adopt the latter approach, and renormalise our fractions by those of physical orbits,
\begin{equation}
\label{eq:f_phys}
F_{\rm phys} = \frac{2}{\pi} \int \frac{l_{\theta,\rm max}(\theta)}{l_{\rm c}} {\rm d}\theta
\end{equation}
where we have (without loss of generality) sampled over $0 \leq \theta \leq \frac{\pi}{2}$ and $0 \leq l_\theta \leq l_{\rm c}$.


The fractional abundance of different orbit types can be analysed across the space of possible $\epsilon$ and $l_z$, as we show in Figure \ref{fig:divingfractiontimes}. Here we sample random orbits in $R_{\rm c}$ and $l_z/l_{\rm c}$ space, to find the fraction 
of orbits which are chaotic and the fraction that dive (the latter being a subset of the former, though the difference is often minor). At very small and large radii, where the potential is almost perfectly spherical, both fractions are zero.  This is also the case at high $l_z/l_{\rm c}$ ($\gtrsim$ 0.1). For intermediate radii and low $l_z$, $F_{\rm chaos}$ and $F_{\rm dive}$ are high, between $20 - 50 \%$ generally. There are clear vertical strips of reduced diving fraction (most notably at a few pc, around 10 pc and at a few 100 pc) which also correspond to a higher fraction of non-diving chaotic orbits (middle panel). This suggests some regular orbit region is present at these $R_{\rm c}$ values which splits the chaotic region into two separate (high and low $|l_\theta|$) chaotic regions.

We see that the fraction of non-diving chaotic orbits (as shown in the lower panel) is small, at most around a third, but only reaching these values at large $R_{\rm c}$ ($> 100$ pc). Thus we can conclude that the vast majority of chaotic orbits lead to (or, as discussed in section \ref{sec:chaotic_orbits}, stem from) arbitrarily close periapse passages.

Crucially for the rest of our analysis, we see that for $l_z/l_{\rm c} \lesssim 0.01$ there is no visible variation with lower $l_z$. Any orbit that we might hope to reach the tidal radius will have lower $l_z$ than this value, therefore $F_{\rm dive}$ can be calculated in the limit of $l_z=0$, as a function of $R_{\rm c}$ alone.

Figure \ref{fig:fractions} show the physical, chaotic and diving fractions for orbits of different $R_{\rm c}$, at $l_z=0$. Again we see a significant diving fraction with substantial radial variation; typically above $20\%$. We use this measure of $F_{\rm dive}$, interpolated to smooth over random fluctuations, as shown in the Figure, in order to calculate the subsequent probabilities and rates of disruptions in the next sections.

\subsection{Varying the NSC model}
\label{ap:axisratio}

As we will show in the next sections, our derived rates are dominated by chaotic diving orbits from within or near the sphere of influence of the MBH. The MBH itself is effectively a point potential with a well-measured mass. It is therefore the properties of the NSC, the other important component of the GC model at these radii, that will have the most effect on our conclusions. In this section, we vary the NSC model as a numerical experiment to test the sensitivity of our results to different Milky Way GC models, and to some extent models of different galaxies' central regions.

A point mass alone produces closed Keplerian orbits. Adding a sub-dominant spherical mass distribution (or including relativistic effects) will generate precession, but will not produce chaotic behaviour. It is the non-sphericity of our simulated GC that gives rise to chaos, and thus in this section we primarily investigate the effect of the degree of flattening of the NSC. This is encapsulated in the parameter $q_{ \rm NSC}$.

The central MBH is defined as a Plummer sphere
\begin{equation}
\Phi_{\rm BH} = -G M_{\rm BH} \left(x^2 + y^2 + z^2 + a_{\rm BH}^2 \right)^{-\frac{1}{2}},
\end{equation}
where $M_{\rm BH}=4\cdot 10^6 \ M_\odot$ is the BH mass and $a_{\rm BH}=10^{-3}$ pc is the scale radius (i.e. the BH potential is slightly softened for numerical stability, but only on small scales below the range most of our orbits probe\footnote{We note that $a_{\rm BH}$ is often larger than the tidal separation radius for the Hills mechanism, and is always much larger than the disruption radius for TDEs, but remind the concerned reader that our orbital integrations are performed primarily to characterize the topology and density of low-$l_z$ chaotic orbits, not to simulate disruptive processes directly.}).  The NSC as in the fiducial model of S22: as a density profile (based on a fit to data from \citealt{Chatzopoulos15}) which is then converted to a potential by AGAMA. More specifically, it is defined via the axisymmetric generalisation of a Dehnen profile \citep{Dehnen93,Merritt96}
\begin{equation}
\rho_{NSC}(\tilde{r}) = \frac{M_{\rm NSC} (3-\gamma)}{4 \pi q_{\rm NSC}} \tilde{r}^{-\gamma} (1+\tilde{r})^{-(4+\gamma)},
\end{equation}
where $M_{\rm NSC}=6\cdot 10^7 \ M_{\odot}$, $\gamma =0.7$, and $q_{\rm NSC}$ (=0.71 in S22) is the axis ratio which defines
\begin{equation}
a_{\rm NSC} \tilde{r}^2 = x^2 + y^2 + \frac{z^2}{q_{\rm NSC}^2}.
\end{equation}
with $a_{\rm NSC}=6$ pc.
Here we simplify our model of the NSD and treat it as a Miyamoto-Nagai disk \citep{Miyamoto75} following
\begin{equation}
\Phi_{\rm NSD} = -G M_{\rm NSD} \left(x^2 + y^2 + (a_{\rm NSD} + \sqrt{z^2 + b_{\rm NSC}})^2\right)^{-\frac{1}{2}}
\end{equation}
where $M_{\rm NSD}=10^9 \ M_\odot$, $a_{\rm NSD}=90$ pc and $b_{\rm NSD}= 30$ pc is a scale height.

The resulting, simplified central potential is broadly consistent with (though less precise than) the empirical model of S22.  As we are interested in the inner few parsecs, we do not include a bulge potential.
In the remainder of this section, we fix all parameters aside from $q_{\rm NSC}$, which we vary from close to zero (very oblate) to close to 1 (spherical). Values above 1 give valid prolate potentials, but we do not consider them here. 

We repeat an analysis similar to the one used in Figure \ref{fig:fractions}. We limit our interest to orbits with $l_z=0$. For each $R_{\rm c}$, we initialise 100 orbits with random $l_\theta$ and $\theta$ at periapse, and follow them for $\approx$400 radial periods. We check whether orbits are physical (as some regions of $l_\theta,\theta$ space are not accessible in flattened potentials) and chaotic, equivalent to $F_{\rm phys}$ and $F_{\rm chaos}$ from the previous section.

\begin{figure}
    \centering
    \subfloat{
    \includegraphics[clip,width=0.95\columnwidth]{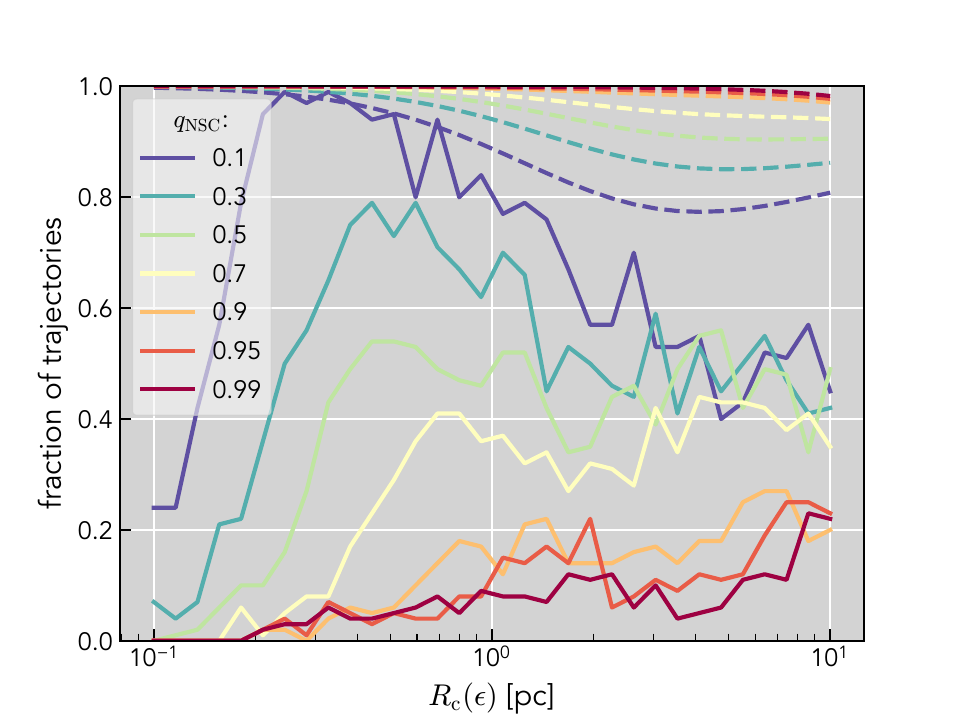}%
    }
    \\
    \subfloat{
      \includegraphics[clip,width=0.95\columnwidth]{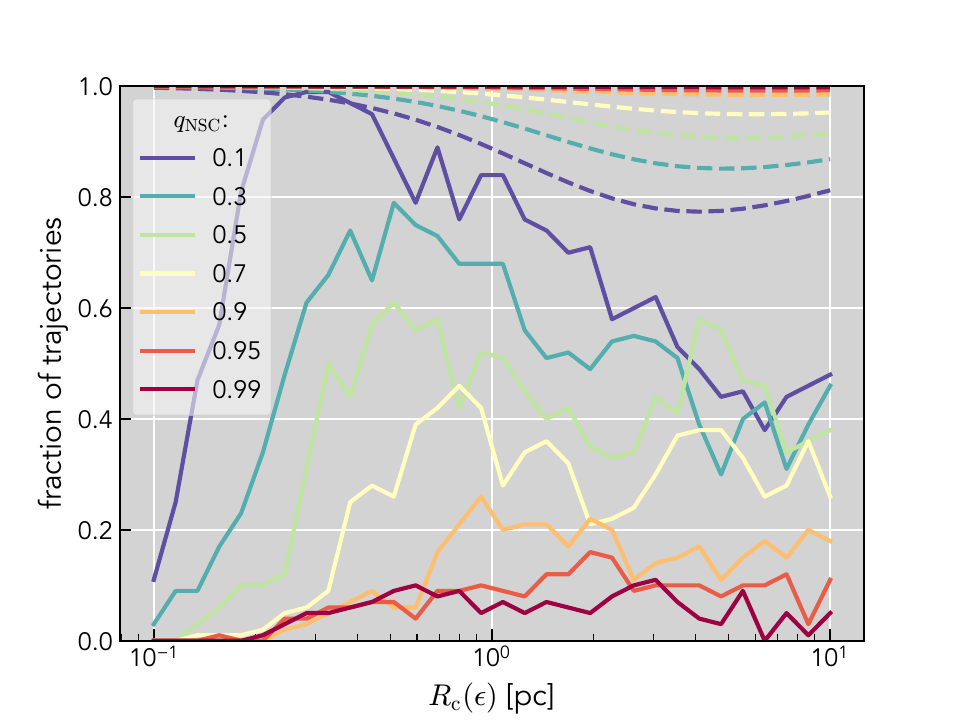}%
    }
    \caption{Top panel: The fraction of chaotic orbits ($F_{\rm chaos}$, solid line) and physical orbits ($F_{\rm phys}$, dashed line) for a given NSC axis ratio ($q_{\rm NSC}$) as a function of $R_{\rm c}$. Bottom panel: The same but in a GC model with only BH and NSC (no NSD).  We see that realistically flattened NSCs produce order-unity chaotic fractions, and even nearly spherical NSCs retain some chaotic orbits.  The NSD increases the chaotic fraction at distances $R_{\rm c} \gtrsim 5$ pc.}
    \label{fig:axisratio}
\end{figure}

We show the results in Figure \ref{fig:axisratio}. More flattened models ($q_{\rm NSC}\rightarrow 0$) exhibit a higher degree of chaos, and a slightly diminished fraction of orbits that are physical at large radii. For $q_{\rm NSC}=0.1$ there is a range of radii ($0.2 \lesssim R_{\rm c} \lesssim 0.4$ pc) where almost all orbits are chaotic. The chaotic fraction reduces at very small radii, where the NSC potential becomes completely subdominant to that of the BH, and at larger radii, where much of the total mass of the NSC is enclosed, and its quadrupole moment is thus of diminished importance. 

We also show the same results excluding the (axisymmetric) NSD, which shows that for systems close to spherical ($q_{\rm NSD} > 0.7$) the NSD slightly increases the chaotic fraction at larger radii ($R_{\rm c} \gtrsim 5$ pc). However, even without the NSD, there is significant chaos observed for all $q_{\rm NSC}$ values; even a few percent of orbits are chaotic for $q_{\rm NSC}=0.99$. A very basic heuristic would be to say $F_{\rm chaos} \sim (1-q_{\rm NSC})$, although this actually underestimates the chaotic fraction for $q_{\rm NSC}$ close to 1.

This numerical result agrees well with that presented in \citet{Vasiliev13} but extends beyond the regime where their analytic calculation holds.  For large values of $l_{\rm t}$ (appropriate for the Hills mechanism and some TDEs), it is necessary to consider larger radii where the aspherical NSC potential has a non-perturbative effect (i.e.  to move towards and beyond the sphere of influence).


\section{Probabilities, properties and timescale of diving orbit disruptions}
\label{sec:prob}

We wish to find the probability that a system disrupts, the properties of such a disruption, and the associated timescale. It is natural to write these as functions of the energy and $z$-component of angular momentum, which control the behaviour of the orbit. The internal properties of the system vulnerable to disruption (size, mass, spin, etc.) set the tidal radius and thus equivalently the tidal angular momentum, but systems with the same $l_{\rm t}$ should behave otherwise identically.

If we are considering the effect of diving orbits alone, $l_z$ can be taken as fixed. This ignores the effect of scattering, or assumes that it operates over timescales so distinct as to be considered separately. For example, if the timescale over which $l_z$-diffusion operates is very short, then whilst one particular particle may drift in $l_z$, the number of particles with a given $l_z$ will remain roughly constant. In the opposite case, where diffusion happens over much longer timescales than disruption, the $l_z$ of a given orbit will remain nearly constant until it disrupts.

For a trajectory to exhibit diving behaviour 
it needs to access a certain region of $l_{\theta,\rm p},\theta_{\rm p}$ space. In theory, our timescales could thus be dependant on the state of the particle, with disruption either certain (eventually) or impossible depending on whether $l_{\theta,\rm p},\theta_{\rm p}$ at any time falls within that region. These regions however are complex, as would be introducing these two extra parameters to any distribution, so instead we average over all states for a given energy and $z$-angular momentum, assuming implicitly equal occupation of $l_{\theta,\rm p},\theta_{\rm p}$ space. This means that we take the probability of an orbit being able to disrupt as equal to $F_{\rm dive}$ normalised by $F_{\rm phys}$, which we have already defined and numerically found in section \ref{sec:orbits}.

The strong assumption of ergodicity, uniform occupation of trajectories across $l_{\theta,\rm p},\theta_{\rm p}$ space, may not in general be true (see for example \citealt{Thomas04}), even as it remains true for any single chaotic trajectory. Brief numerical experiments show that the DFs of our GC model prefer low $|l_{\theta, \rm p}|$ and $\theta_{\rm p}$ close to $\frac{\pi}{2}$. This would boost the effective $F_{\rm dive}$ as this is the region where diving orbits are most frequently found. This makes our simplifying assumption a conservative one.

We define two useful rescalings of any specific angular momentum, $l_i$:
\begin{equation}
\lambda_i=\frac{l_i}{l_{\rm c}},
\end{equation}
and
\begin{equation}
\gamma_i=\frac{l_i}{l_{\rm t}}.
\end{equation}
Depending on the choice of $l_i$, these can either be positive definite, or also include negative values. The normalised angular momentum $\lambda_{i} \le 1$, while $\gamma_{i}$ can have much larger values, but for any plausible disruption is likely to be $\sim 1$.  When used without a subscript, $l$ (or $\lambda$ or $\gamma$) signifies the magnitude of the total specific angular momentum.

It is important to note that $\lambda_{\rm t}$, the ratio of the critical tidal angular momentum to the maximum $l_{\rm c}$, is usually $\ll 1$ for stars or binaries orbiting on galactic scales. Thus even after having simulated many thousands of orbits we might not expect any to have $\lambda \le \lambda_{\rm t}$ (i.e. to have come close to disruption). The small size of the loss region in angular momentum space is why we fold the results from our numerical experiments into an analytical framework, so that we can calculate the very small probabilities associated with disruptions without having to follow trajectories over a commensurately large number of orbits.

For a given potential, $l_{\rm c}$ is just a function of the specific energy $\epsilon$, here parametrized by $\rho = \log_{10}(R_{\rm c}(\epsilon) / {\rm pc})$. The behaviour also depends on the z-angular momentum $l_z$. The tidal specific angular momentum, $l_{\rm t}$, depends on the system of interest (e.g. TDEs versus the Hills mechanism) and thus is a third independent parameter. Therefore we write the probability distribution of some parameter $\chi$, at any given periapse passage, as $p(\chi|\rho,\gamma_z,l_{\rm t})$.  We already observed, in Section \ref{sec:filling}, that the pericenter of a chaotic orbit is equally likely to fall anywhere within the accessible region of $l_\theta$ and $\theta$ space. This means that many of the probabilities associated with disruptions can be approximated analytically.

In this section it is convenient to work in terms of the parameters
\begin{equation}
x=\frac{\left| \theta_{\rm p} - \frac{\pi}{2}\right|}{\frac{\pi}{2}},
\end{equation}
where $x=0$ corresponds to the equatorial plane and $x=1$ to the pole, and
\begin{equation}
y=\left|\frac{l_{\theta, \rm p}}{l_{\rm c}}\right|,
\end{equation}
which also runs from 0 to 1 . They are both uniformly sampled by periapse passages. In many cases it is expedient to use the transformed coordinate
\begin{equation}
\kappa=\cos\left(\frac{\pi x}{2}\right)=\sin\left(\theta_{\rm p}\right),
\end{equation}
which is 0 at the poles and 1 at the equator. In this notation we can write
\begin{equation}
l_{\rm p}=\sqrt{l_{\theta, \rm p}^2 + \frac{l_z^2}{\sz_{\theta_{\rm p}}^2}} = l_{\rm t} \sqrt{\Big(\frac{y}{\lambda_{\rm t}}\Big)^2 + \Big(\frac{\gamma_z}{\kappa}\Big)^2}.
\end{equation}
Because periapsides are uniformly distributed over the chaotic region of $x$ and $y$, the probability of any variable of interest taking a certain value for given $l_{\rm c}$, $l_z$ and $l_{\rm t}$ is simply proportional to the area of the chaotic region associated with that value. Reframing this mathematically, the cumulative probability that $\chi$ falls within some small range of values $C$ to $C+{\rm d}C$ is
\begin{equation}
\label{eq:p_chi_general}
p(\chi=C|\rho,\gamma_z,l_{\rm t}){\rm d}C \propto {\rm d}A_C = \int {\rm d}x_C {\rm d}y_C.
\end{equation}
If we can define $y_C(\kappa)=y(\kappa|\chi=C)$, i.e, the curve where $\chi=C$ as a function of $\kappa$,  then ${\rm d}y_C(\kappa) = \partial_C y (\kappa) {\rm d}C = \frac{\partial y}{\partial \chi}\big|_{\kappa,\chi=C} {\rm d}C$. Using ${\rm d}x = -\frac{2}{\pi}\frac{{\rm d}\kappa}{\sqrt{1-\kappa^2}}$ we can rewrite equation \ref{eq:p_chi_general} as
\begin{equation}
\label{eq:p_chi}
p(\chi=C|\rho,\gamma_z,l_{\rm t}) \propto -\frac{2}{\pi}\int_{\kappa_-}^{\kappa_+} \frac{ \partial_C y (\kappa) d\kappa}{\sqrt{1-\kappa^2}},
\end{equation}
where $\kappa_\pm$ are the limits of valid $\kappa$ for which $y_C(\kappa)$ takes positive real values.

We have seen throughout our numerical experiments, and will assume from here on, that at low $|l_z|$ the chaotic region extends from $0<x\lesssim1$ (and thus also from $0\lesssim\kappa<1$). In almost all cases of interest $\kappa_+$ is thus equal to 1, while $\kappa_-$ still takes non-trivial values that must be evaluated for the case at hand.

For the next part of this section we will use the shorthand that $p(C)$ implies $p(\chi=C|\rho,\gamma_z,l_{\rm t})$, although including these conditionals will of course be important elsewhere when marginalising over $\rho,\ \gamma_z$ and $l_{\rm t}$ to derive rates and average properties.  We will also use the shorthand of writing the variable, $\chi$, in place of the specific value, $\chi=C$, as in all places the meaning should be unambiguous and this avoids introducing and defining extra unnecessary constants.

\subsection{Distribution of diving factors}

We define the diving factor as
\begin{equation}
\beta = \left( \frac{l_{\rm t}}{l_{\rm p}}\right)^2 = \gamma_{\rm p}^{-2}.
\end{equation}
Note that this is similar to but distinct from the common definition of $\beta=r_{\rm t}/r_{\rm p}$ in the TDE literature \citep[e.g.][]{Carter82}. As long as the radii of interest are small, the definitions are approximately equivalent (as near the BH $r_{\rm p} \propto l^2$). However, as the scale of interest becomes larger, the two definitions diverge. We choose this convention so that many of the quantities of interest, which depend directly on the angular momentum, are expressed exactly.  In terms of the parameters defined above we can write
\begin{equation}
\beta = \left(\left(\frac{y}{\lambda_{\rm t}}\right)^2 + \left(\frac{\gamma_z}{\kappa}\right)^2\right)^{-1}.
\end{equation}
Thus
\begin{equation}
y_\beta = \lambda_{\rm t} \sqrt{\frac{1}{\beta} - \left(\frac{\gamma_z}{\kappa}\right)^2},
\end{equation}
and
\begin{equation}
\partial_\beta y(\kappa) = -\frac{\lambda_{\rm t}}{\beta^2 \sqrt{\frac{1}{\beta} - \left(\frac{\gamma_z}{\kappa}\right)^2}}.
\end{equation}
In the function $y_\beta$, the elements of the codomain assume real positive values for $\sqrt{\frac{\gamma_z^2}{\beta}}<\kappa<1$ , and thus has no solutions for $\beta>\frac{1}{\gamma_z^2}$. Using these to evaluate equation \ref{eq:p_chi} we find
\begin{equation}
p(\beta) \propto \frac{2 \lambda_{\rm t}}{\pi}\beta^{-\frac{3}{2}}\int_{\sqrt{\gamma_z^2/\beta}}^1 \kappa (\kappa^2 - \beta \gamma_z^2)^{-\frac{1}{2}}(1-\kappa^2)^{-\frac{1}{2}} d\kappa,
\end{equation}
for $\beta<\frac{1}{\gamma_z^2}$ and $0$ otherwise.

This integral, while superficially imposing, can be written in terms of a standard constant integral (variations of which will appear elsewhere in this section):
\begin{equation}
\label{eq:nastyintegral}
\int_a^1 x (x^2 - a^2)^{-\frac{1}{2}}(1-x^2)^{-\frac{1}{2}} dx =\frac{\pi}{2} 
\end{equation}
for $a < 1$.

Thus the important result is
\begin{equation}
\label{eq:p_beta}
p(\beta) = \begin{cases}
\alpha \beta^{-\frac{3}{2}} & \beta<\frac{1}{\gamma_z^2}\\
0 &\text{otherwise},
\end{cases}
\end{equation}
where $\alpha$ is a normalization constant (yet to be derived). From this we can see that smaller $\beta$ (i.e. shallower, more grazing disruptions) are significantly preferred, and that for larger $\gamma_z=l_z/l_{\rm t}$, only shallow encounters are possible. It's also interesting to compare this to the \textit{geometric} result: $p(\beta) \propto \beta^{-1}$. This is the result we get from considering the area of annulus of a given $r_{\rm p} \underset{\sim}{\propto} \beta^{-1}$ - i.e. the distribution we would expect for random isotropic trajectories near the BH. Our result is substantially steeper, giving a higher fraction of systems passing at larger distances (smaller $\beta$).

\subsubsection{Moderate encounters and partial disruptions}

One issue not directly addressed by our formalism is the role of repeated weak tidal encounters.  For every severe encounter ($\beta \gtrsim 1$) in the full (or ``pinhole'')  regime, we would expect a greater number of moderate ($0 < \beta \lesssim 1$) encounters to have preceded it.  When considering the single star (TDE) loss cone, these weak encounters can excite internal oscillations \citep{Li13} or result in partial disruptions \citep{MacLeod13}; when considering the Hills mechanism loss cone, these can perturb the internal properties of the binary or in some cases drive the binary to merger \citep{Bradnick17}.  Secular oscillations produced by the eccentric Kozai-Lidov mechanism may also be of importance for the progenitors of hypervelocity stars \citep{Petrovich17}.  Any of these ``weak tidal interactions'' will exchange energy between the (outer) orbit and internal degrees of freedom, such that $\rho$, $\gamma_z$ and $\lambda_{\rm t}$ may vary, potentially altering the rate at which stars or binaries are ultimately disrupted by the central MBH. 

It would be possible to investigate these effects further,
for example by taking probabilistic sequential draws of $\beta$ and applying some $\Delta \rho$, $\Delta \gamma_z$ and $\Delta \lambda_{\rm t}$ as a function of $\beta$ at each step, as was done in e.g. \citet{Broggi24} for the TDE problem.  Further investigation is beyond the scope of this paper, but we note it here as an interesting avenue for future study.


\subsubsection{Normalizing $p(\beta)$}

We now return to the normalizing factor in equation \ref{eq:p_beta}. The probability, and thus any normalizing factor, are unbound if $\beta \rightarrow 0$. While there is a hard lower limit on $\beta$, which cannot be less than $\lambda_{\rm t}^2$ (i.e. $l \leq l_{\rm c}$), this is not useful in practice, as $\beta \sim \lambda_{\rm t}^2$ corresponds to quasi-circular orbits irrelevant for tidal disruption or separation. 
A larger but still impractical lower limit on $\beta$ arises from the regular substructures that emerge for moderate $\lambda \leq 1$
(see Section \ref{sec:orbits}).



Astrophysically, however, this divergence will be regulated by restricting our attention to values of angular momentum small enough to produce disruptions (with at least some probability).  This will provide us with a practical minimum value of $\beta$, and is the subject of the next subsection.
For the remainder of this paper, we treat disruption as a binary process (yes or no) and define a binary flag, D, for this purpose.  We make the simplifying assumption that
$\beta \geq 1$ is a necessary and sufficient condition for disruption. Now we can write down a modified version of equation \ref{eq:p_beta} conditional on a disruption having occurred:
\begin{equation}
\label{eq:p_beta_D}
p(\beta|D) = \frac{1}{2(1-|\gamma_z|)}\beta^{-\frac{3}{2}} \qquad  1<\beta<\gamma_z^{-2},
\end{equation}
where we require $\gamma_z \leq 1$, i.e. disruptions cannot occur if $l_z > l_{\rm t}$.  Equation \ref{eq:p_beta_D} still strongly prefers lower $\beta$ encounters. For $l_z=0$ half of all encounters have $\beta \leq 4$.


Here we have assumed a single definite $l_{\rm t}$ that can separate between which encounters lead to disruptions. These simplifying assumption can be complicated for Hills mechanism separations, where disruptions can occur for initially moderate $\beta$ (e.g. $\beta\gtrsim 0.5$), and a minority of systems may survive encounters with $\beta \gg 1$ (see \citealt{Sari10} and \citealt{Sersante25}). 


A fuller description of the loss-cone dynamics could instead take a probabilistic approach, but that is beyond the scope of this work. 


\subsection{Other encounter probabilities}

We can write other parameters of interest in terms of $\beta$, and hence model their distribution with transformations of relationships we have already described. The inclination of the encounter relative to the azimuthal plane of the potential can be expressed as
\begin{equation}
\cz_i = \frac{l_z}{l}= \gamma_z \beta^{\frac{1}{2}},
\end{equation}
where the sign of $\gamma_z$ indicates whether the encounter is prograde or retrograde. Similarly, the eccentricity (valid only close to the BH where the potential is approximately Keplerian) follows
\begin{equation}
e^2 - 1 = \frac{2 \epsilon_0 l^2}{G^2 M^2} = \frac{2 \epsilon_0 l_{\rm c}^2}{G^2 M^2} \lambda_{\rm t}^2 \beta^{-1}.
\end{equation}
Here $\epsilon_0 = \epsilon - \Phi_0$ and $\Phi_0$ is the potential excluding the BH as we approach the center, which we assume (and in our models know) to be constant. In other words $\epsilon_0$ is the energy associated just with orbits near the BH, ignoring the surrounding galaxy.

As $\lambda_{\rm t}$ is necessarily very small for an encounter of interest, $e$ is very close to one and we can define the rescaled deviation from a parabolic orbit
\begin{equation}
\delta = \frac{G^2 M^2}{\epsilon_0 l_{\rm c}^2}\lambda_{\rm t}^{-2} (e-1) \approx \beta^{-1}.
\end{equation}
Note that $\delta>0$ always, and the sign of $e-1$ matches the sign of $\epsilon_0$. 

These expressions are valid whether or not we invoke disruptions for $\beta \geq 1$, but in that case we can derive
\begin{equation}
\label{eq:dist_ci}
p(\cz_i|D)=\left| \frac{d\beta}{d \cz_i}\right| p(\beta|D) = \frac{|\gamma_z|}{1-|\gamma_z|}\cz_i^{-2} \qquad |\gamma_z| < |\cz_i| < 1,
\end{equation}
and
\begin{equation}
\label{eq:dist_delta}
p(\delta|D)= \left| \frac{d\beta}{d \delta}\right| p(\beta|D) = \frac{1}{2(1-|\gamma_z|)}\delta^{-\frac{1}{2}} \qquad \gamma_z^2 < \delta < 1.
\end{equation}
Both of these distributions prefer smaller values of the parameter of interest (close to the lower limit set by $|\gamma_z|$).  
The $\cz_i$ distribution is very steep, favouring highly misaligned orbits, with half of all encounters having $|\gamma_z|<|\cz_i| \le \frac{2|\gamma_z|}{1+|\gamma_z|}$.  The distribution of $\delta$ is less steep but still prefers values closer to the minimum value of $\gamma_z^2$.

Thus if $|\gamma_z| \rightarrow 0$ misaligned ($c_i \rightarrow 0$) parabolic ($\delta \rightarrow 0$) encounters are preferred. If instead $|\gamma_z|$ is close to 1 only aligned or anti-aligned ($|\cz_i|\rightarrow 1$) encounters are possible. In this case whether the orbit is hyperbolic ($e>1$) or elliptical ($e<1$) depends on the sign of $\epsilon_0$. Systems with $R_{\rm c}$ greater than a few parsecs have positive $\epsilon_0$ (see figure \ref{fig:galproperties}) and we would thus expect marginally hyperbolic encounters, and conversely for $R_{\rm c} < r_{\rm inf}$ would have $\epsilon_0<0$ and parabolic encounters.

Thus depending on which region of $(\epsilon, l_z)$ space dominate the rate of disruptions we can expect different characteristics and outcomes of the encounter.


\subsection{Probability and timescales}

The probability that a system can disrupt over its whole lifetime is simply
\begin{equation}
\label{eq:P_d_D}
P(D|\rho,\gamma_z) = \frac{F_{\rm dive}(\rho)}{F_{\rm phys}(\rho)}H(1-|\gamma_z|)
\end{equation}
where $H$ is the Heaviside step-function, enforcing here that only orbits with $|l_z| < l_{\rm t}$ can disrupt. As discussed at the beginning of this Section, this assumes that the initial ($l_\theta,\theta$) are chosen at random from the full space of accessible values (dependant on $F_{\rm phys}$), and that any trajectory that can dive eventually will (though the timescale may be long).

To find the associated timescale we can start by writing the probability of a disruption at \textit{a single} periapse passage, as
\begin{equation}
p(D)=p(\beta \ge 1)=\frac{A_{\beta \ge 1}}{F_{\rm dive}(\rho)},
\end{equation}
where $A_{\beta \ge 1}$ is the area of the x,y plane for which $\beta \geq 1$.
We cannot work directly from $p(\beta)$ because, as was previously discussed, it cannot be robustly normalised. However we can find $A_{\beta \ge 1}$ directly by integrating
\begin{equation}
\begin{split}
\label{eq:areaD}
A_{\beta \ge 1} =& \int_{\beta \ge 1} d x_\beta d y_\beta  =\int_0^{\kappa(x)=|\gamma_z|} y_\beta(x,\beta=1) d x \\
=& - \frac{2 \lambda_{\rm t}}{\pi} \int_{|\gamma_z|}^{1}  \sqrt{\frac{1-\left(\frac{\gamma_z}{\kappa}\right)^2}{1-\kappa^2}} d\kappa \\
=&  \lambda_{\rm t}(1-|\gamma_z|)
\end{split}
\end{equation}
assuming $|\gamma_z|\le 1$ and where we have used the identity (similar to equation \ref{eq:nastyintegral})
\begin{equation}
\int_a^1 x^{-1} (x^2 - a^2)^{\frac{1}{2}}(1-x^2)^{-\frac{1}{2}} dx = \frac{\pi}{2}(1-a),
\end{equation}
for $a \le 1$. Thus we can write, now explicitly including all assumed parameters,
\begin{equation}
\label{eq:p_D}
p(D)=p(D|\rho,\gamma_z,l_{\rm t})= \begin{cases}
\lambda_{\rm t} \frac{1-|\gamma_z|}{F_{\rm dive}(\rho)} & |\gamma_z| \le 1\\
0 &\text{otherwise}
\end{cases}
\end{equation}
Note that as we expect $\lambda_{\rm t} \ll 1$, this probability is small. Also this suggests orbits with $|\gamma_z|$ close to one are less likely to disrupt, though note that any eventual disruption rates will also depend on the number of systems with a given $\gamma_z$.

Given that the orbit is chaotic, we expect that it can visit the tidal region on any given radial period with a probability $p(D)$. Assuming $p(D) \ll 1$, we expect on average a disruption to happen after $1/p(D)$ orbits. Thus the time taken for a tidal disruption to occur is
\begin{equation}
\label{eq:t_d_D}
t_D(\rho,\gamma_z,l_{\rm t}) = 
\frac{\langle T_r \rangle}{p(D)} \approx \frac{T_{\rm c}}{\lambda_{\rm t}} \frac{F_{\rm dive}}{1-|\gamma_z|},
\end{equation}
where we have made the approximation that the average radial period $\langle T_r \rangle$ is approximately the same as the period of a circular orbit of the same energy $T_{\rm c}(\rho)$.

Again, the very small value of $\lambda_{\rm t}$ leads to potentially very long timescales - some large multiple of the orbital timescale, $T_{\rm c}$, which is of around one Myr at 100 pc scales, and reduces with smaller Galactocentric radius. 

\section{Rates of disruption for a given system}
\label{sec:rates}
In this section, we use our earlier dynamical results to estimate the rate at which systems (e.g. stars or binaries) are delivered to small pericenters where they may be disrupted.  In a spherically symmetric potential, the small 3-dimensional volume of $\mathbf{l}$-space (with $l \le l_{\rm t}$; \citealt{Frank76}) vulnerable to disruption is called the ``loss cone,'' which may be either empty or full depending on the ratio of a collisional relaxation time to the orbital time. In an axisymmetric potential such as our model of the Milky Way, the larger 1-dimensional volume of $\mathbf{l}$-space vulnerable to disruption is called the ``loss wedge'' (defined by $|l_z| \le l_{\rm t}$; \citealt{Magorrian99}) which in turn may be either empty or full (as detailed in section \ref{sec:fullemptydiving}). Diving orbits, collisionless  trajectories that pass through the loss wedge, can be efficient at delivering objects to small pericenters.  While this motion is sometimes due to libration along regular orbits \citep{Vasiliev13}, our earlier results (section \ref{sec:orbits}) show that for a realistic Milky Way model, chaotic orbits overwhelmingly dominate collisionless disruption rates from the loss wedge.

Our calculations so far have taken $l_{\rm t}$ as a given. To fully calculate rates we need to know not just what happens to a system with fixed $l_{\rm t}$, but also how numerous such systems are compared to other systems with different $l_{\rm t}$. We save this multi-species calculation for an upcoming paper (Penoyre et al. 2025, in prep.) where we will consider both primordial populations of systems and the condition that they must survive long enough to disrupt. For the rest of this paper we present the calculation of rates \textit{assuming} a single value of $l_{\rm t}$, thus giving a qualitative 
understanding of the total rate of disruptions we might expect, as well as quantitative predictions for the {\it relative} rates of disruptions from various mechanisms. In particular we will calculate and compare a simplified collisionless rate, from diving orbits, and collisional rate, from scattering events.

$\Gamma_{\rm i}$ is the rate of disruption from a given mechanism (where $\rm i$ is either $\rm d$ for diving orbits or $\rm s$ for scattering). Any rate depends on some set of properties, which we can express as a vector $\vec{q}_{\rm i}$ of length $N$, and can be written as
\begin{equation}
\label{eq:rate_general}
\frac{d^N \Gamma_{\rm i}}{d \vec{q}_{\rm i}} = \frac{n(\vec{q}_{\rm i}) P_{\rm i}(D|\vec{q}_{\rm i})}{t_{\rm i}(\vec{q}_{\rm i})}
\end{equation}
where $d\mathcal{N} = n(\vec{q}_{\rm i}) d \vec{q}_{\rm i}$ is the differential number of systems with properties in the range $\vec{q}_{\rm i}$ to $\vec{q}_{\rm i} + d \vec{q}_{\rm i}$, and $P_{\rm i}(D|\vec{q}_{\rm i})$ and $t_{\rm i}(\vec{q}_{\rm i})$ are the probability and the expected timescale respectively that such a system disrupts by mechanism $\rm i$. To find the total disruption rate, we would need to integrate over each individual property in $\vec{q}_{\rm i}$ in turn. For diving orbits we have shown that the relevant properties in $\vec{q}_{\rm d}$ are $(\epsilon,\ l_z,\ l_{\rm t})$ (or transformations thereof). For scatterings 
the equivalent properties are $\vec{q}_{\rm s} = (\epsilon,\ l,\ l_{\rm t})$, dependent on the total angular momentum not just the z-component. 

We have already found $P_{\rm d}(D|\vec{q}_{\rm d})$, in the form of equation \ref{eq:P_d_D}, and we might expect that $t_{\rm d}(\vec{q}_{\rm d})$ is similarly given by equation \ref{eq:t_d_D}. However, as we will show, this is the relevant timescale only in some cases, as we need to consider the rates of phase space depopulation and repopulation to know which timescale dominates any steady state solution.

Given a distribution function we can find $n(\epsilon,l_z)$ (and similar for $l$ in the scattering case) and integrate over $\epsilon$ and $l_z$, holding $l_{\rm t}$ fixed. Integrating over $l_{\rm t}$ would require defining $n(\epsilon, l_z, l_{\rm t})$. We can however continue to work with a given value of $l_{\rm t}$ to find the \textit{relative} rate:
\begin{equation}
\bar{\Gamma}_{\rm i}(l_{\rm t}) = \int \int \frac{n(\epsilon,l_{\rm i}) P_{\rm i}(D|\epsilon,l_{\rm i},l_{\rm t})}{t_{\rm i}(\epsilon,l_{\rm i},l_{\rm t})} d\epsilon d l_{\rm i}. \label{eq:Gamma}
\end{equation}
In this form, $\bar{\Gamma}$ is equivalent to the rate of disruptions that would be observed if every system in the model had this single identical value of $l_{\rm t}$.

There is in general no direct relationship between $\bar{\Gamma}_{\rm i}$ and $\Gamma_{\rm i}$ (or any of their derivatives). We might hope that $\bar{\Gamma}_{\rm i}(l_{\rm t}) = a(l_{\rm t}) \frac{d \Gamma_{\rm i}}{d l_{\rm t}}$ (i.e. that we could find the true rate, $\Gamma_{\rm i}$, by integrating some function of $\bar{\Gamma}_{\rm i}$ over $l_{\rm t}$) but there are two strong conditions that must be met for this to be possible.
\begin{itemize}
\item If we can write $n(\epsilon,l_{\rm i},l_{\rm t})$ as $a(l_{\rm t}) n(\epsilon,l_{\rm i})$, then 
\begin{equation}
\label{eq:seperability}
\frac{d^2}{d\epsilon d l_{\rm i}}\left( \frac{d\Gamma_{\rm i}}{d l_{\rm t}}\right) = a(l_{\rm t})\frac{d^2\bar{\Gamma}_{\rm i}}{d\epsilon d l_{\rm i}}.
\end{equation}
This condition means that there is an equal chance of finding a system with a given $l_{\rm t}$ at any energy and angular momentum.
\item The integral over equation \ref{eq:seperability} spans the range $|l_{\rm i}| \le l_{\rm t}$, thus for these integrals to not introduce further irreconcilable factors of $l_{\rm t}$ they must depend negligibly on their upper limit. 
In other words, almost all disruptions have $|l_{\rm i}| \ll l_{\rm t}$
\end{itemize}

In general these assumptions do not hold, and hence our quantitive comparisons of relative rates can lead to only qualitative comparisons of the true rates. If the first alone is satisfied we can compare quantitively the factors in equation \ref{eq:seperability}. If this condition is not satisfied no quantitive comparison is possible (even when the latter is satisfied).

To give an example of where the first might break down, wide binaries or giant stars (with comparatively large $l_{\rm t}$) may not survive the high density and velocity dispersion environment of the inner Galactic center (e.g. orbits with large negative $\epsilon$) and hence $n(\epsilon,l_{\rm t})$ cannot be separated. Nevertheless, these relative rates are simply calculable and give us an initial intuition about the behaviour of the actual rates of disruption we would expect to see in a full model.



\subsection{Relaxation times}

The rate of both diving and scattering disruptions depends on the relaxation time for angular momentum, $t_{\mathrm{rel},l_z}$, the timescale over which two body interactions redistribute angular momentum. It may seem initially surprising that $\Gamma_{\rm d}$ depends implicitly on scattering events. This is not usually the case for following a single diving trajectory to disruption, but scattering is important in finding a steady-state solution. The relaxation time limits the repopulation rate for the region of phase space that can disrupt via diving orbits.

To calculate and compare $t_{\mathrm{rel},l_z}$ we will use a series of crude approximations. We start with the energy relaxation time,
\begin{equation}
\label{eq:trel_epsilon}
t_{\mathrm{rel},\epsilon}(\mathbf{r}) = \frac{0.34 \sigma_{\rm s}^3}{G^2 m_{\rm s}^2 n_{\rm s} \ln\Lambda},
\end{equation}
where $\sigma_{\rm s}(\mathbf{r})$ is the local velocity dispersion of scatterers, $m_{\rm s}$ their mass (assumed for simplicity to have a single value), $n_{\rm s}(\mathbf{r})$ their local number density, and $\ln \Lambda$ is the Coulomb logarithm, usually taken to have values between 15 and 20 (see \citealt{Spitzer87} or \citealt{Binney08} equation 7.106). This formula assumes 
an isotropic spherical system. While this is not true of the Galactic center, where the NSD is axisymmetric and disk-like (with relatively ordered velocities rotating around one axis) it is likely reasonable over small scales, and provides at least an order-of-magnitude estimate at larger scales.

Perhaps a larger concern is that this is a local measure, at some position $\mathbf{r}$, while we would ideally like to compare to the value for a given trajectory (characterised by energy $\epsilon$) whose orbit may take it through markedly different local conditions over the timescale in which relaxation acts. 
However, scattering is generally most effective at apoapse (where the orbit spends more time and velocity distributions are lower).\footnote{This is not always true: for example, very steep density profiles will may have enough stars at small radii to make periapse dominate the scattering process \citep{Stone+18}. However the assumption holds well for most observed stellar systems and it is possible that systems for which it is not true will evolve until it is true (as effective scattering at small radii will lead to redistribution of energy outwards, moving stars to larger radii and rebalancing the central excess).}

Our trajectories do not have a single apoapse, 
but 
their typical 
apoapse is of order the (cylindrical) circular radius associated with their energy. Thus we calculate the number densities at $R=R_{\rm c}(\epsilon)$ (and $z=0$, ignoring the slight decrease in density away from the disk plane) and similarly we approximate the velocity dispersion as the circular velocity at that radius, i.e.
\begin{equation}
\sigma_{\rm s} \sim v_{\rm c} = \frac{2 \pi R_{\rm c}}{T_{\rm c}}.
\end{equation}
This actually leads to a slight overestimate of $t_{\mathrm{rel},\epsilon}$ in the disk, where depending on the degree of ordered rotation $\sigma_{\rm s}$ could be significantly less than $v_{\rm c}$. In contrast away from the disk plane the density is reduced and the assumed factor of $n_{\rm s}$ can lead to an underestimate, but all orbits must pass through the disk at least once per radial period. We take the value of $\ln \Lambda$ to be 20 and $m_{\rm s} = 1 M_\odot$.

\begin{figure}
\includegraphics[width=0.95\columnwidth]{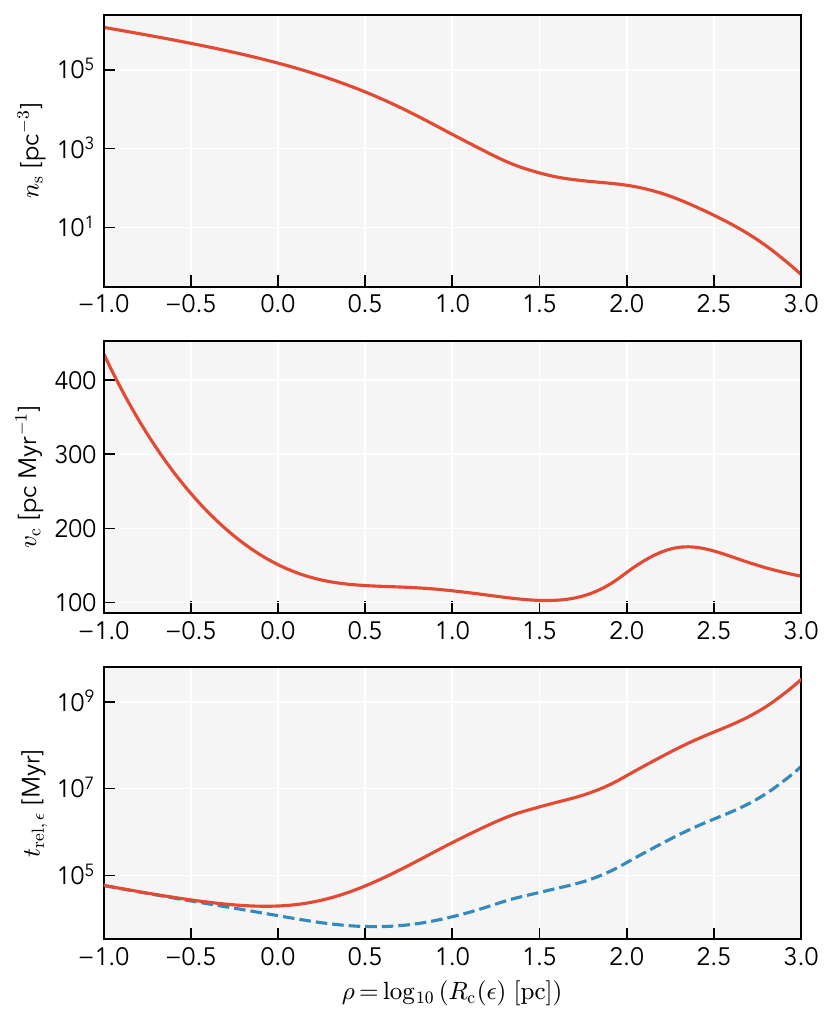}
    \caption{Properties of our GC model needed for calculating the relaxation time (equation \ref{eq:trel_epsilon}), shown as a function of (log) circular radius. From top to bottom we show: the number density of stars (assuming $m=M_\odot$ for all), the local circular velocity (which we use as an estimate for the velocity dispersion), and finally the energy relaxation time itself. These quantities are actually functions of position (rather than energy) and thus the values shown and used are those at $R=R_{\rm c}(\epsilon)$ and $z=0$. In the final panel, the dashed line shows the energy relaxation time modified by our simple prescription of massive perturbers at larger radii (see equation \ref{eq:trel_mps}) which can dominate two body scatterings and significantly speed up redistribution of energy and angular momentum.}
    \label{fig:trel}
\end{figure}

In Figure \ref{fig:trel} we show the properties of our Galaxy model relevant for calculating $t_{\mathrm{rel},\epsilon}$: the densities, velocities, and relaxation times themselves. Though we work in terms of energy these are local spatial measures for which we give approximate values calculated at $R=R_{\rm c}$, and $z=0$.
Most relevant quantities change by many orders of magnitude over the range of $\rho$ (from 0.1 pc to 1 kpc) and thus when comparing quantities like timescales it is generally safe to assume that one is $\ll$ than the other (and that the transition regions are small and can be ignored in our simplified calculations).

We also show an alternate evaluation of the energy relaxation time including a simple treatment of the presence of massive perturbers in the outer GC (for example, giant molecular clouds of masses > $10^5 M_\odot$ which might significantly increase the rate of energy redistribution). This effect is detailed fully in section \ref{sec:mps}.

This timescale of energy redistribution is generally long (of order Gyr at parsec scales and rapidly increasing with radius) but the more relevant quantity, the angular momentum relaxation time, can be significantly shorter:
\begin{equation}
\label{eq:trel}
t_{\mathrm{rel},l_z}(\epsilon,l_z) \sim \left( \frac{l_z}{l_{\rm c}}\right)^2 t_{\mathrm{rel},\epsilon} = \gamma_z^2 \lambda_{\rm t}^2 t_{\mathrm{rel},\epsilon}
\end{equation}
(see for example \citealt{Vasiliev13} section 4.1). Only orbits with $|l_z| \le l_{\rm t}$ can dive (and thus need to be repopulated) and thus $|\gamma_z| \le 1$ for all cases of interest, whilst the factor $\lambda_{\rm t}$ is generally extremely small and decreases with increasing radius (as $l_{\rm t}$ is fixed for a given system, and $l_{\rm c}$ increases).

\subsection{Full and empty diving regimes}
\label{sec:fullemptydiving}

After a disruption by a diving orbit has occurred, we might expect the occupation of that part of $(\epsilon, \ l_z)$ space to be reduced, and in extreme cases fully depopulated. However, because scattering redistributes stars and binaries via a random walk in angular momentum $\mathbf{l}$, depleted regions are eventually repopulated.

We can 
consider two different limits by comparing the diving time $t_{\rm D}$ and the angular momentum relaxation time $t_{\mathrm{rel},l_z}$:
\begin{itemize}
\item Full loss wedge: $t_{\mathrm{rel},l_z} \ll t_D$. If the system can relax over comparatively short timescales then it can repopulate the stars lost to diving orbits. In this case, the ``rate limiting step'' is the time necessary for a chaotic orbit to acquire $l<l_{\rm t}$ (i.e. $t_D$).
\item Empty loss wedge: $t_{\rm D} \ll t_{\mathrm{rel},l_z}$. If instead diving depopulates parts of parameter space more quickly than the galaxy can repopulate them via angular momentum relaxation, then we effectively run out of systems that can dive, and need to wait a time $\sim t_{\mathrm{rel},l_z}$ for them to be replaced and then disrupted almost immediately again.  In this limit, the rate limiting step is angular momentum relaxation.
\end{itemize}
In the empty loss-wedge case, typical disrupted systems will have $|\gamma_z|$ close to one, as they would not have had time to diffuse to lower values. A fuller calculation would include the depletion of low $l_z$ systems in $n(\epsilon,l_z)$, which as shown in \citet{Magorrian99} reduces the empty loss-wedge rate by a factor of order $F_{\rm dive}$. However, as we will go on to show, the total number of disruptions is dominated by the full loss wedge and thus this correction should only have a small effect.

This picture can be complicated if the age of the underlying galaxy model, $t_{\rm gal}$, is much shorter than either dynamical timescale (for example, if both diving and relaxation have timescales longer than the age of the galaxy, or at least the age of its current configuration). In this case, $t_{\rm D}$ is the timescale of interest regardless of $t_{\mathrm{rel},l_z}$, as there has not yet been enough time for the loss wedge to depopulate. 

Thus the relevant timescale for the steady-state rate of disruptions by diving orbits depends on $t_{\rm max}=\mathrm{max}(t_{\rm D},t_{\mathrm{rel},l_z})$ and follows
\begin{equation}
\label{eq:t_d}
t_{\rm d} = \begin{cases}
t_{\rm max} & t_{\rm max} \ll t_{\rm gal},\\
t_D & \text{otherwise}.
\end{cases}
\end{equation}
It is also possible for angular momentum redistribution to move stars in $l_{\theta}$, thus allowing a higher fraction of orbits than $F_{\rm dive}$ to disrupt, as previously stable orbits may be scattered onto chaotic trajectories. We do not include this effect in our models, but note that if it is maximally efficient it will lead to $F_{\rm dive} \rightarrow F_{\rm phys}$ and thus increase the rates by at most a factor of a few.

Although we do not consider time evolution in this paper, if we wanted to think of the galaxy as dynamic, we might initially start with a fully populated DF emptying at a rate set by $t_{\rm D}$, but as the loss wedge starts to empty out, the timescale for further disruptions asymptotes to $t_{\mathrm{rel},l_z}$.

\subsection{Disruption rates from diving orbits}

The final ingredient needed to calculate the collisionless disruption rate from diving orbits is the number of systems that we can expect to have a given energy and angular momentum.  We calculate this by numerically sampling the distribution functions, as detailed in appendix \ref{ap:n}. Now we can combine the approximated number densities in table \ref{tab:dffits} with equations \ref{eq:P_d_D} and \ref{eq:t_d} to write down the relevant form of equation \ref{eq:rate_general}:
\begin{equation}
\label{eq:partial_diving}
\frac{\partial^2}{\partial \rho \partial \gamma_z}\bar{\Gamma}_{\rm d}(l_{\rm t}) = \frac{P(D|\rho, \gamma_z, l_{\rm t}) n(\rho,\gamma_z)}{t_{\rm d}(\rho,\gamma_z,l_{\rm t})}
\end{equation}
where
\begin{equation}
n(\rho,\gamma_z) = \sum_{\rm I} \bar{N}_{\rm I} \frac{\partial \lambda_z}{\partial \gamma_z} p_{\rm I}(\rho,\lambda_z) = \lambda_{\rm t} \sum_{\rm I} \bar{N}_{\rm I}  p_{\rm I}(\rho,\lambda_z)
\end{equation}
where $\bar{N}_{\rm I}$ is the total number of systems in component I. $N_{\rm I}= M_{\rm I}/\langle m_{\rm s} \rangle$ where $M_{\rm I}$ is the total mass of the component, and we have already taken the average mass of test particles to be $\langle m_{\rm s} \rangle= 1 M_\odot$. The only components that we consider here are the NSC and NSD, and we will see that even the latter contributes little to the rates (further justifying ignoring the contribution of systems in the bulge).

\begin{figure*}
\includegraphics[width=0.98\textwidth]{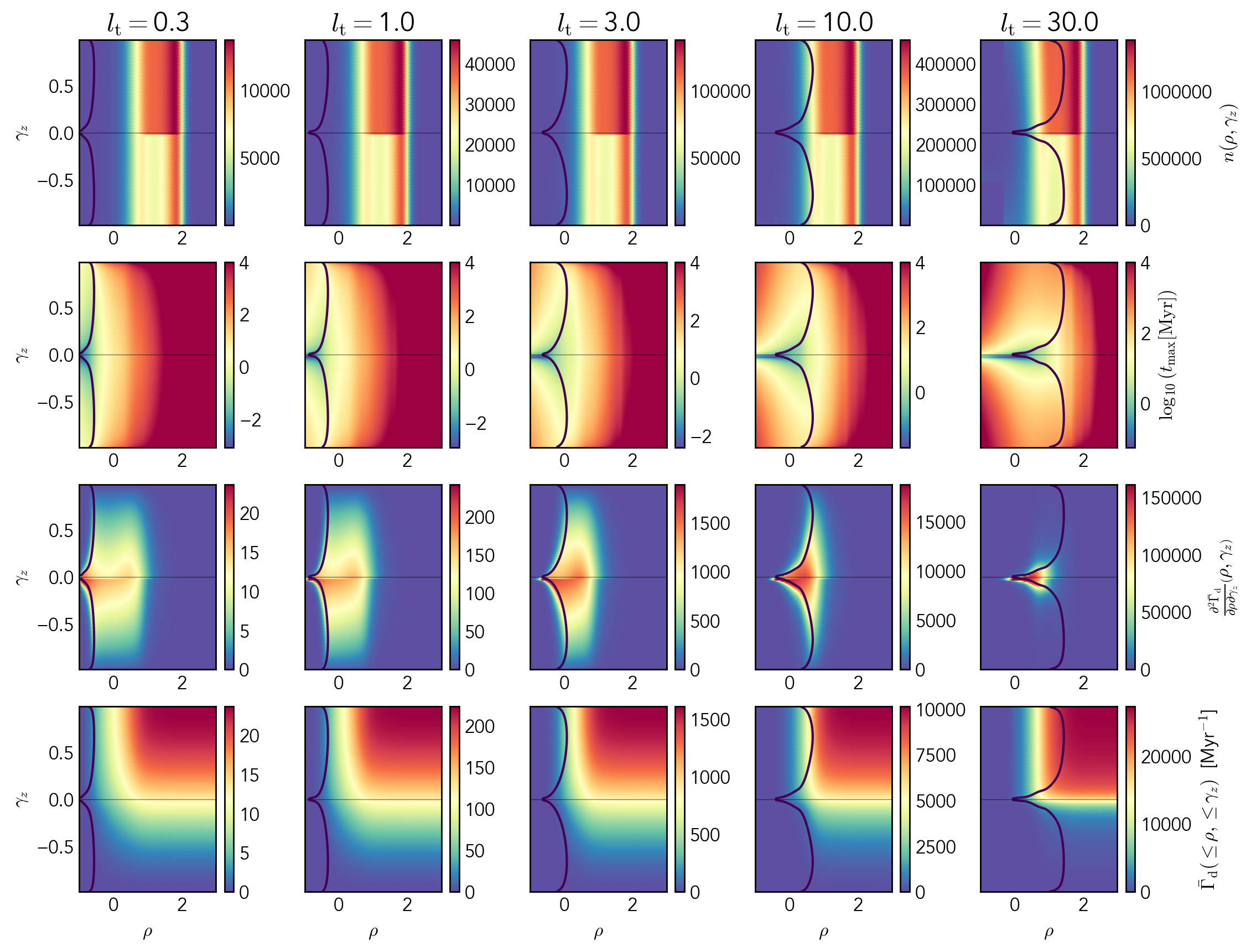}
    \caption{Dimensionless stellar concentrations (top row), rate-limiting timescales (second row), differential contribution to relative rates (third row) and cumulative relative rates (fourth row) for {\it collisionless} disruptions by diving orbits for a GC model composed entirely of systems of a given $l_{\rm t}$.  Different $l_{\rm t}$ values are labelled in units of [pc$^{2}$ Myr$^{-1}$], and are shown in different columns. All panels plot their respective quantities against $\rho=\log_{10}(R_{\rm c}(\epsilon)/{\rm pc})$ and $\gamma_z = l_z / l_{\rm t}$.  Note that the colour bar legends vary from panel to panel.  The solid black lines show the transition from empty to full loss region (with the full regime, $t_{\mathrm{rel},l_z} < t_{\rm d}$, predominating to the right of the black lines at 
    large $\rho$). Though the stellar concentrations show very little variation with $|\gamma_z|$, the total number with $|\gamma_z|\leq 1$ is proportional to $\lambda_{\rm t}$, and this increase leads systems with larger $l_{\rm t}$ to have a significantly larger rate. The final row shows the cumulative rate for all values of $\rho$ and $\gamma_z$ less than or equal to that at each point, thus the total rate is the value at the top right corner. From the final row, we see that  disruptions come overwhelmingly from the full loss-wedge regime.
    }
    \label{fig:rates_diving}
\end{figure*}

In Figure \ref{fig:rates_diving}, we show the stellar concentrations\footnote{We note that the dimensionless function $n(\rho, \gamma_z)$ gives the number density of stars in the dimensionless space of ($\rho, \gamma_z$).  To avoid confusion with the true (dimensional) 3D number density of stars $n_{\rm s}$ (as is e.g. plotted in figure \ref{fig:trel}), we term $n(\rho, \gamma_z)$ to be the stellar ``concentration.''} $n(\rho, \gamma_z)$, timescales $t_{\rm max}$ and relative disruption rates $\bar{\Gamma}_{\rm d}$ for five values of $l_{\rm t}$ (using $t_{\rm gal}=10$ Gyr). These can be mapped to approximate values of the tidal disruption radius $r_{\rm t}$ through equation \ref{eq:lperi}, and range from $\approx$0.5 AU to $\approx$5000 AU (in multiples of $\sim$10). Thus for a $\sim 10^6 M_\odot$ BH, the smallest $l_{\rm t}$ shown is appropriate for the tidal disruption of a Sun-like star, the middle value for a relatively tight ($\sim$ 0.5 AU) binary or a giant star, and the largest for a wide ($\sim$ 50 AU) binary or a typical triple star system.

The 
stellar concentration $n(\rho, \gamma_z)$ is primarily a function of radius, with a discontinuity above and below $\gamma_z=0$ corresponding to a break in the distribution function of the NSC (see Figure \ref{fig:pNSC_rho_lambdaz}). The total number of stars for which $|\gamma_z|\le 1$ increases with $l_{\rm t}$, but as $\lambda_{\rm t} \ll \ 1$ 
even for the largest $l_{\rm t}$ used, there is little visible variation with increasing $\gamma_z$. This suggests that a model that estimates $n(\rho,\gamma_z=\pm 0)$ (taking into account the discontinuity between co- and counter- rotating populations) might be sufficient to model the observed behaviour without having to further quantify the dependence on $|\gamma_z|$. It is also important to note that the concentration is small below $\rho \lesssim 0.5$ (this transition corresponds to peaks that we will soon see in the overall rates).

For most of the parameter space shown, the longer (and thus rate-limiting) timescale is $t_{\rm D}$, the time taken for a single diving orbit to come sufficiently close to disrupt. We show the line for which $t_{\rm D} = t_{\mathrm{rel},l_z}$ in black, highlighting that we are in the full loss-wedge regime for all but smallest radii and largest $|\gamma_z|$. We can see why the empty regime dominates at small radii by comparing the relatively shallow increase in the relaxation time with radius (see figure \ref{fig:trel}) while $t_{\rm D} \propto T_{\rm c}$ increases rapidly in comparison.  We know that $t_{\mathrm{rel},l_z} \propto \gamma_z^2 \rightarrow 0$ as $|\gamma_z|\rightarrow 0$. Meanwhile, $t_{\rm D} \propto \frac{1}{1-|\gamma_z|}$ gets very large as $|\gamma_z| \rightarrow 1$. This explains the shape of the transition line in these two extremes, and thus why the empty regime does not extend to low $|\gamma_z|$.

The smallest $t_{\rm max}$ are generally near the transition between the two regions (where both timescales are comparable) and at low $|\gamma_z|$. The minimum $t_{\rm max}$ shown in this parameter space is around 1 Myr, though note that there may be smaller values still for $\rho \lesssim -1$, which are not shown because of the negligible number of systems there. The largest values quickly exceed 10 Gyr (which we could take as an upper limit on $t_{\rm gal}$) beyond 10 to 100 pc. Larger $l_{\rm t}$ have smaller $t_{\rm D}$ and thus bigger systems have larger limiting $\rho$ below which they can disrupt. This also leads to the full/empty transition shifting to larger $\rho$ with increasing $l_{\rm t}$.

Generally, the peaks in the differential rate $\frac{\partial^2 \bar{\Gamma}_{\rm d}}{\partial \rho \partial \gamma_z}$ correspond to combinations of high $n(\rho,\gamma_z)$ and low $t_{\rm max}$. There is an effective maximum $\rho$ ($\sim$ 1) above which $t_{\rm max}$ is prohibitively large, and the differential rate falls off quickly. The total rate, which can be seen in the $\bar{\Gamma}_{\rm d}$ panels of Figure \ref{fig:rates_diving}, is generally dominated by systems in the full regime, 
and prefers systems from moderate ($\rho \sim 0.5$) radii for which $n$ is significant. For most values of $l_{\rm t}$, the total disruption rate is overwhelmingly dominated by the full loss-wedge regime.  

Almost all disrupted systems have low $|\gamma_z|$ (due to the shorter $t_{\rm max}$ values for low $l_z$).  While a wide range of disruption parameters, such as the inclination and orbital eccentricity (see equations \ref{eq:dist_ci} and \ref{eq:dist_delta} respectively) are possible, the most likely scenario will be a near-parabolic, grazing disruption strongly misaligned with the disk plane.\footnote{Both photometric \citep{Nishiyama13} and dynamical \citep{Sormani22} studies show that the NSD, and hence the whole GC, is well aligned with the Milky Way stellar disk.} We therefore expect to see more ejecta (tidal debris for TDEs; hypervelocity stars for the Hills mechanism) directed along the polar directions orthogonal to the galactic plane. 

This notable injection anisotropy would have been reversed had we found that most disruptions occur in the empty loss-wedge regime; in this case, almost all disruptions would occur with $|\gamma_{z}|$ close to 1, and $l_{\theta, \rm p}=0$, limiting ejecta to fly out within the equatorial plane of the galaxy.


\subsection{Disruption rates from scattering}
\label{sec:scattering}

Systems can also acquire low angular momentum orbits by the {\it collisional} mechanism of scattering alone (i.e. ignoring any collisionless evolution in $l$ produced by asphericity of the potential). This rate 
now depends on a much smaller region of phase space: $l<l_{\rm t}$ instead of $l_z<l_{\rm t}$ (i.e. a small volume in 3 dimensions, instead of just 1). 
Collisional scattering of stars into the loss cone of a spherical star system has been widely studied in the TDE literature \citep{Frank76, Cohn78, Magorrian99,Wang04,Stone16} and thus for comparison we'll present a brief and simple version of the calculation, mostly drawn from \citet{Merritt13book} chapter 6. In this calculation there are analogous full- and empty-loss cone regimes, now dependent on whether the diffusive change in angular momentum over one period is more (full) or less (empty) than $l_{\rm t}$.

We will here assume spherical symmetry. Though this is evidently
an oversimplification for the significantly axisymmetric NSC/NSD system, it allows us to simply produce a rough estimate likely accurate to within a factor of a few.

\subsubsection{Full loss cone}
\label{sec:flc}

We begin with the full loss-cone case where radial periods, $T_{r}$, are much longer than the timescale over which angular momentum is redistributed, $t_{\mathrm{rel},l}$.  The particular angular momentum relaxation time we care about is that at the loss-cone boundary, where $l = l_{\rm t}$.  The full loss-cone regime is attained at larger radii, where the loss cone shrinks to a small ``pinhole'' in angular momentum space.  In this regime, neglible depletion occurs, and the relevant version of equation \ref{eq:rate_general} (for a given $l_{\rm t}$) is
\begin{equation}
\label{eq:rate_full}
\frac{d^2 \bar{\Gamma}_{\rm s,F}}{d \rho d \gamma} \approx \frac{n(\rho,\gamma)}{T_{r}(\rho,\gamma)} \ \mathrm{ for } \ \gamma \leq 1,
\end{equation}
where we have assumed that $P(D)=1$ for $\gamma \leq 1$ and 0 otherwise. 
As before, we approximate $\langle T_{r} \rangle$ as being $\approx T_{\rm c}(\rho)$ (the period of a circular orbit of the same energy).

It is usually assumed that $n(\rho,l) \propto l \cdot n(\rho)$ (see e.g. \citealt{Merritt13book} equation 6.10a). 
However, as detailed in Appendix \ref{ap:n}, we do not find this approximation robust for our GC model and 
so we integrate $n(\rho,\gamma) = \lambda_{\rm t} n(\rho,\lambda)$ directly. 
More concretely, our detailed GC model specifies the angular momentum anisotropy in each bin of orbital energy for both the NSC and the NSD.  We parametrize this energy-dependent anisotropy using numerical fitting functions illustrated in Figs. \ref{fig:pNSD_rho_lambda} and \ref{fig:pNSC_rho_lambda}.  In brief, the NSD transitions from a quasi-isotropic state at small radii ($\rho \lesssim 1.75$) to a tangential bias at larger radii (Fig. \ref{fig:pNSD_rho_lambda}).  Conversely, the NSC is tangentially biased at small radii ($\rho \lesssim 1.25$) and becomes radially biased at larger radii (Fig. \ref{fig:pNSC_rho_lambda}).

In Appendix \ref{ap:n}, we approximate the probability that a system in component $I$ has a given $\rho$ and $\lambda$ to be
\begin{equation}
p_{\rm I}(\rho,\lambda)=10^{a(\rho)}\lambda^{b(\rho)}.
\end{equation}
We can integrate this over angular momenta to find the number of systems that are susceptible to disruption,
\begin{equation}
\label{eq:n_full}
n_{\rm I,t}(\rho) = \int_0^1 n_I(\rho,\gamma) d\gamma = \bar{N}_{\rm I} \int_0^{\lambda_{\rm t}} p_{\rm I}(\rho,\lambda) d\lambda = \bar{N}_{\rm I} \frac{10^a \lambda_{\rm t}^{b+1}}{b+1},
\end{equation}
where $\bar{N}_{\rm I}$ is the total number of systems in component I (the above assumes $b>-1$, which is true for our fitted models). To compare directly with the empty loss-cone rate we can integrate equation \ref{eq:rate_full} over angular momenta to find
\begin{equation}
\label{eq:rate_sF}
\frac{d \bar{\Gamma}_{\rm s,F}}{d \rho} \approx \sum_{\rm I}\frac{1}{T_{\rm c}(\rho)} n_{\rm I,t}(\rho).
\end{equation}
This can similarly be compared to rates from diving orbits by integrating over $\gamma_z$.

\subsubsection{Empty loss cone}

In the empty loss-cone regime, systems are disrupted faster than they are repopulated (i.e. $T_{r} \ll t_{\mathrm{rel},l}$). We therefore need to account for the fact that the relevant stellar concentration is reduced, i.e. $n_{E}(\rho,\gamma) \leq n(\rho,\gamma)$. The classic result in loss-cone theory is to find a stationary solution to the Fokker-Planck equation in angular momentum space \citep{Cohn78}.  In spherical symmetry, this solution features logarithmic depletion of the distribution function at smaller values of $l$.  In the limit of $T_{r} \ll t_{\rm rel, l}$, the distribution function goes to zero at $l=l_{\rm t}$.  In other words, stars random walk through angular momentum space, but do so with such small steps that they are destroyed on the first orbit for which $l \le l_{\rm t}$.

Once the stationary solution in the empty loss-cone regime is established, the disruption rate becomes
\begin{equation}
\frac{d \bar{\Gamma}_{\rm s,E}}{d \rho} \approx \frac{1}{2|\ln(\lambda_{\rm t})|} \frac{n(\rho)}{t_{\mathrm{rel},\epsilon}(\rho)}.
\end{equation}
One way of interpreting this result is to say that for a fixed energy, a fraction $1/|\ln(\lambda_{\rm t}^2)|$ of all systems pass into the loss cone over each energy relaxation time $t_{\mathrm{rel},\epsilon}$.
For practical evaluation of this formula, we find $n(\rho)$ by sampling from the DF in AGAMA and summing over components. This procedure is detailed in appendix \ref{ap:n}.

At the transition point between the full and empty loss-cone regimes (where $t_{\mathrm{rel}, \epsilon} \sim T_{\rm c}$) the rate should smoothly interpolate between the full and empty limits, but for simplicity here we will use
\begin{equation}
\frac{d \bar{\Gamma}_{\rm s}}{d \rho} = \mathrm{min}\left( \frac{d \bar{\Gamma}_{\rm s,F}}{d \rho},\frac{d \bar{\Gamma}_{\rm s,E}}{d \rho}\right).
\end{equation}
In general, the empty loss-cone regime dominates at small radii (where orbital periods are small) and the full loss-cone regime comes into effect at larger radii.

\subsection{Comparing rates}

\begin{figure*}
\includegraphics[width=0.98\textwidth]{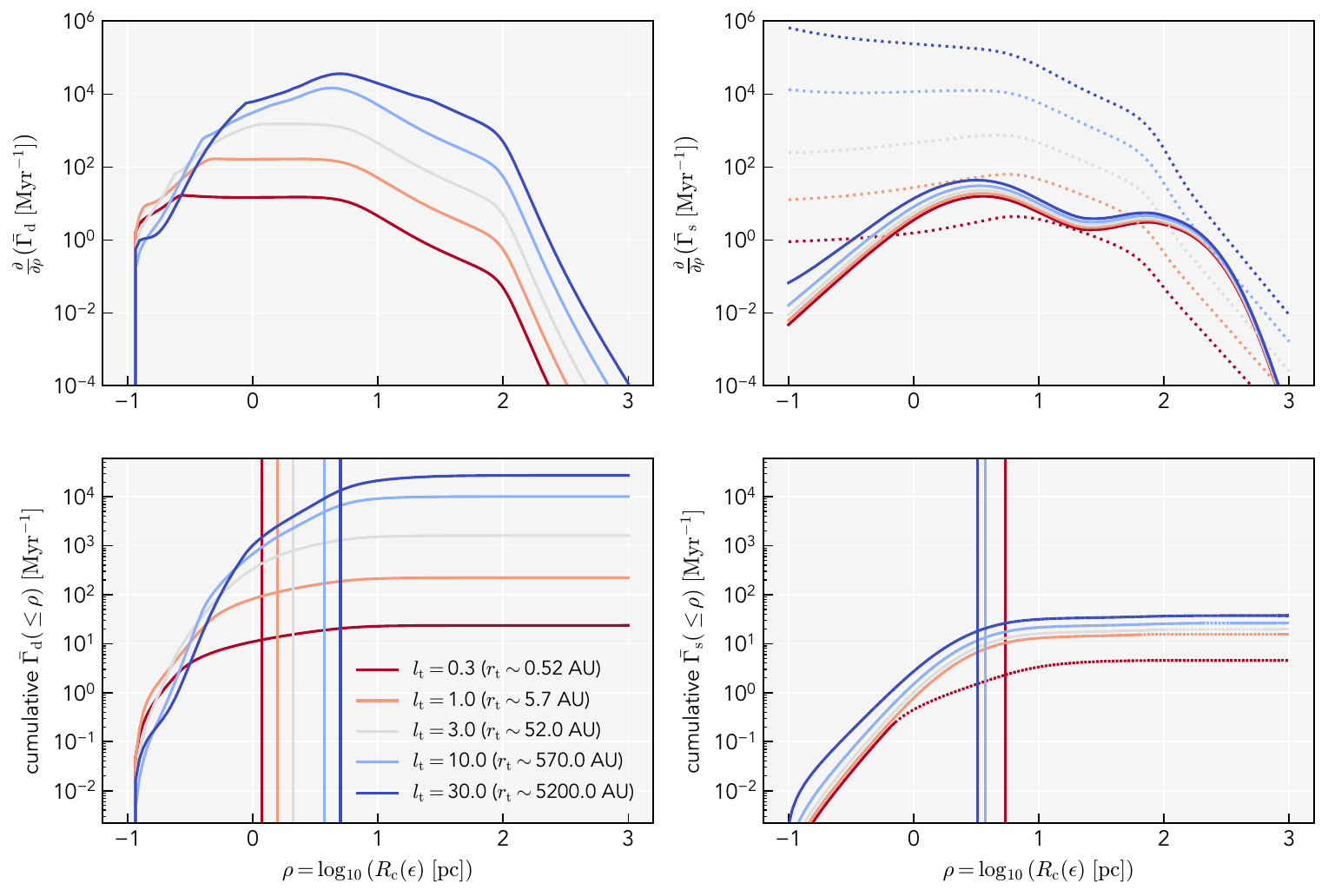}
    \caption{The contribution to the relative rate of disruption (top panels) and cumulative relative rates (bottom panels) for disruptions by diving orbits (left panels) and scatterings (right panels) - for systems of a given $l_{\rm t}$ (in units of [pc$^{2}$ Myr$^{-1}$]). We also give the approximate tidal radius ($r_{\rm t} \sim l_{\rm t}^2(2 G M_{\rm BH})^{-1}$ where $M_{\rm BH}=4\cdot 10^6 M_\odot$). For the scattering calculations we show the empty loss-cone rate (solid line) and the full loss-cone rate (dotted). We take the true rate of disruptions by scattering to be whichever is smaller, and in the cumulative rate the solid or dotted regions correspond to which rate is relevant in that region. Vertical lines show the $\rho$ which contributes half of the relative rate (giving a rough idea of where we expect disrupted systems to originate from).}
    \label{fig:rates_full}
\end{figure*}

In Figure \ref{fig:rates_full}, we now compare the total contribution from scattering to that from diving orbits. We see that diving orbits have a higher relative rate of disruptions, marginally so for low $l_{\rm t}$ but boosted by more than an order of magnitude for $l_{\rm t} \gtrsim 1$.

To give an intuitive explanation for this we can evaluate both rates under the assumption of isotropy: $n(\rho,l)\propto l n(\rho)$ and $n(\rho,l_z) \propto n(\rho)$. The number of stars in the loss cone (integrating up to $l_{\rm t}$) is then $n_{\rm LC, ISO}\sim \lambda_t^2 n(\rho)$ and similarly for the loss wedge $n_{\rm LW, ISO}\sim \lambda_{\rm t} n(\rho)$. The full loss-cone rate under this assumption is approximately
\begin{equation}
\frac{\partial \bar{\Gamma}_{\rm LC, ISO}}{\partial \rho}  \sim \frac{n_{\rm FLC, ISO}}{T_{\rm c} }  = \frac{\lambda^2_{\rm t}}{T_{\rm c} } n(\rho).
\end{equation}
and similarly the full loss-wedge rate is
\begin{equation}
\frac{\partial \bar{\Gamma}_{\rm FLW, ISO}}{\partial \rho}  \sim F_{\rm dive}\frac{n_{\rm LW, ISO}}{t_{ D} }  = \frac{\lambda^2_{\rm t}}{T_{\rm c} F_{\rm phys}} n(\rho),
\end{equation}
i.e. the two are essentially the same (as $F_{\rm phys}$ is generally close to 1).
The full regime (in both cases) represents the maximum possible rate, and can only be maintained if the rate of refilling matches or exceeds the rate of depletion.

One difference between the two mechanisms of disruption comes from the fact that $t_{\rm D}$ is long compared to $t_{\mathrm{rel},l_z}$ down to small radii, and thus the full loss wedge extends to more systems (lower $\rho$) than the full loss cone. Another way to state this is that both mechanisms are able to deplete systems at the same rate, but the loss wedge represents a larger volume and surface are of parameter space and thus is refilled quicker.

The second difference comes from deviations from isotropy. This is most clearly seen in the power law dependence of $n(\rho,l)$ on $l$, with a tangential bias (fewer radial orbits) leading to a steeper power law and thus potentially $n_{\rm LC}\ll n_{\rm LC,ISO}$. A tangential bias is seen in our GC model (see section \ref{sec:flc}) and confirmed in recent observations of the GC \citep{FK25} and has been explored in other contexts as a way to reduce the full loss-cone TDE rate \citep{Lezhnin15, Teboul24}. Anisotropy affects the loss wedge less strongly as $n(\rho,\lambda_z)$ is normally not a strong function\footnote{One exception to this is our NSD model at larger $\rho$ (see figure \ref{fig:pNSD_rho_lambdaz}), where the disk-like nature of the DF leads to the majority of orbits having $l_z \sim l_{\rm c}$, though at such large $\rho$ ($\gtrsim 2$) these systems would contribute little to our total rates even if the DF was isotropic.} of $\lambda_z$ near $|\lambda_z|\sim 0$.

It is also interesting to compare our scattering rates to previous work. We predict a substantially lower overall rate for scattering-driven TDEs, $10^{-5}$ yr$^{-1}$, than commonly reported values of order $10^{-4}$ yr$^{-1}$ (see for example \citealt{Generozov18}). We also see that we are only in the full loss-cone regime for small $l_{\rm t}$ and intermediate $\rho$, whilst it's normally assumed that at large $\rho$ we should always be in the full loss-cone regime. Both of these differences stem mostly from the anisotropy of our measured $n(\rho,l)$ distributions. The fact that we are not in the full loss-cone regime at larger $\rho$ also comes from our neglecting scatterings from bulge stars (which should dominate at $\rho \gtrsim 1$) though these should have little impact on the overall rate.

\subsubsection{Massive perturbers}
\label{sec:mps}

We consider one more effect that could change the balance of relative rates. As argued in \citet{Perets07}, massive perturbers such as Giant Molecular Clouds (GMCs) may actually dominate the rate of scattering (which scales according to $\langle n m^2 \rangle$) at large radii, decreasing relaxation times by potentially orders of magnitude, and increasing disruption rates by shifting regions where the loss cone is empty into the full regime.

The transition point between the empty and full loss-cone regimes depends on $l_{\rm t}$, with larger values leading to a 
larger region where the empty loss-cone regime dominates. As shown in \citet{Perets07} and figure \ref{fig:rates_full}, including these massive perturbers thus has little effect on the disruption of individual stars (with small $l_{\rm t}$), but can significantly boost collisional disruption rates for giants and binary systems.


GMCs can reside in the NSD, but not in the NSC, as tidal forces would tear these relatively dilute self-gravitating bodies apart.  We incorporate this into our models by applying a significant decrease to the relaxation time around and beyond 10 pc via
\begin{equation}
\label{eq:trel_mps}
t'_{\mathrm{rel},\epsilon} = t_{\mathrm{rel},\epsilon}\left(1+\frac{100}{1+e^{5(1-\rho)}}\right)^{-1}
\end{equation}
which starts to deviate from $t_{\mathrm{rel},\epsilon}$ at about a few pc and reaches a constant value of $0.01\ t_{\mathrm{rel},\epsilon}$ for all radii above a few 10's of pc.  We note that this is a relatively optimistic choice for GMCs, as a tidal disruption radius of 10 pc corresponds to a GMC density of $\sim 10^4$ particles ${\rm cm}^{-3}$, well above the typical range in the galaxy (though compatible with the highest density GMCs, which indeed are observed in the GC).  

\begin{figure*}
\includegraphics[width=0.98\textwidth]{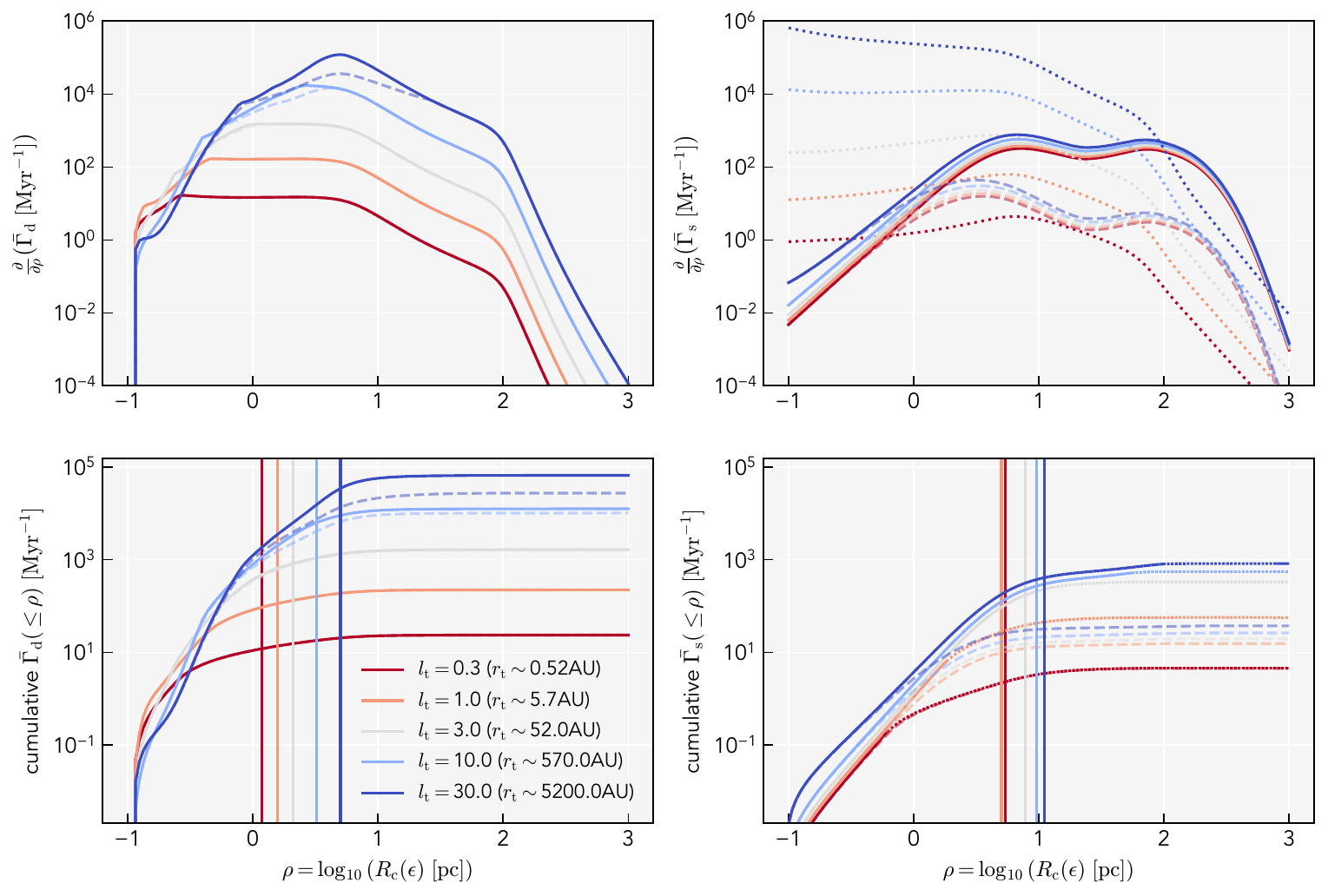}
    \caption{The equivalent of figure \ref{fig:rates_full} but now including a simple prescription for the impact on the relaxation time of massive perturbers (such as GMCs in the NSD). The previous rates (from figure \ref{fig:rates_full}) are shown with dashed lines (the full loss-cone rates are unchanged). The diving rates have been increased by a factor of a few for larger $l_{\rm t}$ ($\gtrsim$ 10 pc$^{2}$ Myr$^{-1}$) while the scattering rates have increased by orders of magnitude for all but the smallest $l_{\rm t}$ shown.}
    \label{fig:rates_full_mps}
\end{figure*}

The equivalent version of figure \ref{fig:rates_full} with the GMC-augmented relaxation rate is shown in figure \ref{fig:rates_full_mps}. The rates from diving orbits are only slightly enhanced, and only for high $l_{\rm t}$ systems. At lower $l_{\rm t}$, the empty loss region does not extend to large enough radii to be affected by the change. At larger $l_{\rm t}$ the rates for disruption by scatterings are markedly increased. The increased scattering rate enlarges the range of radii where the full (and limiting) loss-cone rate is the relevant one. Total rates are now boosted by an order of magnitude or more, and the median radii from which disrupted systems originate is now significantly larger (of order 10 pc or greater).

The relative rate from scattering is still lower, though almost comparable for $l_{\rm t} = 1$ pc$^2$ Myr$^{-1}$. This suggests that for the smallest systems (main-sequence stars or very tight binaries) the nuances brushed over by our simplified estimates, especially the survivability of such systems at small galactic radii, may be important in making any robust quantitative statement about which rate dominates. However for larger systems, e.g. binaries, our previous qualitative conclusions are still likely robust: chaotic diving orbits likely dominate the rate of disruptions, possibly giving orders of magnitude more disruption events.


As a caveat, we note that in this calculation, 
our use of standard relaxation time formulae implicitly assumed that the GMCs are moving isotropically. 
In reality, observed GC GMCs reside mostly in the disk and corotate with it. This configuration may reduce the effectiveness of GMCs at refilling depleted parameter space (since isotropic scatterings in their frame will be tangentially biased in the GC frame). For the comparison in this section, we did not attempt to include this nuance, and only compared to the nominal effect as discussed in \citet{Perets07}.  If indeed it is a disk-like distribution of GMCs that dominate collisional loss-cone repopulation, this may introduce an injection anisotropy into the hypervelocity star population, though of a different nature than the collisionless anisotropy we explored previously. Such collisional injection anisotropies were first found in \citet{Hamers17}, where nuclear spiral arms were considered as the dominant massive perturber.

\begin{figure}
\includegraphics[width=0.98\columnwidth]{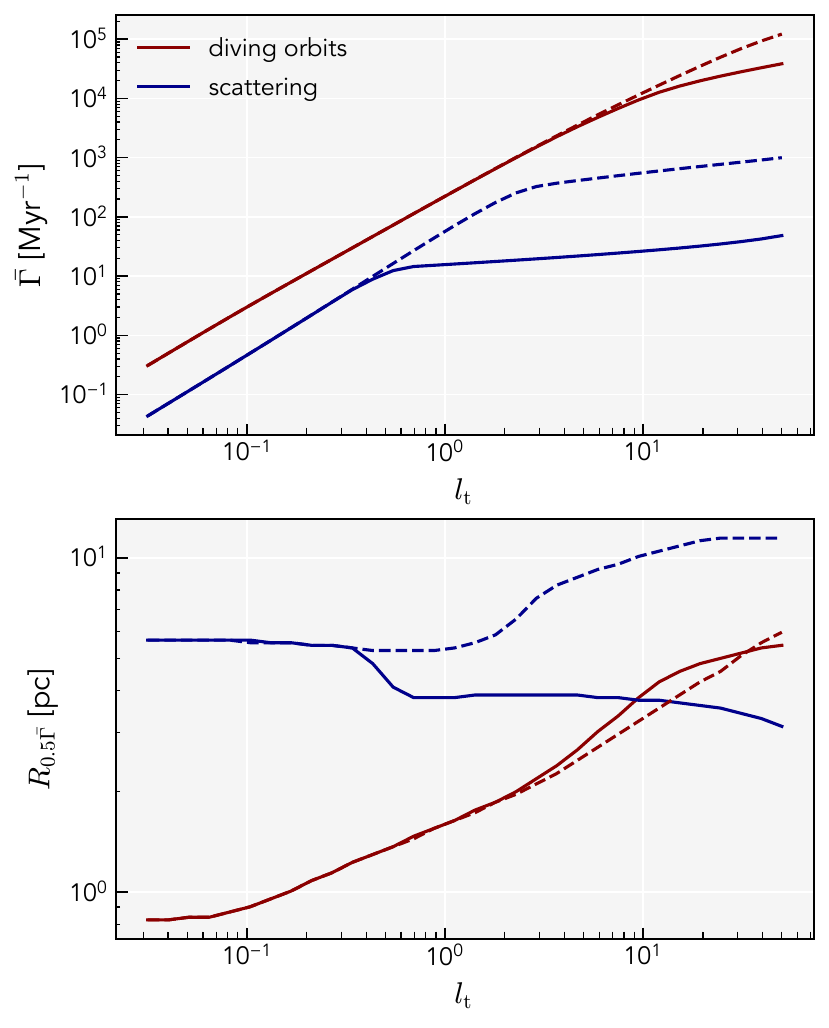}
    \caption{The total relative rate of disruptions, $\bar{\Gamma}$, for an idealized population composed entirely of systems with a given tidal angular momentum $l_{\rm t}$. We show both the implied rate from diving orbits (the collisionless channel) and 2-body scatterings (the collisional channel). The bottom panel shows the $R_{\rm c}(\epsilon)$ below which half of the total rate is produced. In both panels the dashed lines show the equivalent results for the massive perturber model.}
    \label{fig:total_rates}
\end{figure}

\subsubsection{Comparing across $l_{\rm t}$}
\label{sec:compare}

Finally, in Figure \ref{fig:total_rates}, we show the relative rates and the characteristic $R_{\rm c}$ from which disruptions occur, as functions of $l_{\rm t}$. In all cases the collisionless channel dominates, by a factor of a few for low $l_{\rm t}$ up to orders of magnitude for high $l_{\rm t}$. This remains true with the inclusion of massive perturbers, though the range of $l_{\rm t}$ over which the rates are comparable is increased.

The collisionless channel is dominated by disruptions of systems from lower $\rho$ (except at $l_{\rm t}>10$). For $l_{\rm t}\lesssim 1$, i.e. values appropriate for tidal disruption of a main sequence star, most disruptions come from within $r_{\rm inf}$ and are thus expected to be marginally elliptical. In contrast, at larger $l_{\rm t}$, appropriate for the disruption of stellar giants or binary systems, most disruptions come from outside of $r_{\rm inf}$ and are expected to be marginally hyperbolic.

One way to understand the different rates from the two mechanism is to compare the form of the full and empty rates, remembering the useful approximate heuristic that the regime we're in is that which gives the lower of the two rates. The full loss rate scales as $\propto \lambda_{\rm t}^2$, and the empty loss-wedge rate as $\propto \lambda_{\rm t}^0$ whilst the empty loss-cone rate scales much more moderately as $\propto |\ln (\lambda_{\rm t})|^{-1}$ which becomes insensitive to $l_{\rm t}$ as $\lambda_{t}\rightarrow 1$.\footnote{The (non)dependence of the empty loss-wedge rate on $\lambda_t$ can be seen, for example, by expanding out MT99 equation 50 when $q \ (\propto \lambda_{\rm t}^{-2})$ is $\ll1$.} Thus the collisionless rate increases more rapidly with increasing $l_{\rm t}$. We see this in figure \ref{fig:total_rates} between $0.5 \lesssim l_{\rm t} \lesssim 10$ (or, in our massive perturber model, $1 \lesssim  l_{\rm t} \lesssim 50$) where the boost to the empty loss wedge overlaps with significant density of NSC stars.

Whilst we cannot translate these relative rates directly into observable predictions, we can perform a very simple comparison by assuming that to first order the true rate may be $\Gamma \sim f(l_{\rm t})\bar{\Gamma}$, where $f(l_{\rm t})$ is the fraction of systems in the GC with a given $l_t$. For example, if we were to say that the whole GC was composed of 
Sun-like MS stars with $l_{\rm t} \approx 0.3$ pc$^2$ Myr$^{-1}$ (such that $f(0.3)\sim 1$), we would predict a total TDE rate of a few $10^{-5}$ yr$^{-1}$.  As mentioned previously, this rate is significantly below past (collisional) TDE rate estimates made assuming isotropic velocities \citep{Generozov18}; we find a far lower collisional rate due to tangential bias in the NSC, and this is only partially made up for by the higher collisionless rates.  

We can also use known constraints on observed disruption rates to give rough estimates of $f(l_{\rm t})$. For example we can use the rate of binary disruptions implied by observed HVS and S-stars from \citet{Verberne25} of $10^{-5}$ yr$^{-1}$. If we were to approximate all these systems as having come from 0.5 AU binaries ($l_{\rm t} \sim 3$ pc$^2$ Myr$^{-1}$) the relative rate we calculate would imply $f(3) \sim 10^{-2}$.

These simple approximations could be further improved by integrating $f(l_{\rm t})$ over a range of $l_{\rm t}$. However, as detailed at the start of section \ref{sec:rates} a full calculation of expected rates requires taking into account the survival probability of a given system as a function of its properties and environment, a calculation we will save for a follow up work, Penoyre et al. 2025 (in prep.).

\subsection{Comparing to Magorrian and Tremaine 1999}
\label{sec:MT99}

Our analysis is partly informed by the calculations presented in \citet{Magorrian99} (MT99), especially their Section 4, where they assess the rate of disruptions caused by precession of orbits in axisymmetric potentials. Our results broadly agree, though there are some differences to the methodology and hence the results which we thought it valuable to discuss in detail here. 

Firstly, whilst MT99 also produce and draw inferences from a SoS, they only examine the state at apoapse. Unlike the SoSs used in our work, which we calculate from direct orbit integration, theirs comes from a discrete, orbit-averaged (Chirikov-like) map. This follows the method\footnote{Originally this method was developed for slightly triaxial systems to model precession in the azimuthal angle, but MT99 apply it to the polar angle in the axisymmetric case.} from \citet{Touma97} to calculate the successive $\theta_{\rm a}, \ l_{\theta,\rm a}$ values. We have compared the two approaches (direct integration versus the orbit-averaged discrete map) and found that while the MT99 approach works well in the limit of perturbative axisymmetry, it breaks down for non-perturbative deviations from sphericity. The orbit-averaged method assumes that the variation in state between successive apoapses are small, implicitly relying on the assumption of mild axisymmetry, which is not satisfied in our GC model. They also consider regular orbits precessing into the loss wedge, though in our experiments we find that any trajectory that comes close enough to be disrupted is chaotic. 

They find a critical value of $l_\theta$ below which orbits are assumed to be chaotic, which they term $J_{\rm l}$. To do this they perform their stepwise method on a trajectory starting from $l_{\theta,\rm a}=0$ and a range of initial $\theta_{\rm a}$. They then average over all the trajectories and take $J_{\rm l}=2\langle l_{\theta,\rm a} \rangle$. We do not find the equivalent of $J_{\rm l}$ directly, but note that it should be of order $F_{\rm dive}$ with some bias depending on the particular shape of the chaotic sea. Looking at figure \ref{fig:bigmap} we can see two limitations of this approach - firstly that at apoapse not all $l_{\theta, \rm a}=0$ trajectories are chaotic, and secondly that the distribution of $l_{\theta, \rm p}$ (marginalised over $\theta$) does not in general match the distribution of $l_{\theta, \rm a}$. In general we would expect the differences between this part of their method and ours to be small, with $J_{\rm l}\sim F_{\rm dive}$ and both being used equivalently (comparing their equation 42 with our equation \ref{eq:p_D}).

Our relative rates agree well with their calculation at low $l_{\rm t}$, appropriate for MS stars and hence TDEs, showing collisionless rates a factor of $\sim 2$ times larger than the collisional. The difference between the two rates is slightly larger in our results ($\sim$ a factor 10) because of the measured anisotropies of the DF, but otherwise consistent. However they do not explore larger values of $l_{\rm t}$, appropriate for the disruption of multi-body systems (and some giant stars) and it is here where the collisionless rate can dominate over the collisional one by 2 or 3 orders of magnitude (as discussed in section \ref{sec:compare}).

\section{Discussion and Conclusions}
\label{sec:discuss}

In this paper we have quantified the behaviour of low 
angular momentum orbits in the axisymmetric potential of the inner Milky Way, focusing especially on those orbits with small values of $l_z$ (the conserved component of angular momentum). We have shown that chaotic trajectories are common, accounting for around 30\% of the total parameter space at low $l_z$. These chaotic trajectories can, and generally do, bring stars and binaries as close to the central MBH during extreme periapse passages as allowed by their fixed $l_z$. While some regular trajectories experience substantial angular momentum oscillations, we find that all of the trajectories that reach their minimum permitted pericenter are chaotic (this is in contrast to the situation deep inside the sphere of influence, where deeply plunging trajectories are regular; \citealt{Vasiliev13}).

Referring to these trajectories as 'diving orbits,' we calculated the associated timescales and properties of the disruptions that they can produce, accounting for both TDEs and hypervelocity stars by parameterizing disruptions with the tidal angular momentum $l_{\rm t}$. For systems with energies corresponding to a circular orbit of $\lesssim$10 pc, a disruptively close periapse passage can be expected over timescales of 10-100 Myr or less. These disruptions strongly prefer orbits with very small $z$-angular momentum, only a fraction of $l_{\rm t}$. This implies that these encounters, which are generally near-parabolic, preferentially approach the MBH from a direction close to perpendicular to the disk plane. This further implies that any post disruption remnant (for example a hypervelocity star from a disrupted binary, or a debris stream from a tidally disrupted star) will preferentially be ejected towards the galactic poles. At the relevant $l_{\rm t}$ for TDEs encounters come from within the sphere of influence and will thus be mildly elliptical, while for the disruption of binaries the inverse is true and the encounters will be mildly hyperbolic.

We can call disruptions by diving orbits  '\textit{collisionless}', in contrast with the '\textit{collisional}' disruptions of loss-cone repopulation predicted from two-body scattering calculations. We compared the collisionless and collisional \textit{relative} rates for our GC model, under the numerically simplifying assumption that all systems in the model have a single tidal angular momentum, $l_{\rm t}$ (i.e. the same tidal radius). We found that collisionless relative rates dominate for high $l_{\rm t}$ systems (e.g. massive evolved stars and all but the tightest binaries), often by one or more orders of magnitude.  
For moderate $l_{\rm t}$ (e.g. tidal disruption of a single main sequence star), the collisionless relative rates still dominate, but now only by a factor of a few, consistent with previous studies of the TDE ``loss wedge'' \citep{Magorrian99, Vasiliev13}. 

These results are robust whether or not the origin of angular momentum diffusion in the collisional channel is star-star scattering or star-GMC scattering. In agreement with past work \citep{Perets07}, we find that introducing a population of massive perturbers such as GMCs can easily increase the collisional disruption rate for binaries by an order of magnitude.  However, because these massive perturbers are themselves vulnerable to tidal disruption, they cannot increase relaxation rates below a certain minimum galactocentric radius.  In our model of the Milky Way GC, the bulk of the collisionless rate comes the NSC (where orbital timescales are much lower) rather than the NSD, and thus from scales of a few pc, well inside the tidal disruption radius for GMCs.  

Though we focus on the Milky Way, these calculations can be extended to any axisymmetric model for the central region of other galaxies. In particular, we show (Section \ref{ap:axisratio}) that the dominant variable determining whether diving orbits are common (and thus whether they significantly contribute to disruption rates) is the degree of flattening of the innermost stellar component. Even modestly flattened systems still display diving behaviour and thus we expect this process to be ubiquitous across a wide range of galaxies.

Though not shown here, over the course of our investigation we have also experimented with similar potentials where other individual components are removed or altered (for example removing the bulge, or central MBH). In all cases the general qualitative analysis and conclusions of the following work are unchanged, and quantitative changes are modest (at most of order a few) across the regions of interest. The largest difference is seen in models without a central MBH, which still show equivalent behaviours but with regular orbits now possible for $l_{\theta, \rm p}=0$ - suppressing but not eliminating diving orbits. We chose our galactic center model to represent observed reality as closely as we could, but note that many of the results are likely generic to any similar combinations of disk and spheroidal potentials, not some particular quirk of our own galaxy's center.

There are a few caveats and limitations to our work that we recapitulate here:
\begin{itemize}
\item {\it The survival of high $l_{\rm t}$ systems at small radii}: our calculations have focused on fixed values of $l_{\rm t}$ without acknowledging the number of systems of a given $l_{\rm t}$ is dependent on energy (and thus radii).
We might expect softer binaries (larger $l_{\rm t}$) to be collisionally ionized or while harder binaries may be driven to merger in the dense and dynamically hot inner GC, suppressing collisional and collisionless disruption rates. The peak of the rates for the collisionless mode is at smaller radii than the collisional mode, and thus if the binaries of interest were strongly suppressed on scales of a few parsecs, the collisionless rate could be significantly diminished.
\item {\it Stellar lifetimes vs the timescales of collisionless disruptions}: the collisionless mode has a long associated timescale (up to 100 Myrs for systems of interest) but acts on a much larger population of systems. Comparatively the collisionless mode acts on a small population but with a shorter timescale. In balance the collisionless mode seems to dominate, but if we are interested in short-lived systems (such as B-type stars which currently are the most frequent HVS candidates \citealt{Brown18}) they may not survive long enough and again the collisional rate may be diminished. Similarly ongoing star formation of short lived stars may bias the DF, for example if they are mostly formed on disk-like circular orbits and need time to relax to the overall DF.
\item {\it Deviations from axisymmetry}: while an axisymmetric model for the GC is more realistic than a spherical one, the inner parts of the MW likely have at least some triaxiality \citep{FeldmeierKrause17}, which may further increase the importance of the collisionless channel for disruption \citep{Merritt04}.
\item {\it Massive perturber populations}: in this paper we have considered quite idealized models of massive perturber distributions, and in principle the collisional channel could regain dominance over collisionless binary disruptions if massive perturber populations extended to smaller radii.  However, our choice of $\approx 10$ pc for the innermost GMCs is ultimately generous to the collisional channel, as this represents the tidal disruption radius for a quite extreme GMC ($\sim 10^4$ particles ${\rm cm}^{-3}$).  The crudeness of our GMC model matters less at large radii, where the most interesting loss-cone fluxes (for e.g. binaries tight enough to produce hypervelocity stars) are already ``maxed out'' at the full loss-cone rate.  Considering other massive perturbers, we note that the innermost star clusters in the MW (Arches and Quintuplet) have projected radii of $\approx 30$ pc and are thus unlikely to be important.
\item {\it Accuracy of observed models}: the collisionless rates we calculate are dominated by, and thus sensitive to, the population of NSC stars with $R_{\rm c}\sim {\rm few} \times {\rm pc}$ at $l_z \sim 0$. While we our chosen DF is reasonably well motivated from the most recent available data, it is important to note that our results are very sensitive to the anisotropy of the inner GC.
\end{itemize}
The first two of these caveats stymie the translation of robust conclusions from our relative rates to actual expected rates of disruptions. We will address these in an upcoming work (Penoyre et al. 2025, in prep.), incorporating a treatment for the population and survivability of stars and binaries as a function of age, energy and environment.




Our conclusions on TDE rates are in qualitative agreement with past investigations of the axisymmetric loss wedge \citep{Magorrian99, Vasiliev13}, finding a typically modest (factor of $\approx 2-3$) enhancement to TDE rates.  In contrast to these works, we have also investigated binary disruptions, for which the collisionless mode can produce much greater enhancements over the collisional mode.  We have also 
found a different scaling of $p(\beta)$, (equation \ref{eq:p_beta}), which emerges from the fact that we consider the state of the system at periapse rather than averaged over the whole orbit.

In summary, analyzing the orbital dynamics in a state-of-the-art model for our Galactic Center suggests that disruptions in the Milky Way are primarily collisionless, not driven by star-star or star-GMC scatterings.  Applying this result to the Hills mechanism, the immediate implication is a novel injection anisotropy, which should preferentially funnel hypervelocity stars into the direction orthogonal to the NSC/NSD equatorial plane.  Such anisotropy is important both in estimating the completeness of hypervelocity star samples but also in attempting to apply hypervelocity star populations to measure the large-scale geometry of the MW potential.  Applied to TDEs, our result suggests that asphericity of extragalactic NSCs may be an important ingredient in TDE rate calculations, potentially increasing TDE rates by a factor of a few  and also altering their $\beta$ distribution.

\section*{Data Availability}
The data underlying this article will be shared on reasonable request to the corresponding author.

\section*{Acknowledgements}
We first and foremost would like to thank Eugene Vasiliev, for their expertise and for producing and advising on the \texttt{AGAMA} code and its many relevant uses in this work. We thank the referee for their insights and comments. We also thank Mattia Sormani for discussions, references, and detailed explanations of the GC model implemented in \texttt{AGAMA}.  We also thank Re'em Sari for multiple insightful discussions. We would also like to thank John Magorrian and Em Sandford for providing useful comments on the work. ZP and EMR acknowledge
support from European Research Council (ERC) grant number:
101002511/project acronym: VEGA P.  NCS gratefully acknowledges support from the Binational Science Foundation (grant Nos. 2019772 and 2020397) and the Israel Science Foundation (Individual Research Grant Nos. 2565/19 and 2414/23).

\bibliographystyle{mnras}
\bibliography{bib}
\bsp

\appendix

\section{Estimating probabilities from distribution functions}
\label{ap:n}

For our calculations we need to know the probability (or equivalently the number density) of orbits with a given energy (characterised by $\rho=\log_{10}(R_{\rm c}(\epsilon) \ \mathrm{[pc]})$) and angular momentum.

In this section we'll work in terms of
\begin{equation}
\lambda_z = \frac{l_z}{l_{\rm c}}
\end{equation}
where $l_{\rm c}(\rho)$ is the circular (maximum) angular momentum of an orbit with the corresponding energy. We will similarly use $\lambda=l/l_{\rm c}$ where appropriate.

We seek to find $p(\rho,\lambda_j)$ (or a transformable equivalent) from our pre-defined distribution functions. As these DFs are expressed in terms of different parameters (the radial, vertical and azimuthal actions, which themselves are not conserved over an orbit in our axisymmetric potential) we employ a relatively brute force method here. 

We use \texttt{AGAMA}'s functionality to sample random $(\mathbf{x},\mathbf{v})$ pairs from the DF and from these calculate a distribution of $\rho$, $\lambda_z$ and $\lambda$. We do this for both the NSC and NSD with 1,000,000 random draws each. We then approximate the functions we need numerically.

\subsection{Useful functions}

There is a useful analytic functions that we will make use of throughout this section for numerically approximating our probabilities - the integral of a series of logistic functions.

If we have a curve which has $N$ segments with relatively constant gradient, and smooth transitions between them, we can write the gradient as
\begin{equation}
\frac{dy}{dx}=\sum_{i=0}^N \frac{\alpha_{i+1}-\alpha_i}{1+e^{-k_i(x-x_i)}}
\end{equation}
where $\alpha_0=0$ and $x_0=-\infty$. Here $\alpha_i$ is the gradient of the $i^{th}$ segment, $x_i$ is the location of the transition to the next gradient, and $k_i$ (>0) dictates the sharpness of the transition.

Integrating this gives
\begin{equation}
\label{eq:sigint}
y=\kappa + \alpha_1 x +\sum_{i=1}^N \frac{\alpha_{i+1}-\alpha_i}{k_i}\ln(1+e^{k_i(x-x_i)}),
\end{equation}
a very versatile function of relatively few parameters that can be tuned to fit most continuos curves.

We will use versions of this function with $N=2$ or $3$. We find a good fit by eye, starting by taking $k_i$ to be large (to give sharp transitions) and approximating $\alpha_i$ and $x_i$. Under the assumption that $k_i (x_{i+1}-x_i), k_i (x_{i}-x_{i-1}) \gg 1$ (i.e. that the transitions are sharp) we can use the approximations
\begin{equation}
y_j =y(x_j) \approx \kappa + \alpha_1 x + \sum_{i=1}^{j-1} (\alpha_{i+1}-\alpha_i) (x_{i+1}-x_i) + \frac{\alpha_{j+1}-\alpha_{j}}{k_j}\ln(2).
\end{equation}
This expression can be inverted to find approximations of $k_j(y_j,\kappa)$ (bearing in mind $y_j$ can easily be read from the data) meaning that we only need to find a value of $\kappa$ that fits well.



\begin{table}
\centering
\begin{tabular}{ c|cccccccc }
& $\alpha_1$ & $\alpha_2$ & $\alpha_3$ & $x_1$ & $x_2$ & $k_1$ & $k_2$ & $\kappa$ \\
\hline
\multicolumn{9}{l}{$p_{NSD}(\rho)=10^{a(\rho)}$} \\
\hline
$a$ & 2.4 & -8 & & 2.45 & & 5.5 & & -4.6 \\
\hline
\multicolumn{9}{l}{$p_{NSD}(\rho,\lambda_z)=10^{a(\rho) + b(\rho)\lambda_z}$}\\
\hline
$a$ & 2 & -5 & & 2 & & 20 & & -4.2 \\
$b$ & 1 & 6 & & 1.8 & & 10 & & -1 \\
\hline
\multicolumn{9}{l}{$p_{NSD}(\rho,\lambda)=10^{a(\rho)}\lambda^{b(\rho)}$}\\
\hline
$a$ & 2.6 & 0 & -3.6 & 1.85 & 2.3 & 9.2 & 11& -4.7 \\
$b$ & 0 & 2.5 & & 2.0 & & 10 & & 1 \\
\hline
\multicolumn{9}{l}{$p_{NSC}(\rho)=10^{a(\rho)}$}\\
\hline
$a$ & 2.5 & -0.375 & -2 & 0.78 & 1.9 & 5.8 & 17.7 & -1.85 \\
\hline
\multicolumn{9}{l}{$p_{NSC}(\rho,\lambda_z)=10^{a_\pm(\rho) + b_\pm(\rho)\lambda_z}$}\\
\hline
$a_+$ & 3.5 & 0.25 & -2.5 & 0.8 & 1.95 & 5.9 & 11.8 & -2.95 \\
$b_+$ & -3 & -1 & & 1.5 & & 2.85 & & 4 \\
$a_-$ & 3.5 & -0.25 & -3.5 & 0.8 & 1.95 & 12 & 6.9 & -3.2 \\
$b_-$ & 6 & -2 & & 1.15 & & 8.5 & & -3 \\
\hline
\multicolumn{9}{l}{$p_{NSC}(\rho,\lambda)=10^{a(\rho)}\lambda^{b(\rho)}$}\\
\hline
$a$ & 2.3 & -0.25 & -2.5 & 0.8 & 2 & 8.4 & 6.4 & -1.4 \\
$b$ & -0.8 & -0.2 & & 1.1 & & 3 & & 1.8 \\
\end{tabular}
\caption{Parameters for fits of the form given in equation \ref{eq:sigint} to the distributions of particles implied by our NSD and NSC potentials and distribution functions.}
\label{tab:dffits}
\end{table}

\subsection{NSD}

\begin{figure}
\includegraphics[width=0.95\columnwidth]{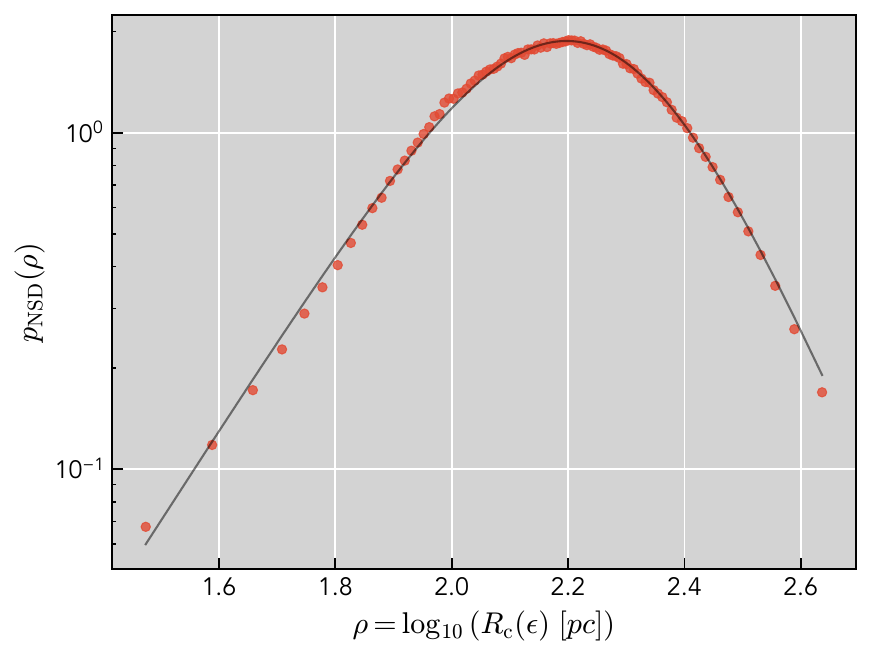}
    \caption{The probability of a system in the NSD having a particular $\rho=\log_{10}(R_{\rm c}(\epsilon) \ [\mathrm{pc}])$. Red points show the distribution of 1,000,000 systems sampled for the DF and potential using AGAMA (in bins of equal number). Black line shows the simple approximating function (following equation \ref{eq:sigint}).}
    \label{fig:pNSD_rho}
\end{figure}

We can start with $p_{NSD}(\rho)$ (i.e. the distribution of energy marginalised over angular momentum) as shown in figure \ref{fig:pNSD_rho}.

We use bins containing equal numbers of systems and thus varying widths. The probability of a source having a given $\rho$ is the number in that bin, divided by the width of the bin and the total number of samples.

The resulting distribution is shown in red. This has a relatively simple form and we approximate it directly as following equation \ref{eq:sigint} with $N=2$ (with parameters reported in table \ref{tab:dffits}).

\begin{figure*}
\includegraphics[width=\textwidth]{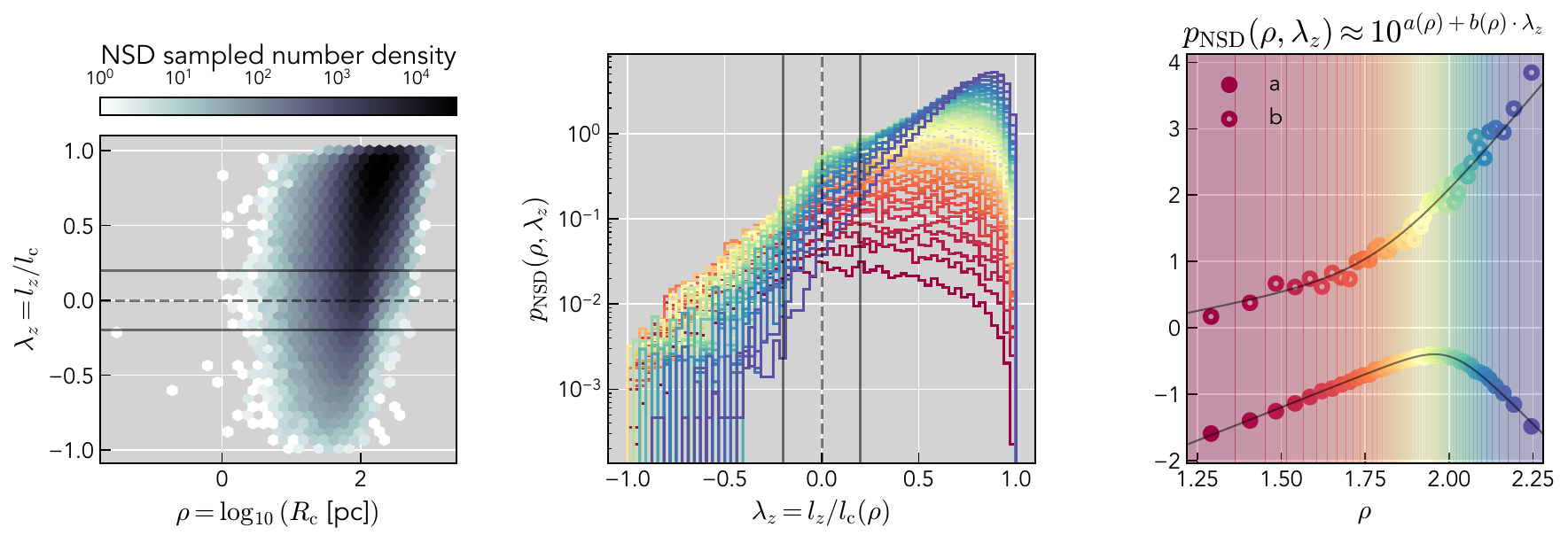}
    \caption{The two dimensional equivalent of figure \ref{fig:pNSD_rho} in $\rho$ and $\lambda_z = l_z/l$. Left panel shows the distribution of sampled points. Middle panel shows the distribution in $\lambda_z$ in narrow bins of $\rho$ (containing equal numbers of systems). Right panel shows the numerical fits to these distributions (filled and empty points) and the approximate fitting functions in black. The fitting is performed over the regions between the solid lines of constant $\lambda_z$ in the first two panels, separated into two regions by the dashed line. The regions which define the colours used are shown in the third panel.}
    \label{fig:pNSD_rho_lambdaz}
\end{figure*}

We then move to the joint distribution of $\rho$ and $\lambda_z$ - noting that for $l_{\rm t} \ll l_{\rm c}$ (as will be true for essentially all systems of interest in this paper) we only need our approximation to be robust at low $|\lambda_z|$. Looking at the distribution of our 1,000,000 samples in bins of $d\rho$ and $d\lambda_z$ (left panel and middle panels of figure \ref{fig:pNSD_rho_lambdaz}) it seems sensible to define our estimate of the form
\begin{equation}
\label{eq:pNSDapprox}
p_{NSD}(\rho,\lambda_z) = 10^{a(\rho) + b(\rho)\cdot \lambda_z}
\end{equation}
(i.e. the log of the probability appears approximately linear in $\lambda_z$).

A simple robust estimate for $a$ and $b$ can be found by finding the number of systems between some small $\lambda_0$ and 0, which we'll denote as $N_+$, and similarly the number between 0 and $-\lambda_0$, $N_-$. Integrating equation \ref{eq:pNSDapprox} over these two intervals and rearranging we get
\begin{equation}
b(\rho)=\frac{1}{\lambda_0} \log_{10}\left(\frac{N_+}{N_-}\right)
\end{equation}
and
\begin{equation}
A(\rho) = 10^{a(\rho)} =\frac{b \ln(10)}{\Delta \rho}\left(\frac{N_+ N_-}{\bar{N}(N_+ -N_-)}\right)
\end{equation}
where $\bar{N}$ is the total number of samples used (1,000,000).

The values of $a$ and $b$ implied by these relationships, using $\lambda_0 =0.2$, are shown in the right panel of figure \ref{fig:pNSD_rho_lambdaz}. We fit both of these with functions of the form of equation \ref{eq:sigint} with $N=2$ and parameters reported in table \ref{tab:dffits}.

We apply a similar formalism to fitting for the distribution of $\rho$ and $l$, with a slight modification required by the constraint that $l \ge 0$ always.

\begin{figure*}
\includegraphics[width=\textwidth]{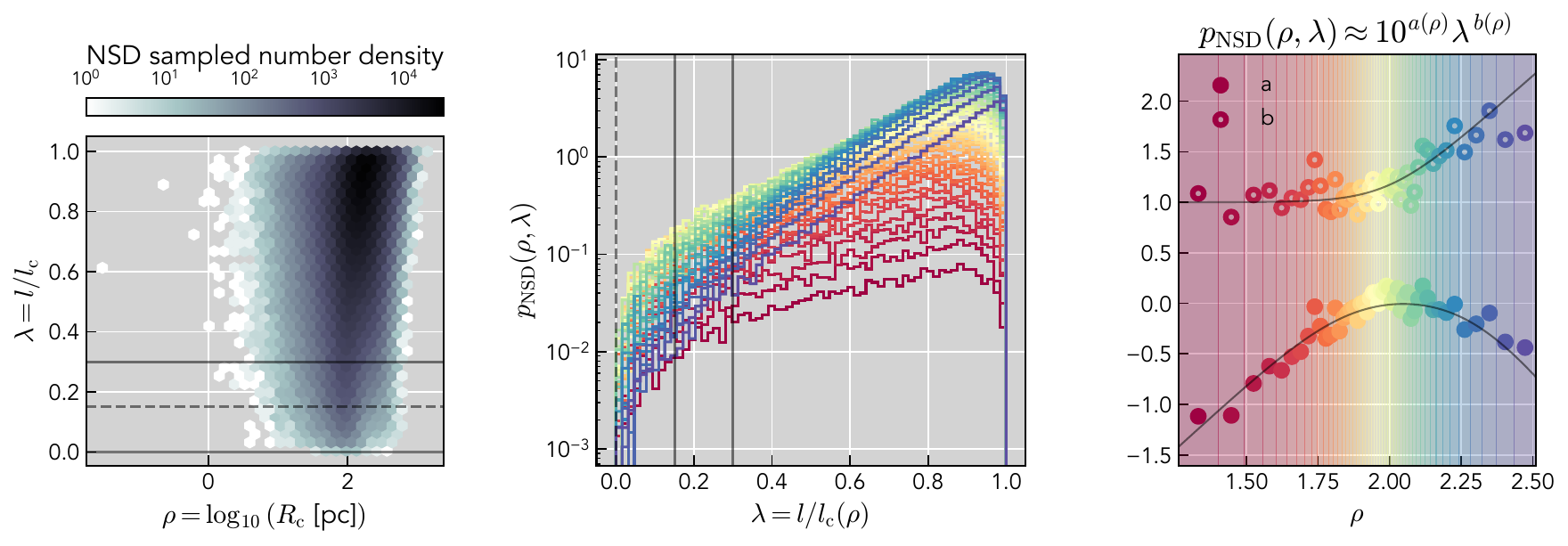}
    \caption{Similar to figure \ref{fig:pNSD_rho_lambdaz} for $\lambda$ (which is always $\ge 0$) instead of $\lambda_z$.}
    \label{fig:pNSD_rho_lambda}
\end{figure*}

Examining the distribution (figure \ref{fig:pNSD_rho_lambda}) we suggest a form 
\begin{equation}
\label{eq:pNSD_rho_lambda}
p_{NSD}(\rho,\lambda) = 10^{a(\rho)}\lambda^{b(\rho)}.
\end{equation}
We use a similar method to the previous fit, but now using $N_1 = N(0<\lambda<\lambda_0)$ and $N_2 = N(\lambda_0<\lambda<2\lambda_0)$.

Integrating equation \ref{eq:pNSD_rho_lambda} over these intervals and rearaanging gives
\begin{equation}
b(\rho)=\log_2\left(\frac{N_2}{N_1}+1 \right)-1
\end{equation}
and
\begin{equation}
A(\rho) = 10^{a(\rho)} =\frac{N_1}{\bar{N}}\frac{b+1}{\Delta \rho}\lambda_0^{-(b+1)}
\end{equation}

Again the parameters of the fitted function, using $\lambda_0=0.15$, are reported in table \ref{tab:dffits}.

It is interesting to compare our fitted values of $b$ to the conventional approximation, asumming spherical symmetry, that at low lambda $p(\epsilon,\lambda) \approx p(\epsilon) \cdot \lambda$ (i.e. that $b=1$, see for example \citealt{Merritt13book} equation 5.159). This does seem to be approximately true at low $\rho$ but steepens significantly at larger $\rho$ suggesting a dearth of high energy, low angular momentum orbits implied by our GC model.

\subsection{NSC}

\begin{figure}
\includegraphics[width=0.95\columnwidth]{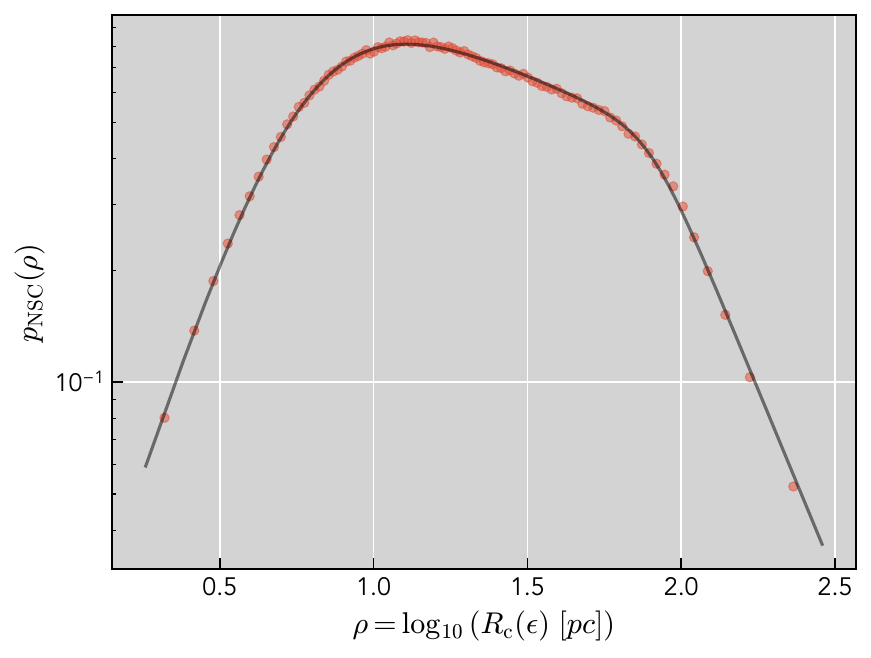}
    \caption{The equivalent of figure \ref{fig:pNSD_rho} for the NSC.}
    \label{fig:pNSC_rho}
\end{figure}

\begin{figure*}
\includegraphics[width=\textwidth]{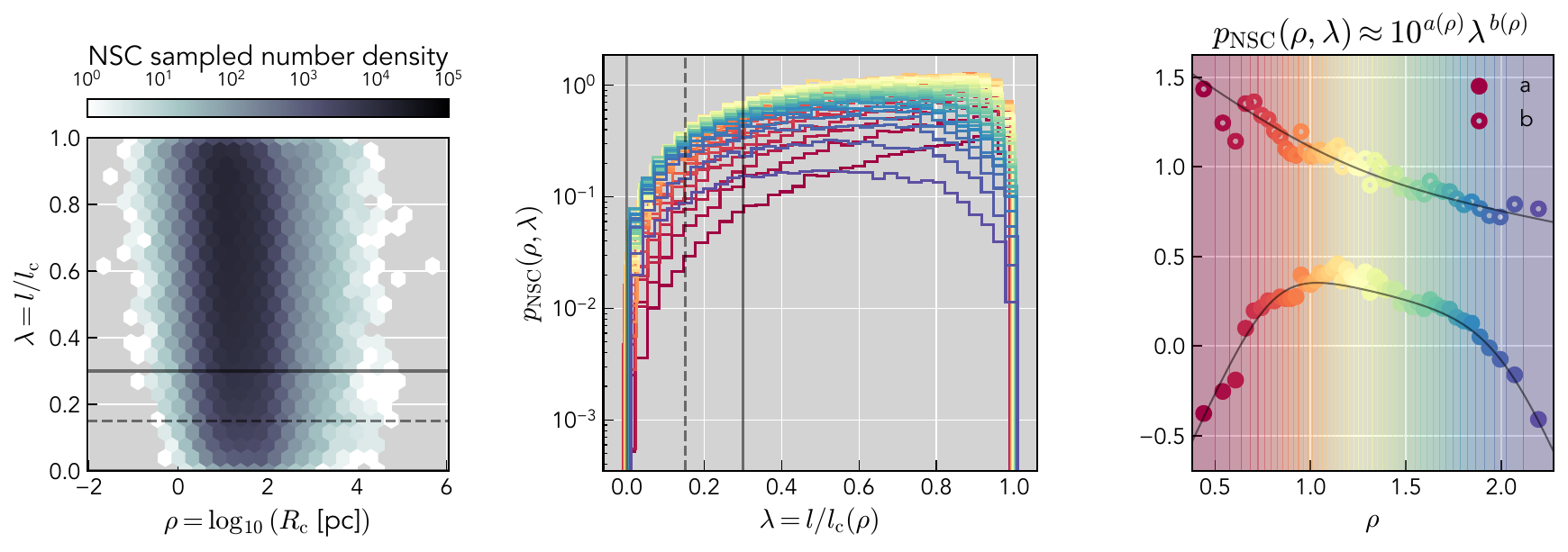}
    \caption{Similar to figure \ref{fig:pNSD_rho_lambda} for the NSC.}
    \label{fig:pNSC_rho_lambda}
\end{figure*}

We can repeat essentially the same procedure for the NSC as was used in the last section for the NSD. One small change is that we tend to have to use more segments ($N=3$) for NSC components. Otherwise the procedure for $p_{NSC}(\rho)$ (figure \ref{fig:pNSC_rho}) and $p_{NSC}(\rho,\lambda)$ (figure \ref{fig:pNSC_rho_lambda}) are essentially identical, with parameters reported in table \ref{tab:dffits}.

It is important to note that unlike the $p_{NSD}(\rho,\lambda)$, where $b\approx 1$ seemed like a reasonable assumption, this is far from true for $p_{NSC}(\rho,\lambda)$ - with the steepness decreasing across all $\rho$ (though still centred near $b=1$). This apparent tangential bias at small radii suggests that the assumption that $p_{NSC}(\epsilon,\lambda) \propto \lambda$ may not be appropriate and more care needs to be taken when using this distribution in calculations of scattering rates.

\begin{figure*}
\includegraphics[width=\textwidth]{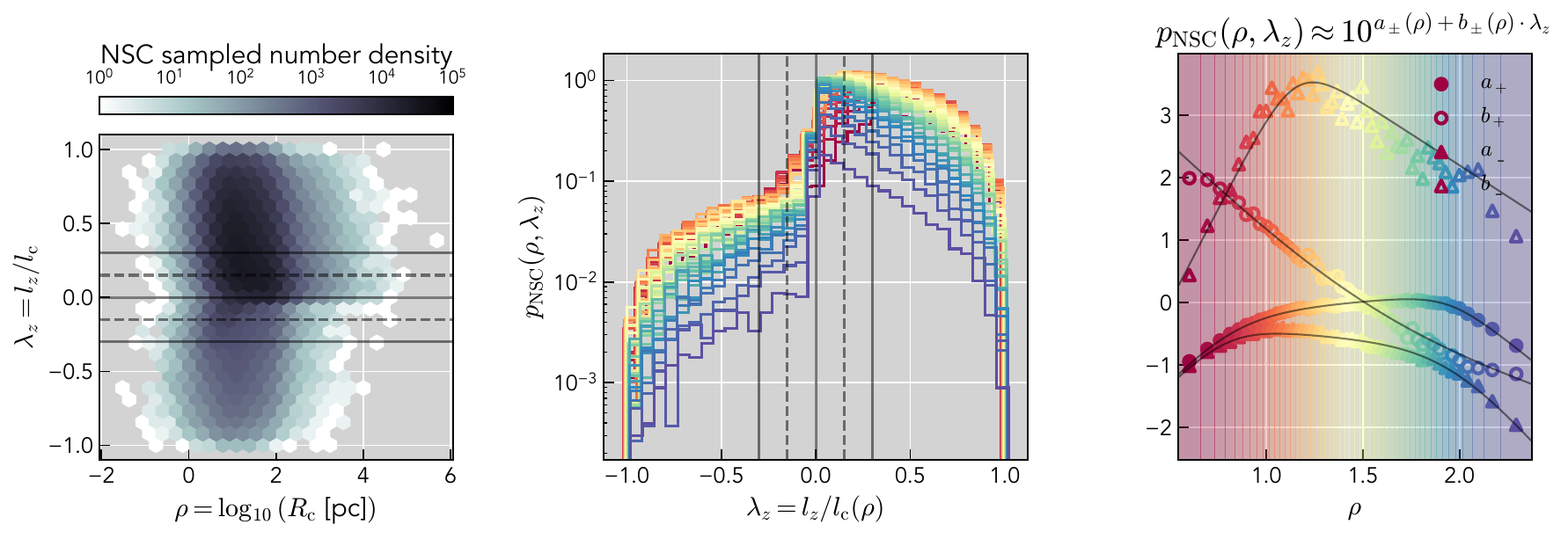}
    \caption{The equivalent of figure \ref{fig:pNSD_rho_lambdaz} for the NSC. The clear break in the behaviour of the distribution over $\lambda_z=0$ motivates a separate fit for positive and negative $\lambda_z$, as shown in the third panel.}
    \label{fig:pNSC_rho_lambdaz}
\end{figure*}

There is a large difference in calculating $p_{NSC}(\rho,\lambda_z)$ - which can be understood with reference to figure \ref{fig:pNSC_rho_lambdaz}. Here we see that the distribution has a sharp break at $\lambda_z=0$, with counter-rotating orbits much less common than corotating orbits.

This obliges us to define two sets of functions ($a_+,\ b_+$ and $a_-,\ b_-$) depending on the sign of $\lambda_z$, i.e.
\begin{equation}
p_{NSC}(\rho,\lambda_z) = 10^{a_\pm(\rho)+b_\pm(\rho)\lambda_z}
\end{equation}
where the form used matches the sign of $\lambda_z$.

To fit these we apply a similar though slightly modified version of the previous calculation, defining $N_1,\ N_2,\ N_{-1}$ and $N_{-2}$ where $N_{\pm n}=N\left((n-1)\lambda_0<\pm \lambda_z< n\lambda_0\right)$.

From these we can derive
\begin{equation}
b_\pm = \frac{\pm\log_{10}\left(\frac{N_{\pm 2}}{N_{\pm 1}}\right)}{\lambda_0}
\end{equation}
and
\begin{equation}
A_\pm=10^{a_\pm} = \pm \frac{b_\pm \ln(10)}{\Delta \rho}\frac{N_{\pm 1}^2}{\bar{N}(N_{\pm 2}-N_{\pm 1})}.
\end{equation}

The parameters for the fits to these coefficients, with $\lambda_0=0.15$, are given in table \ref{tab:dffits}.

\label{lastpage}

\end{document}